\documentclass[12pt]{iopart}
\usepackage{graphics,iopams,mathrsfs}

\makeatletter
\def\@fnsymbol#1{^{\thefootnote}\relax}
\makeatother

\begin{document}

\topical{Combining DFT and Many-Body Methods to Understand
Correlated Materials}

\author{I V Solovyev}

\address{Computational Materials Science Center,
National Institute for Materials Science,
1-2-1 Sengen, Tsukuba, Ibaraki 305-0047, Japan}
\ead{solovyev.igor@nims.go.jp}

\begin{abstract}
The
electronic and magnetic properties of many strongly-correlated
systems are controlled by a limited number of
states, located near the Fermi level and well isolated
from the rest of the spectrum. This opens a formal way for combining
the methods of
first-principles electronic structure calculations,
based on the density-functional theory (DFT), with many-body models,
formulated in the restricted Hilbert space of states close
to the Fermi level. The core of this project is the
so-called ``realistic modeling''
or the construction of the model many-body Hamiltonians
entirely from the first principles.
Such a construction should be able to go beyond the
conventional local-density approximation (LDA),
which typically supplements the density-functional theory,
and incorporate
the physics of Coulomb correlations.
It should also provide a transparent
physical picture for the low-energy properties of strongly correlated
materials.
In this review article, we will outline the basic ideas of such a
realistic modeling, which consists of the following
steps:
(i) The construction of the complete Wannier basis set for the low-energy LDA band;
(ii) The construction of the one-electron part of the model Hamiltonian
in this Wannier basis;
(iii) The calculation of the screened Coulomb interactions for the low-energy bands
by means of the constrained DFT. The most difficult part of this project is the evaluation
of the screening caused by outer bands, which may have the same
(e.g., the transition-metal $3d$) character as the low-energy bands.
The latter part can be efficiently done by combining the
constrained DFT with the
random-phase approximation
for the screened Coulomb interaction.
The entire procedure will be illustrated on the series of examples,
including the distorted transition-metal perovskite oxides,
the compounds with
the inversion symmetry breaking caused by the defects, and the alkali hyperoxide KO$_2$,
which can be regarded as an
analog of strongly-correlated systems
where the localized electrons reside on the \textit{molecular orbitals} of the O$_2^-$ dimer.
In order to illustrate abilities of the realistic modeling,
we will also consider solutions of the obtained low-energy models
for a number of systems, and argue that it can be
used as a powerful tool for the exploration and understanding of
properties of strongly correlated materials.
\end{abstract}


\maketitle

\section{Introduction}

  Many successes of modern condensed-matter physics and chemistry are
related with the development of the density-functional theory (DFT),
which is designed for the exploration of the ground state
properties
of various substances
and based on the minimization
of the total energy functional $E[\rho]$ with respect to the electron density
$\rho$ \cite{HohenbergKohn,KohnSham,ParrYang}. For practical
applications, DFT resorts to iterative solution of one-electron
Kohn-Sham equations
\begin{equation}
\left( -\frac{\hbar^2 }{2m} \nabla^2 + V \right) \psi_i = \varepsilon_i \psi_i,
\label{eqn:KS}
\end{equation}
together with the equation for the electron density
\begin{equation}
\rho = \sum_i f_i |\psi_i|^2,
\label{eqn:rho}
\end{equation}
defined in terms of
eigenfunctions ($\psi_i$), eigenvalues ($\varepsilon_i$), and
occupation numbers ($f_i$) of Kohn-Sham quasiparticles.
The potential $V$ can be divided into the Coulomb ($V_{\rm H}$),
exchange-correlation ($V_{\rm XC}$), and the
external parts ($V_{\rm ext}$), which are the functional derivatives of corresponding contributions to the
total energy with respect to the electron density.
Formally speaking, this procedure is fully \textit{ab initio} and free of any
adjustable parameters.

  However, the
form of the
exchange-correlation potential is generally unknown.
For practical purposes, it is typically treated in the local-density
approximation (LDA), which employs an analytical expression
borrowed from the theory of
homogeneous electron gas
in which the density of the electron gas
is replaced by the local density of the real system.
LDA is far from being perfect and
there are many examples of the so-called strongly correlated materials
where the conventional LDA fails in describing the
excited-
as well as the ground-state properties \cite{IFT}.

\begin{figure}[h!]
\begin{center}
\resizebox{10cm}{!}{\includegraphics{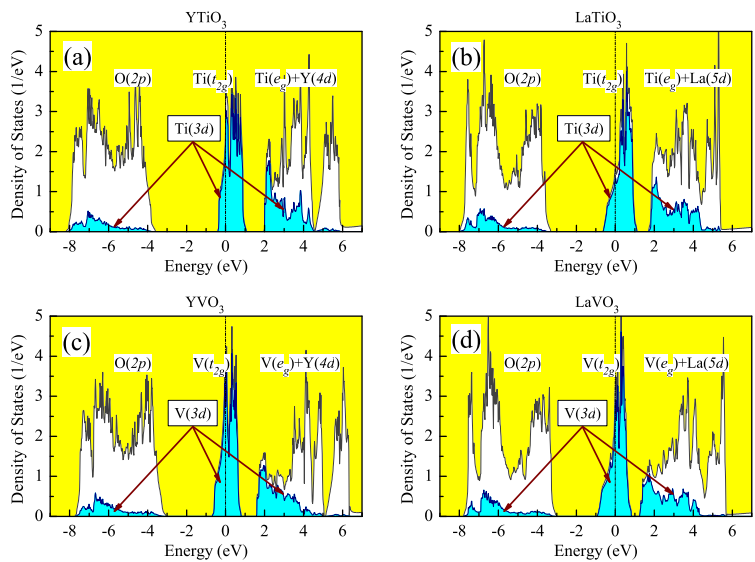}}
\end{center}
\caption{\label{fig.PerovskiteDOS} Total and partial densities of states
of YTiO$_3$ (a), LaTiO$_3$ (b), YVO$_3$ (c, orthorhombic phase),
and LaVO$_3$ (d) in the local-density approximation \protect\cite{PRB06b}.
The shaded area shows the contributions of transition-metal $3d$-states.
Other symbols show the positions of the main bands.
The Fermi level is at zero energy.}
\end{figure}

  In the strongly correlated materials, the state of each electron strongly
depends on the state of other electrons of the system, which are
coupled (or correlate with each other) via the Coulomb interaction. Thus,
this is the real many-body problem, and the situation
is
very different from the behavior of the homogeneous electron gas.
The canonical example of strongly correlated materials is the
transition-metal oxides \cite{IFT}. A typical example of the electronic
structure of the transition-metal oxides in the local-density
approximation is shown in Figure \ref{fig.PerovskiteDOS} for the
series of distorted perovskite compounds.\footnote{
The properties of the distorted perovskite oxides will be discussed in
details in Section \ref{sec:DistortedPerovskites}.
}
We would like to emphasize two points.
\begin{enumerate}
\item
The common feature of many strongly correlated systems
is the existence of
a limited group of states,
located near the Fermi level and well isolated from the rest of the spectrum.
In the following, these states will be called the as ``low-energy states'' or the
``low-energy bands'' or, simply, the $L$-bands.
In the case of perovskite oxides depicted in Figure \ref{fig.PerovskiteDOS}
these are the narrow transition-metal $t_{2g}$ bands. From this point of view, the
theoretical
description of the strongly correlated systems is feasible, and this is certainly a good sign.
For example, if we are
interested in electronic or magnetic properties, that are mainly controlled by the
states close to the Fermi level, we can mainly concentrate on the behavior of this
group of
states and disengage ourself from other details of the electronic structure.
Since the number of such states is limited, the problem can be solved,
at least numerically.
\item
However, the bad points is that in order to solve this problem we should inevitably
go beyond the local-density approximation, which
greatly oversimplifies the physics of Coulomb correlations.
For example, the systems depicted in Figure \ref{fig.PerovskiteDOS}
are metals within LDA, while in practice all of them are Mott insulators \cite{IFT}.
\end{enumerate}

  The insulating behavior is frequently associated with the excited state properties,
which are not supposed to be reproduced by the Kohn-Sham equations designed for the
ground state. However, the problem is much more serious. Suppose that we are
interested in the behavior of interatomic magnetic interactions, which are the ground state properties.
For the narrow-band compounds, the main contributions to the magnetic interactions
can be typically identified with the double exchange and superexchange
mechanisms \cite{PRL99,Springer}.
The double exchange operates in the metallic systems.
It is always ferromagnetic and proportional to the kinetic-energy gain, $t$, caused by
free electron hoppings between different sites of the lattice \cite{deGennes}.
The superexchange dominates in insulating compounds, where the double exchange vanishes \cite{PRB03}.
The superexchange can be either antiferromagnetic or ferromagnetic, depending on the number
of electrons and the orbital state of the system \cite{KugelKhomskii}.
It is proportional to $t^2$ and inversely proportional to the parameter of on-site Coulomb
interaction ${\cal U}$ \cite{PWA}.
Now, suppose that because of the limitations of LDA, our system is metallic
rather than insulating. Then, the ferromagnetic double exchange will
clearly
dominate and we may have a totally wrong answer for the interatomic magnetic interactions
as well as for the magnetic ground state. Therefore, if the system is insulating,
an energy gap should be present already in the spectrum of Kohn-Sham eigenvalues.
If it is not, the approximations underlying these Kohn-Sham equations
must be corrected.

  Because of these difficulties,
for the long time
the strongly correlated compounds
have been
almost exclusive prerogative of the model condensed-matter physics, which has
accumulated a great knowledge about treating and solution of this many-body problem but only for
model systems, described in terms of a limited number of model parameters.
The model analysis has indeed provided a useful and insightful information about many
properties of the strongly correlated systems. However, as the complexity of such systems rose,
it inevitably posed a question about the correct choice of the model parameters, and in a number
of cases -- of the model itself. For example, the typical questions are how to incorporate
the information about the chemical signature of elements composing the lattice
into the model or how to
treat lattice distortions? These questions gave rise to the new direction, which can be called as
the ``realistic modeling''. The basic idea
of the realistic modeling
is to construct the model Hamiltonian
entirely from the
first principles, and make it free of any adjustable parameters, and then to solve this model
Hamiltonian
by using modern many-body methods. This was the main motivation in the area of
model condensed-matter physics.

  Then, why do we need the realistic modeling in the
area of computational physics dealing with the first-principles
electronic structure calculations?
\begin{enumerate}
\item
Presumably, it is needless to say that the solution of the many-body problem
for many realistic systems is a tremendous computational task. Therefore, one
would always like to formulate it rigorously only in a \textit{restricted} Hilbert space,
which would pick
up only those states that are primarily responsible for the physics of either of
the considered phenomenon, and include the effect of other states into the
renormalization of the model parameters in the primary Hilbert space.
\item
It is true that the methods of first-principles electronic structure calculations
are currently on the rise. The standard computations within the density-functional theory
become a matter of routine and frequently regarded as a useful tool for the
numerical modeling or the numerical experiment in the materials science.
However, we will always need some additional tools for the analysis and
interpretation of the
obtained data.
Our final goal is not to obtain or reproduce the experimental trend for some complex compounds.
Our goal is provide
some transparent physical interpretation for this trend
on the microscopic level
and come up with some useful
suggestions about how this trend can be further improved.
In this sense, the realistic modeling will continue to play an important role as the tool
for the analysis and interpretation of data obtained in the standard electronic structure
calculations.
\end{enumerate}

  Why should the realistic modeling work? It is not an idle question, because
eventually we would like to
start with the electronic structure in the local-density approximation, construct the model
by relying on this electronic structure, and apply this model for the strongly correlated systems.
In the other words, we start with something what was claimed to be wrong for the strongly correlated systems
and try to
find its refinements by continuing to stay within this disputable picture and relying on these disputable
approximations. Nevertheless, there are several reasons to believe that such a construction is possible
and propose a meaningful strategy for improving LDA by staying within LDA and, at the same time,  bypassing its
limitations and shortcomings.
\begin{enumerate}
\item
By the construction, the Kohn-Sham theory is the one-electron theory \cite{KohnSham,ParrYang}.
Therefore, it should provide a good starting point for the one-electron part of the model, no matter
whether it is supplemented with LDA or not. Moreover, all contributions
to the Kohn-Sham Hamiltonian coming from the exchange and correlations are \textit{local} in the
local-density approximation. Therefore, they can be easily subtracted
in the process of construction of the model
Hamiltonian
in order to avoid the
double-counting problem.
\item
The Coulomb correlations is certainly a weak point of LDA. However, LDA is an
approximation in the density-functional theory, which is formulated for the ground state.
Hence,
it should be able to provide a good estimate for the total energy of the system \cite{HohenbergKohn,KohnSham,ParrYang}.
Then, there is always a chance to derive the effective Coulomb interactions
from the total energy difference
(or any other quantity uniquely related with the total energy)
by applying certain constraint conditions, which would
simulate the redistribution of certain order parameters
(typically, the electron density of the system).
\end{enumerate}

  The goal of this review article is
to outline the main ideas, principles, and methods of the realistic modeling as well as
to illustrate
abilities and
perspectives of this direction
for the solution of several practical questions
related with the understanding
of the real physical properties
of transition-metal oxides
and
other
strongly correlated systems.
After specifying the model in Section \ref{sec:Model},
in the next two Sections
we will discuss how the
parameters of this model can be derived from the first-principles electronic
structure calculations.
Section \ref{sec:Method} will be devoted to the construction of
the one-electron Hamiltonian starting from the LDA band structure, while
Section \ref{sec:Coulomb} will deal with
the problems of screening and
calculation of the effective Coulomb
interaction.
We will try not only to illustrate the method, but also to present a clear physical
picture for underlying ideas and approximation.
Then,
Section \ref{sec:Solution} will briefly summarize
the methods of solution of the model Hamiltonian.
In Section \ref{sec:applications} we will give practical examples and
show applications for realistic materials.
We will derive the parameters of the model Hamiltonian
and discuss what these parameters mean for
understanding the
electronic and magnetic properties
of the considered compounds.
Section \ref{sec:Summary} will contain brief summary and
concluding remarks.

\section{\label{sec:Model}Model Hamiltonian}

  First of all, we would like to specify our model for the
the low-energy bands. We will be mainly dealing
with the effective
multi-orbital Hubbard model,
\begin{equation}
\hat{\cal{H}}= \sum_{{\bf R}{\bf R}'} \sum_{\alpha \beta} h_{{\bf
R}{\bf R}'}^{\alpha \beta}\hat{c}^\dagger_{{\bf
R}\alpha}\hat{c}^{\phantom{\dagger}}_{{\bf R}'\beta} + \frac{1}{2}
\sum_{\bf R}  \sum_{\alpha \beta \gamma \delta} U_{\alpha \beta
\gamma \delta} \hat{c}^\dagger_{{\bf R}\alpha} \hat{c}^\dagger_{{\bf
R}\gamma} \hat{c}^{\phantom{\dagger}}_{{\bf R}\beta}
\hat{c}^{\phantom{\dagger}}_{{\bf R}\delta},
\label{eqn:Hmanybody}
\end{equation}
where $\hat{c}^\dagger_{{\bf R}\alpha}$ ($\hat{c}_{{\bf R}\alpha}$)
creates (annihilates) an electron in the Wannier orbital
$\tilde{W}_{\bf R}^\alpha$ of (typically, the transition-metal)
center ${\bf R}$, and $\alpha$ is
a joint index, incorporating all remaining (spin and orbital)
degrees of freedom, unless it is specified otherwise.
The one-electron Hamiltonian $\hat{h}_{{\bf R}{\bf R}'}$$=
$$\| h_{{\bf R}{\bf R}'}^{\alpha \beta} \|$
usually includes the following contributions:
the site-diagonal part (${\bf R}$$=$${\bf R}'$) describes
the local level-splitting, caused by the crystal field and/or the
relativistic
spin-orbit interaction, whereas the off-diagonal part
(${\bf R}$$\neq$${\bf R}'$) stands for transfer integrals,
describing the kinetic energy of electrons.
$$
U_{\alpha \beta \gamma \delta}
=
\int d{\bf r} \int d{\bf r}' \tilde{W}_{\bf R}^{\alpha \dagger}({\bf r})
\tilde{W}_{\bf R}^\beta({\bf r}) v_{\rm scr}({\bf r},{\bf r}')
\tilde{W}_{\bf R}^{\gamma \dagger}({\bf r}') \tilde{W}_{\bf R}^\delta({\bf r}')
\equiv \langle \tilde{W}_{\bf R}^\alpha \tilde{W}_{\bf R}^\gamma | v_{\rm scr} |
\tilde{W}_{\bf R}^\beta \tilde{W}_{\bf R}^\delta \rangle
$$
are
the matrix elements of the \textit{screened} Coulomb interaction
$v_{\rm scr}({\bf r},{\bf r}')$, which are supposed to be diagonal with
respect to the site indices $\{ {\bf R} \}$.
In principle, $U_{\alpha \beta \gamma \delta}$ can also depend on the
site-index ${\bf R}$. Nevertheless, for the sake of simplicity of
our notations, here and throughout in this paper we drop the index ${\bf R}$
in the notation of the Coulomb matrix elements.
The intersite matrix elements of $U_{\alpha \beta \gamma \delta}$ are typically
small in comparison with the on-site ones.

\section{\label{sec:Method}One-Electron Hamiltonian and Wannier functions}

  The one-electron part of the model Hamiltonian (\ref{eqn:Hmanybody})
is typically identified with the Kohn-Sham Hamiltonian in the basis of Wannier functions
representing the low-energy part of the spectrum \cite{Wannier}.
Therefore, the concept
and definition of the
Wannier functions is one of the key parts of the methods, and we
would like to start our discussion by making several general comments about the
relationship between Wannier functions and localized atomic orbitals,
which represent the basis of many computational schemes.

  Let us assume that there is a certain set of localized
orbitals $\{ \tilde{\chi}_{\bf R}^\alpha \}$ centered at the atomic sites $\{ {\bf R} \}$
and specified by the orbital indices $\{ \alpha \}$.\footnote{
In the following, a set of nonorthonormalized atomic-like orbitals will be denoted
as $\{ \chi_{\bf R}^\alpha \}$. The orthonormalized orbitals, constructed from $\{ \chi_{\bf R}^\alpha \}$,
are denoted as $\{ \tilde{\chi}_{\bf R}^\alpha \}$. Generally, such an orthonormalization
can be performed numerically.}
The corresponding Kohn-Sham Hamiltonian
in the basis of $\{ \tilde{\chi}_{\bf R}^\alpha \}$
will be denoted as $\hat{H}$.
The orbitals are orthonormalized and form a
complete basis in the valence part of the spectrum,
so that each eigenvector $\psi_i$ of $\hat{H}$
can be expressed as a linear combination of $\{ \tilde{\chi}_{\bf R}^\alpha \}$.
The concrete examples of such bases can be the orthonormalized atomic
orbitals or the muffin-tin orbitals \cite{LMTO1,LMTO2,LMTO3}.

  Since the Wannier functions $\{ \tilde{W}_{\bf R}^\alpha \}$
are also defined as certain set of localized orbitals representing $\{ \psi_i \}$ \cite{Wannier,MarzariVanderbilt},
we immediately recognize that
for the full Hamiltonian $\hat{H}$,
$\{ \tilde{\chi}_{\bf R}^\alpha \}$
can be regarded as
one of the possible (and fully legitimate)
choices for $\{ \tilde{W}_{\bf R}^\alpha \}$.
This is a natural result and advantage of the basis of localized
atomic orbitals.
In the plane-wave basis, the localized Wannier functions can be constructed from
the eigenstates of $\hat{H}$ in the valence part of the spectrum, for example,
by minimizing the square of the position operator
$\langle {\bf r}^2 \rangle$ \cite{MarzariVanderbilt}.
However, we would like emphasize that
this is nothing but an elegant way of constructing the compact atomic-like orbitals from the
extended
plane waves, a step which becomes rather unnecessary if one works
from the very beginning in
the
atomic basis.

  However, what we typically need in the process of construction of the model Hamiltonians
is different. For example,
the solution of the many-body problem
is practically impossible in the Hilbert space of states
$\{ \tilde{\chi}_{\bf R}^\alpha \}$ of
the full Hamiltonian $\hat{H}$.
Instead, one would like to concentrate on the behavior of a small number of $L$-bands,
typically located near the Fermi level, and construct
the Wannier basis only for this group
of bands,
which would be also orthogonal to the
rest of the eigenstates of $\hat{H}$.
This causes an additional
complication because the basis functions $\{ \tilde{\chi}_{\bf R}^\alpha \}$,
though can be regarded as the Wannier functions for the full Hamiltonian $\hat{H}$,
are no longer those for any subspace of $\hat{H}$.

  At present, there are two methods, which are typically used
to circumvent this problem and construct the Wannier functions
for the subspace of $\hat{H}$:
the projector-operator method \cite{MarzariVanderbilt,WeiKu,Anisimov2005,Streltsov}
and the downfolding method \cite{PRB06b,PRB04,Imai,PRB06a}.

\subsection{\label{sec:projector}The Projector-Operator Method}

  In the projector-operator method,
each (nonorthonormalized) Wannier function is generated by
projecting
a trial basis function $| \tilde{\chi}_{{\bf R}}^t \rangle$,
centered at the site ${\bf R}$,
onto the $L$-bands:
\begin{equation}
|W_{{\bf R}}^t \rangle = \hat{P} | \tilde{\chi}_{{\bf R}}^t \rangle,
\label{eqn:Wannier_definition}
\end{equation}
where
\begin{equation}
\hat{P} = \sum_{i \in L} | \psi_i \rangle \langle \psi_i |
\label{eqn:projector}
\end{equation}
is the projector-operator onto the $L$-bands, $\psi_i$ is the eigenstate
of $\hat{H}$, and $i$ is a joint index combining the band index and the position of the
momentum ${\bf k}$ in the first Brillouin zone.
The functions $\{ W_{{\bf R}}^t \}$ can be
numerically orthonormalized,
\begin{equation}
| \tilde{W}_{\bf R}^t \rangle = \sum_{{\bf R}'t'} | W_{{\bf R}'}^{t'} \rangle
[ \hat{\cal S}^{-1/2} ]_{{\bf R}'{\bf R}}^{t't},
\label{eqn:orthonormalization}
\end{equation}
where
$\hat{\cal S}$$=$$\| {\cal S}_{{\bf R}'{\bf R}}^{t't} \|$
is the overlap matrix,
\begin{equation}
{\cal S}_{{\bf R}'{\bf R}}^{t't} = \langle W_{{\bf R}'}^{t'} | W_{\bf R}^t \rangle
\equiv \langle \tilde{\chi}_{{\bf R}'}^{t'} | \hat{P} | \tilde{\chi}_{{\bf R}}^{t} \rangle .
\label{eqn:overlapP}
\end{equation}
Then, the one-electron part of the model Hamiltonian (\ref{eqn:Hmanybody})
is defined by the matrix elements of $\hat{H}$ in the basis of
these
orthonormalized Wannier orbitals:
\begin{equation}
h_{{\bf R}{\bf R}'}^{tt'} = \langle \tilde{W}_{\bf R}^t | \hat{H} | \tilde{W}_{{\bf R}'}^{t'} \rangle.
\label{eqn:HamiltonianP}
\end{equation}

\subsection{\label{sec:downfolding}The Downfolding Method}

  The conventional downfolding method also implies that the atomic basis
can be
divided into two parts:
$\{ \tilde{\chi}_{\bf R} \}$$=$$\{ \tilde{\chi}_{\bf R}^t \}$$\oplus$$\{ \tilde{\chi}_{\bf R}^r \}$,
so that the low-energy part of the spectrum is mainly
represented by the $\{ \tilde{\chi}_{\bf R}^t \}$-states, while $\{ \tilde{\chi}_{\bf R}^r \}$
is the rest of the basis states, which mainly contribute to the higher-energy part.
Then, each eigenstate of $\hat{H}$
can be identically presented as the sum
$|\psi_i \rangle$$=$$|\psi_i^t \rangle$$+$$|\psi_i^r \rangle$,
where $|\psi_i^t \rangle$ and $|\psi_i^r \rangle$
are expanded over the basis states of the ``$t$'' and ``$r$'' types, respectively.
In this case, the Schr\"{o}dinger equation for $| \psi_i \rangle$ takes the following form:
\begin{eqnarray}
( \hat{H}^{tt}-\omega ) | \psi_i^t \rangle  +  \hat{H}^{tr} | \psi_i^r \rangle & = & 0  \label{eqn:seceq1}\\
\hat{H}^{rt} | \psi_i^t \rangle  +  ( \hat{H}^{rr}-\omega ) | \psi_i^r \rangle & = & 0, \label{eqn:seceq2}
\end{eqnarray}
where $\hat{H}^{t(r)t(r)}$ are the blocks of
matrix elements of $\hat{H}$ in the basis of ``$t$''(``$r$'')-states.
The effective $\omega$-dependent Hamiltonian $\hat{H}_{\rm eff}$
is obtained by expressing
$| \psi_i^r \rangle$ from (\ref{eqn:seceq2}),
\begin{equation}
| \psi_i^r \rangle = - ( \hat{H}^{rr}-\omega )^{-1} \hat{H}^{rt} | \psi_i^t \rangle,
\label{eqn:relimination}
\end{equation}
and substituting into (\ref{eqn:seceq1}).
This yields
\begin{equation}
\hat{H}_{\rm eff}(\omega) = (\hat{H}^{tt} - \omega) - \hat{H}^{tr}
(\hat{H}^{rr} - \omega)^{-1}\hat{H}^{rt},
\label{eqn:Heff}
\end{equation}
which formally acts
only on $| \psi_i^t \rangle$.
However, $| \psi_i^t \rangle$ is only a part of the eigenvector, which is not orthonormalized.
Therefore,
$\hat{H}_{\rm eff}(\omega)$ should be additionally transformed to an orthonormal representation:
\begin{equation}
\hat{h}(\omega) = \hat{S}^{-1/2}(\omega) \hat{H}_{\rm eff}(\omega) \hat{S}^{-1/2}(\omega) + \omega,
\label{eqn:TB}
\end{equation}
which is specified by the overlap matrix,
\begin{equation}
\hat{S}(\omega)=1+\hat{H}^{tr}
(\hat{H}^{rr}-\omega)^{-2}\hat{H}^{rt}.
\label{eqn:overlapDF}
\end{equation}
The latter is obtained after the substitution of
(\ref{eqn:relimination}) into
the normalization condition:
$\langle \psi_i^t | \psi_i^t \rangle$$+$$\langle \psi_i^r | \psi_i^r \rangle$$=$$1$.

  In the conventional downfolding method, $\hat{h}$ is typically evaluated in the center of gravity
of the $L$-bands, $\omega_0$.
Although the downfolding method does not explicitly require
the construction of the Wannier functions, they can be formally reconstructed from
$\hat{h}(\omega_0)$ \cite{PRB06a}.

\subsection{\label{sec:downfoldingasprojector}Downfolding as the Projector-Operator Method}

  The conventional downfolding method is exact. However, this property is guaranteed by
the $\omega$-dependence of $\hat{h}$, which
is hardly useful from the
practical point of view.
Formally, for each $\psi_i$, $\omega$ in (\ref{eqn:TB}) should coincide with the
eigenvalue of $\hat{H}$ corresponding to this $\psi_i$.
Moreover, $\hat{h}(\omega)$ retains an excessive information about $\hat{H}$,
so that the full spectrum of
$\hat{H}$
can be formally derived from $\hat{h}(\omega)$.
However, typically we do not need such a redundant information and
would like to use
$\hat{h}$ only for a small group of electronic states
located near the Fermi level, and do it in the most exact form.

  For these purposes, the downfolding method can be
reformulated as a projector-operator method and reduced to it \cite{PRB07}.
The trick
is to replace
the original Hamiltonian $\hat{H}$
in the downfolding method
by a modified Hamiltonian $\hat{H}'$, which has the same set of eigenvalues
$\{ \varepsilon_i \}$ and eigenfunctions $\{ \psi_i \}$ in the region of $L$-bands.\footnote{
This procedure was already used in \cite{PRB06b,PRB04,Imai,PRB06a}.
However, the details have been explained only in \cite{PRB07}.}
The rest of the eigenstates is not important
for the construction of the one-electron part of the model Hamiltonian
and can be placed
to the region of infinite energies.
Hence, we define $\hat{H}'$ in the following form:
\begin{equation}
\hat{H}' = \sum_{i \in L} | \psi_i \rangle \varepsilon_i \langle \psi_i |
+ \epsilon \hat{P}_\perp \equiv \hat{P} \hat{H} \hat{P} + \epsilon \hat{P}_\perp,
\label{eqn:modifiedH}
\end{equation}
where $\hat{P}_\perp$$=$$\hat{1}$$-$$\hat{P}$ is the projector operator to the subspace
orthogonal to the
$L$-bands and $\epsilon$$\rightarrow$$\infty$.
According to the choice of the
basis functions
$\{ \tilde{\chi}_{\bf R}^t \}$ and
$\{ \tilde{\chi}_{\bf R}^r \}$
in the downfolding method, the latter
mainly contribute to the high-energy part of the spectrum. Therefore,
the overlap between $\psi_i$ in the low-energy part and any of $\{ \tilde{\chi}_{\bf R}^r \}$
should be small, so that all eigenvalues of $(\hat{H}')^{rr}$
are of the order of $\epsilon$.
Then, it is intuitively clear that
in the limit $\epsilon$$\to$$\infty$,
the $\omega$-dependence in (\ref{eqn:relimination})
will disappear and
the method will be exact.
This statement
can be proven rigorously, and such a proof can be found in \cite{PRB07}.

   Thus, by introducing the modified Hamiltonian of the form
(\ref{eqn:modifiedH}), the downfolding method can be naturally reformulated as
the projector-operator method.
The advantage of this procedure is that it allows us to go directly to the
construction of the one-electron Hamiltonian $\hat{h}$,
and formally skip the step of the construction of the Wannier functions.

  Finally, we would like to note that (\ref{eqn:modifiedH}) is nothing but a scissor-operator-like
transformation of the original Hamiltonian $\hat{H}$, and this
strategy is different from the order-$N$ muffin-tin orbital method,
which was also used for the construction of the Wannier functions
and
where the basic idea was to
make some
\textit{approximations} for the $\omega$-dependence of the
downfolded Hamiltonian \cite{NMTO}.

\subsection{\label{sec:trial}Choice of Trial Orbitals and Localization of the Wannier Functions}

  At the beginning of this section
we have argued that the basis functions $\{ \tilde{\chi}_{\bf R}^\alpha \}$
can be regarded as the Wannier functions of the full Hamiltonian $\hat{H}$.
Now, let us assume that
each basis function is localized around the central atomic site and
satisfies certain criteria of the ``maximal localization'', such that any
linear combination of $\{ \tilde{\chi}_{\bf R}^\alpha \}$ will be ``less localized''
in comparison with the basis function $\tilde{\chi}_{\bf R}^\alpha$
from the original basis set,
or at least has the same degree of the localization,
if we are dealing with the linear combination of orbitals
centered at the same atomic site.\footnote{
The precise criterion of the maximal localization
is not really important at this stage, because this is merely
a mathematical construction and depending on the considered physical property
one can introduce different criteria of the ``maximal localization''.}
However,
this is not necessarily true if one wants to construct the
Wannier functions
only for some part of the electronic structure, which is specified by the
subspace $L$ of the full Hamiltonian $\hat{H}$.
Due to the additional orthogonality condition to other bands, such a Wannier function
will
inevitably
be a linear combination of $\{ \tilde{\chi}_{\bf R}^\alpha \}$.
Therefore, it will inevitably be
less localized in comparison with the trial
function $\tilde{\chi}_{\bf R}^t$.

  Nevertheless, one can try to ``minimize''
the deviation from the original formulation
for the full Hamiltonian $\hat{H}$ and
ask which single atomic orbital centered at the site ${\bf R}$
will be the best representation for the Wannier orbital. Therefore, we search a new
set of trial
functions in the form:
\begin{equation}
| \tilde{\phi}^t_{\bf R} \rangle = \sum_{\alpha} c_{\bf R}^{\alpha} | \tilde{\chi}_{\bf R}^{\alpha} \rangle,
\label{eqn:newtrial}
\end{equation}
and find the coefficients $\{ c_{\bf R}^{\alpha} \}$, which
maximize the
projection $\langle \tilde{\phi}^t_{\bf R} | W^t_{\bf R} [\tilde{\phi}^t_{\bf R}] \rangle$
of $\tilde{\phi}^t_{\bf R}$ onto
the nonorthonormalized Wannier function
constructed from $\tilde{\phi}^t_{\bf R}$ using the projector-operator technique,
$|W^t_{\bf R} [\tilde{\phi}^t_{\bf R}]\rangle$$=$$\hat{P} | \tilde{\phi}^t_{\bf R} \rangle$.
It will automatically guarantee that $\tilde{\phi}^t_{\bf R}$ is the best
single-orbital representation for $W^t_{\bf R}$ in the projector-operator method
among the trial orbitals of the form (\ref{eqn:newtrial}). By substituting
$W^t_{\bf R} [\tilde{\phi}^t_{\bf R}]$ into the projection
$\langle \tilde{\phi}^t_{\bf R} | W^t_{\bf R} [\tilde{\phi}^t_{\bf R}] \rangle$,
problem is reduced to the maximization of
$$
D = \max_{ \{ c_{\bf R}^{\alpha} \} } \left\{ \langle \tilde{\phi}^t_{\bf R} | \hat{P} | \tilde{\phi}^t_{\bf R} \rangle -
\lambda ( \langle \tilde{\phi}^t_{\bf R} | \tilde{\phi}^t_{\bf R} \rangle - 1 ) \right\}
$$
with respect to $\{ c_{\bf R}^{\alpha} \}$, where the Lagrange multipliers $\{ \lambda \}$
enforce the orthonormality condition for $\{ \tilde{\phi}^t_{\bf R} \}$.
Then, the maximization of $D$ is equivalent to the diagonalization of
$\hat{P}_{\bf RR}$$=$$\| \langle \tilde{\chi}_{\bf R}^{\alpha} | \hat{P} |
\tilde{\chi}_{\bf R}^{\alpha'} \rangle \|$, which is nothing but the site-diagonal part of the
density matrix constructed from the $L$-bands in the basis of atomic orbitals
$\{ \tilde{\chi}_{\bf R}^{\alpha} \}$.
After the diagonalization,
we should simply pick up $n$ eigenstates
$ \{ \tilde{\phi}^t_{\bf R} \}$,
corresponding to maximal eigenvalues $\{ \lambda \}$, where $n$ is the number
of Wannier functions centered at the atomic site ${\bf R}$.\footnote{
This can be paraphrased in a different way \cite{PRB06b}. Of course, any set of the Wannier functions
should be able to reproduce the total density at the site ${\bf R}$. Each Wannier
function consists of the central part (or ``head''), located at the site ${\bf R}$,
and the tail, spreading over the other sites. By identifying $\{ \tilde{\phi}^t_{\bf R} \}$
with eigenstates corresponding to the maximal eigenvalues of the density matrix $\{ \lambda \}$,
we guarantee that the main part of the density at the site ${\bf R}$
is described by the ``heads'' of the Wannier functions. Then, the remaining part of the density,
corresponding to other (small) $\{ \lambda \}$, is described by the tails of the Wannier functions
coming from other sites. This implies that the weights of these tails should be also small.
}
These $ \{ \tilde{\phi}^t_{\bf R} \}$ will maximize $D$.
This procedure has been proposed in \cite{PRB04} without any proof. Then, some
intuitive arguments have been given in \cite{PRB06a}.
The rigorous proof
has been given in \cite{PRB07}.

\subsection{\label{sec:other}Other parameters of model Hamiltonian}

  Electronic structure calculations are typically supplemented with some
additional approximations, like the atomic-sphere-approximation (ASA),
where the LDA potential is spherically averaged inside certain atomic spheres \cite{LMTO1}.
If ASA is used in the process of construction of the model Hamiltonian,
the parameters $\{ h_{{\bf R}{\bf R}'}^{\alpha \beta} \}$
will
include all effects of the
covalent mixing effects between
atomic orbitals. However, there will be other effects,
which are not accounted by
$\{ h_{{\bf R}{\bf R}'}^{\alpha \beta} \}$.
The most important contribution comes from the nonsphericity (n-s)
of the Madelung potential for the electron-ion interactions, which
contributes to the crystal-field splitting \cite{MochizukiImada}.\footnote{
In fact, the contribution of Coulomb interactions to the crystal-field splitting is a
tricky issue. Despite an apparent simplicity of the problem, one should clearly
distinguish different contributions and not to include them twice,
for example,
to the one-electron
and Coulomb interaction parts of the model Hamiltonian (\ref{eqn:Hmanybody}).
In this sense, the use of
full-potential techniques does not automatically guarantee the right answer.
Note that the nonsphericity of \textit{on-site} interactions is explicitly
included into the second part of the model Hamiltonian (\ref{eqn:Hmanybody}).
Therefore, in order to not to include it twice,
one should subtract
the corresponding contributions to the
one-electron part
originating from the Coulomb and exchange-correlation
potentials in LDA.
}
The proper correction to
$\| h_{{\bf R}{\bf R}'}^{\alpha \beta} \|$ can be
computed
in the Wannier basis as:
\begin{equation}
\Delta^{\rm n-s} h_{\bf RR}^{\alpha \beta} = \sum_{{\bf R}'\neq{\bf R}}
\langle \tilde{W}_{\bf R}^{\alpha} |
\frac{- Z^{*}_{{\bf R}'} e^2}{|{\bf R}+{\bf r}-{\bf R}'|}
| \tilde{W}_{\bf R}^{\beta} \rangle,
\label{eqn:CFEI}
\end{equation}
where $Z^{*}_{{\bf R}'}$ is the total charge associated with the
site ${\bf R}'$
(namely, the nuclear charge minus the screening electronic charge encircled by the
atomic sphere),
and ${\bf r}$ is the position of electron in the sphere ${\bf R}$.

  The main idea behind this treatment is based on
certain hierarchy of interactions in solids. It implies that the strongest
interaction, which leads to the energetic separation of the $L$-band from other
bands (Figure \ref{fig.PerovskiteDOS}), is due to the covalent mixing. For example, in many
transition-metal oxides this interaction is responsible for the famous splitting
between transition-metal $t_{2g}$ and $e_g$ bands \cite{Kanamori1957}.
The nonsphericity of the Madelung potential is considerably weaker than this splitting.
However, it can be comparable with the covalent mixing
in the narrow $L$-band.
Therefore, the basic idea is to treat this nonsphericity as a
pseudo-perturbation \cite{LMTO1}, and calculate the matrix elements of the
Madelung potential
in the basis of Wannier functions constructed
for spherically averaged ASA potential.

  The same strategy can be applied to the spin-orbit (s-o) interaction, which
yields the following correction to
$\| h_{\bf RR}^{\alpha \beta} \|$:
$$
\Delta^{\rm s-o} h_{\bf RR}^{\alpha \beta} =
\langle \tilde{W}_{\bf R}^{\alpha} | \frac{\hbar}{4m^2c^2} (\bnabla V \times {\bf p})\cdot \bsigma |
\tilde{W}_{\bf R}^{\beta} \rangle.
$$
Here, $V$ is the self-consistent LDA potential and
$\bsigma$ is the vector of Pauli matrices.

\section{\label{sec:Coulomb}Effective Coulomb Interactions}

  Generally, the matrix elements of the effective Coulomb interaction in
the $L$-band are defined as the energy cost for moving
an electron from one Wannier orbital,
say $\tilde{W}_{{\bf R}'}^\beta$,
populated by
$n_{{\bf R}'\beta}$ electrons, to another orbital,
say $\tilde{W}_{\bf R}^\alpha$, which was initially populated by
$n_{{\bf R}\alpha}$ electrons \cite{Herring}:
\begin{equation}
U_{\alpha \alpha \beta \beta} = E\left[ n_{{\bf R}\alpha} + 1,
n_{{\bf R}'\beta} - 1 \right] -
E\left[ n_{{\bf R}\alpha} , n_{{\bf R}'\beta} \right].
\label{eqn:Coulomb.definition}
\end{equation}
For ${\bf R}$$\ne$${\bf R}'$, the above matrix elements define the on-site
Coulomb
interactions, which are screened by intersite interactions.
In principle, by considering different combinations of ${\bf R}$ and ${\bf R}'$,
one can extract individual parameters of on-site and intersite
interactions \cite{PRB06a}. However, in practice, the intersite interactions
are substantially smaller than the on-site ones, and can be
neglected in many cases.\footnote{
Yet, one exception can be the parameters derived for the single-orbital model.
In this case, the number of variables available for the localization
of the Wannier functions is limited so that the latter can be
rather extended in the real space \cite{PRB06a}.
Therefore,
the \textit{bare} intersite Coulomb and exchange integrals, calculated in the
basis of these Wannier functions, are no longer negligible \cite{WeiKu,Mazurenko}.
Nevertheless, at present it is not entirely clear to which extent these
interactions in the $L$-band will be screened by other bands of the system.
}
The total energy difference
(\ref{eqn:Coulomb.definition}) corresponding to ${\bf R}$$=$${\bf R}'$,
but with different orbital indices $\alpha$ and $\beta$,
define the
nonspherical part of on-site interactions, which
is responsible for Hund's rules.
More generally, one can consider an unitary transformation of $\alpha$ and $\beta$
at each site of the system to a new basis, and repeat the same arguments.
In the original basis, this would correspond to the transfer of an electron from a
linear combination of Wannier orbitals at the site ${\bf R}'$ to
a linear combination
of Wannier orbitals
at the site ${\bf R}$. This procedure will define the full
matrix of screened Coulomb interactions $\hat{U}$$=$$\| U_{\alpha \beta \gamma \delta}\|$.

\subsection{\label{sec:cDFT}Constrained density-functional theory}

  The dependence of the total energy $E[\{ n_{{\bf R}\alpha} \}]$ on the individual
occupation numbers $\{ n_{{\bf R}\alpha} \}$ can be obtained by applying the
constrained density-functional theory \cite{Dederichs,Gunnarsson1989,AnisimovGunnarsson,PRB94a}:
\begin{equation}
E\left[ \{ n_{{\bf R}\alpha} \} \right] = E[\rho] +
\sum_{{\bf R}\alpha} V_{{\bf R}\alpha}
\left( \langle \tilde{W}_{\bf R}^\alpha | \hat{\rho} | \tilde{W}_{\bf R}^\alpha \rangle - n_{{\bf R}\alpha} \right),
\label{eqn:Coulomb.constraint}
\end{equation}
where $\hat{\rho}({\bf r},{\bf r}') = \sum_i f_i \psi_i^\dagger ({\bf r}) \psi_i({\bf r}')$
is the density operator constructed from the Kohn-Sham orbitals,
$\rho({\bf r})= \hat{\rho}({\bf r},{\bf r})$ is the electron density, and $\{ V_{{\bf R}\alpha} \}$
are the external potentials, which play the role of Lagrange multipliers and enforce the
occupations of the Wannier orbitals to be equal to $\{ n_{{\bf R}\alpha} \}$.
As it was already pointed out in the Introduction,
in spite of many limitations for the strongly correlated systems, LDA is formulated
as
an approximation to
the theory of the ground-state. Therefore,
there is always a hope that it will provide a good estimate for
$U_{\alpha \alpha \beta \beta}$ as long as the latter is expressed
in terms of the total energy difference (\ref{eqn:Coulomb.definition})
or any other quantity, which is uniquely related
with this total energy difference.

  The total energy difference (\ref{eqn:Coulomb.definition}) is typically
replaced by the difference of Kohn-Sham eigenvalues
$\{ \varepsilon_{{\bf R} \alpha} \}$ calculated for an intermediate
configuration, by using
Slater's transition state arguments:\footnote{Here, we also use the
reciprocity condition
$\varepsilon_{{\bf R}' \beta}[n_{{\bf R}\alpha}$$+$$\frac{1}{2},n_{{\bf R}'\beta}$$-$$\frac{1}{2}] =
\varepsilon_{{\bf R} \alpha}[n_{{\bf R}\alpha}$$-$$\frac{1}{2},n_{{\bf R}'\beta}$$+$$\frac{1}{2}]$
for two fluctuations around the uniform state with $n_{{\bf R}\alpha} = n_{{\bf R}'\beta}$.}
\begin{equation}
U_{\alpha \alpha \beta \beta} \approx
\varepsilon_{{\bf R} \alpha}[n_{{\bf R}\alpha}+\frac{1}{2},n_{{\bf R}'\beta}-\frac{1}{2}] -
\varepsilon_{{\bf R} \alpha}[n_{{\bf R}\alpha}-\frac{1}{2},n_{{\bf R}'\beta}+\frac{1}{2}].
\label{eqn:SlaterTT}
\end{equation}
It implies the validity of Janak's theorem \cite{Janak}
\begin{equation}
\varepsilon_{{\bf R} \alpha} = \partial E / n_{{\bf R}\alpha}.
\label{eqn:Janak}
\end{equation}
However, in order to use this theorem, $\{ \tilde{W}_{\bf R}^\alpha \}$ must be
the eigenvectors of Kohn-Sham equations (\ref{eqn:KS}).
Otherwise, Janak's theorem is not applicable,
and the eigenvalues $\{ \varepsilon_{{\bf R} \alpha} \}$ are ill-defined.\footnote{
For example, by defining $ \varepsilon_{{\bf R} \alpha} $ as the diagonal matrix element
of the Kohn-Sham Hamiltonian $\hat{H}$,
$\varepsilon_{{\bf R} \alpha}$$=
$$\langle \tilde{W}_{\bf R}^\alpha | \hat{H} | \tilde{W}_{\bf R}^\alpha \rangle$,
does not necessary guarantee that this $\varepsilon_{{\bf R} \alpha}$ is equal to
$\partial E / n_{{\bf R}\alpha}$, and (\ref{eqn:SlaterTT}) is consistent with
the more general definition (\ref{eqn:Coulomb.definition}).}
Of course, this assumption does not work for the regular Wannier functions
$\{ \tilde{W}_{\bf R}^\alpha \}$ constructed for the LDA Hamiltonian $\hat{H}$.\footnote{
Note that if $\{ \tilde{W}_{\bf R}^\alpha \}$
were the eigenvectors of $\hat{H}$,
all transfer integrals, which are defined as off-diagonal matrix elements
of $\hat{H}$ with respect to $\tilde{W}_{\bf R}^\alpha$ and $\tilde{W}_{{\bf R}'}^\beta$
would vanish (see Section \ref{sec:projector}).}

  Nevertheless, there is a transparent and physical solution of this problem, which
rehabilitates the use of Janak's theorem. Indeed, since the kinetic-energy term
is explicitly included into the Hubbard model (\ref{eqn:Hmanybody}), it \textit{should not}
contribute to the total energy difference (\ref{eqn:Coulomb.definition}).
Otherwise,
it would be included twice, and
we would face the problem of the double counting \cite{Gunnarsson1989,AnisimovGunnarsson}.
This can be done by artificially switching off all matrix elements of $\hat{H}$
between $\tilde{W}_{\bf R}^\alpha$ and $\tilde{W}_{{\bf R}'}^\beta$, which
is equivalent to switching off the transfer integrals.
Another solution is to modify the Wannier functions
$\{ \tilde{W}_{\bf R}^\alpha \} \rightarrow \{ \bar{W}_{\bf R}^\alpha \}$
(apparently by considering another boundary conditions)
in order to enforce the property
$\langle \bar{W}_{\bf R}^\alpha | \hat{H} | \bar{W}_{{\bf R}'}^\beta \rangle$$=$$0$ for the
given LDA Hamiltonian $\hat{H}$ \cite{PRB06a}.\footnote{
This can be naturally done by reconstructing the Wannier functions
from the matrix elements $h_{{\bf RR}'}^{\alpha \beta}$
derived from
the downfolded method \cite{PRB06a}.
Nevertheless, it seems that for many strongly correlated systems,
$\{ \bar{W}_{\bf R}^\alpha \}$
do not substantially differ from $\{ \tilde{W}_{\bf R}^\alpha \}$.
The intuitive reason for that is that the transfer integrals in the narrow $L$-band
are typically much smaller in comparison with other effects of the
covalent mixing, which lead to the separation
of the $L$-band from other bands and mainly determine the shape of
the Wannier functions. Another reason is that for strongly correlated systems,
the transfer integrals are typically much smaller than the on-site Coulomb
interactions. Therefore, to certain extent it is unimportant whether
the small transfer integrals are included in the definition of the large Coulomb interactions
or not as they cause only small change of these interactions.
}
These $\{ \bar{W}_{\bf R}^\alpha \}$
can be regarded as the eigenfunctions of $\hat{H}$,
that justifies
the use of Janak's theorem.

  Finally, in the first order of $\pm$$\frac{1}{2}$, (\ref{eqn:SlaterTT}) can be transformed to
\begin{equation}
U_{\alpha \alpha \beta \beta} = d \varepsilon_{{\bf R} \alpha}/d n_{{\bf R} \alpha},
\label{eqn:Coulomb.derivative}
\end{equation}
where the derivative is calculated under the condition
that the total number of electrons is conserved:
$n_{{\bf R} \alpha}$$+$$n_{{\bf R}' \beta}$$=$${\rm const}$.

  It is important that in the process of solution of the Kohn-Sham equations, the Wannier orbitals
(and all remaining electronic structure) are allowed to relax in order to to
adjust the change of the occupation numbers $\{ n_{{\bf R}\alpha} \}$.
This relaxation is an important
channel of screening. As we shall see below, the
values of the screened Coulomb interactions in the $L$-band are defined not only by
the extension of the Wannier functions in the ground state,
but mainly by the change of these functions during the reaction
($n_{{\bf R}\alpha}$,$n_{{\bf R}'\beta}$) $\rightleftharpoons$ ($n_{{\bf R}\alpha}$$+$$1$,$n_{{\bf R}'\beta}$$-$$1$).
Thus, in order to calculate $\hat{U}$, it is not sufficient to know the Wannier
functions in the ground state. Even more important question is how these Wannier functions will be modified
in the process of transfer of an electron from one site
of the system
to another.

\subsection{\label{sec:RPA}Random-Phase Approximation}

  Another way of calculating
the screened Coulomb interaction is to use the
random-phase approximation (RPA), which constitutes
the basis of the GW method \cite{Hedin,FerdiGunnarsson,OnidaReiningRubio}.
In this case, the effective Coulomb interaction is calculated
in terms of the response onto the external perturbation of the electron density,
$\delta \rho_{\rm ext}({\bf r})$,
which is introduced as a probe of the electron system.
RPA has many similarities with the
constrained DFT. It consists of the following steps.
\begin{enumerate}
\item
$\delta \rho_{\rm ext}({\bf r})$ creates the Coulomb potential
$\delta V_{\rm ext}({\bf r}) = e^2 \int d{\bf r}' |{\bf r}$$-$${\bf r}'|^{-1} \delta\rho_{\rm ext}({\bf r}')$, which
is similar to $\{ V_{{\bf R}\alpha} \}$ in the constrained DFT, and
$\delta \rho_{\rm ext}({\bf r})$ itself has a meaning of the electron density,
which is controlled by the orbital
occupations $\{ n_{{\bf R}\alpha} \}$.\footnote{Yet, there is also a
difference: $\delta V_{\rm ext}({\bf r})$ is the local potential, whereas $\{ V_{{\bf R}\alpha} \}$ act
on the individual Wannier functions $\{ \tilde{W}_{\bf R}^\alpha \}$.
This corresponds to the external potential
$\delta V_{\rm ext}({\bf r},{\bf r}') = \sum_{{\bf R} \alpha} V_{{\bf R}\alpha} | \tilde{W}_{\bf R}^\alpha \rangle \langle \tilde{W}_{\bf R}^\alpha |$
in the form of the projector operators,
which are essentially \textit{nonlocal}. This is the one of the basic differences between constrained DFT and RPA.
The role of this nonlocality
in the screening of the Coulomb interactions is not fully understood \cite{PRB05}.}
In principle, the perturbation of the electron density can also depend on time, that would correspond
to the time-dependent DFT \cite{RungeGross,OnidaReiningRubio}.
Although such an extension can be certainly done, we will consider only the static (or time-independent)
limit, strictly following the definition (\ref{eqn:Coulomb.definition}).
\item
$\delta V_{\rm ext}({\bf r})$ is treated as a perturbation.
Then, the updates for the Kohn-Sham orbitals,
$\{ \delta \psi_i \}$, can be easily calculated in the first order of $\delta V_{\rm ext}({\bf r})$,
by using the regular perturbation theory.
The screening of the Coulomb interaction in RPA is entirely related
with the change of $\{ \psi_i \}$ or, in the other words, the relaxation of the Kohn-Sham orbitals.
\item
The change of the electron density is calculated from $\{ \delta \psi_i \}$
in the first order of $\delta V_{\rm ext}({\bf r})$, namely,
$\delta \rho({\bf r}) = \sum_i [ \delta \psi_i^\dagger({\bf r}) \psi_i({\bf r}) +
\psi_i^\dagger({\bf r}) \delta \psi_i({\bf r}) ]$.
\item
$\delta \rho({\bf r})$
results in the
additional change of the Coulomb potential, also in the first order of $\delta V_{\rm ext}({\bf r})$:
$\delta V_{\rm H}({\bf r}) = e^2 \int d{\bf r}' |{\bf r}$$-$${\bf r}'|^{-1} \delta \rho({\bf r}')$.
\item
The steps (ii)-(iv) are repeated for the new potential, $\delta V_{\rm ext}({\bf r})$$+$$\delta V_{\rm H}({\bf r})$.
Then, the potential is updated again,
and so on.
This is equivalent to the iterative solution of the Kohn-Sham equations within DFT.
The constrained DFT also takes into account the change of the exchange-correlation
potential, $V_{\rm XC}({\bf r})$. However, this
change is typically treated in LDA,
where the effect is small.
\item
After summing up an infinite number of updates for $\delta V_{\rm H}({\bf r})$,
the screened Coulomb interaction in RPA is defined as
$U({\bf r},{\bf r}') = \delta V_{\rm ext}({\bf r})/\delta \rho_{\rm ext}({\bf r}') +
\delta V_{\rm H}({\bf r})/\delta \rho_{\rm ext}({\bf r}')$.
\end{enumerate}
Then,
one can easily to show that
$U({\bf r},{\bf r}')$ satisfies the Dyson equation \cite{Hedin,FerdiGunnarsson}:
\begin{equation}
\fl U({\bf r},{\bf r}') = e^2|{\bf r}-{\bf r}'|^{-1} + \int d {\bf r}''  \int d {\bf r}'''
e^2|{\bf r}-{\bf r}''|^{-1}{\cal P}({\bf r}'',{\bf r}''') U({\bf r}''',{\bf r}'),
\label{eqn:Coulomb.RPA.Dyson}
\end{equation}
where ${\cal P}({\bf r},{\bf r}')$ is the polarization function, which is obtained from the
first-order perturbation
theory expansion
for $\{ \delta \psi_i \}$:
\begin{equation}
{\cal P}({\bf r},{\bf r}',\omega) = 2 \sum_{ij}
\frac{(f_i-f_j)
\psi^\dagger_i({\bf r}) \psi_j({\bf r})
\psi^\dagger_j({\bf r}') \psi_i({\bf r}')}
{\omega - \varepsilon_j + \varepsilon_i + i\delta (f_i-f_j)}.
\label{eqn:Polarization_Function}
\end{equation}
The $\omega$-dependence of ${\cal P}({\bf r},{\bf r}',\omega)$ corresponds to the time-dependent
perturbation $\delta v_{\rm ext}({\bf r}) \exp(-i\omega t)$.
The static limit corresponds to
${\cal P}({\bf r},{\bf r}',0) \equiv {\cal P}({\bf r},{\bf r}')$. The prefactor ``2'' stands for
two spin channels in the case of non-spin-polarized LDA.

\subsection{\label{sec:cDFTRPA}Combining Constrained DFT and RPA}

\begin{figure}[h!]
\begin{center}
\resizebox{7cm}{!}{\includegraphics{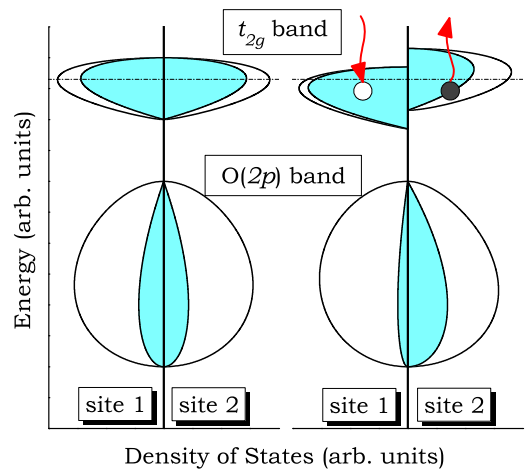}}
\end{center}
\caption{\label{fig.pdhybridization}
A schematic view on the change of the $p$-$d$ hybridization in the
oxygen $2p$ and $t_{2g}$ bands of the transition-metal oxides associated
with the repopulation of the Wannier orbitals at the neighboring
transition-metal (TM) sites
at two sides of the reaction $2(d^n)$$\rightleftharpoons$$d^{n+1}$$+$$d^{n-1}$.
Left panel corresponds to the ground-state configuration, $2d^n$.
In the right panel, the removal (addition) of an electron from (to)
the Wannier orbital in the $t_{2g}$ part of the spectrum is simulated
by the shift of these orbitals relative to the Fermi level
(shown by dot-dashed line). Around
each transition-metal site, it changes the Coulomb potential,
which controls the distribution of the $3d$-states
as well as the hybridization between
transition-metal $3d$ and oxygen $2p$ states.
Generally, the removal of an electron from (or the addition of an electron to)
the Wannier orbital is partially compensated by the change of the
amount of the $3d$-states (shown by shaded area), which is admixed
into the oxygen $2p$ band. This transfer of the spectral weight
works as an efficient channel of
screening of the local Coulomb interactions in the
transition-metal oxides \protect\cite{PRB06a}.
}
\end{figure}

  The screened
Coulomb interactions
in RPA satisfies one very important property \cite{Ferdi04},
which directly follows from the Dyson equation (\ref{eqn:Coulomb.RPA.Dyson}).
Suppose that one can identify two different channels of screening, by
dividing the
full polarization function (\ref{eqn:Polarization_Function}) into two parts:
${\cal P} = {\cal P}_1 + {\cal P}_2$.
Then, the screened Coulomb interaction $U$ can be computed in two steps.
\begin{enumerate}
\item
By replacing total ${\cal P}$ by ${\cal P}_1$
in (\ref{eqn:Coulomb.RPA.Dyson}), one can compute the
effective interaction $U_1$,
which takes into account only the first channel of screening.
\item
Then, the final interaction $U$,
corresponding to ${\cal P}_1$$+$${\cal P}_2$,
can be obtained
from $U_1$
again by using the Dyson equation (\ref{eqn:Coulomb.RPA.Dyson}),
but where the full polarization function
is replaced by ${\cal P}_2$:
$$
\fl
U({\bf r},{\bf r}') = U_1({\bf r},{\bf r}') + \int d {\bf r}''  \int d {\bf r}'''
U_1({\bf r},{\bf r}''){\cal P}_2({\bf r}'',{\bf r}''') U({\bf r}''',{\bf r}').
$$
\end{enumerate}
This property has two important consequences.
\begin{enumerate}
\item
In order to calculate $U$, we would like to start with
the LDA band structure.
However, for Mott insulators, LDA yields an incorrect metallic behavior in the region
of $L$-bands. Therefore, we should get rid of this unphysical metallic screening,
which will always appear in RPA if one starts from LDA.
More generally, RPA may not be a good approximation for treating the electron
correlations in the
narrow
$L$-bands. Therefore, the general strategy is to get rid of the
RPA self-screening associated with all kind of transitions between $L$-bands in the
polarization function (\ref{eqn:Polarization_Function}), and to replace it by
a more rigorous model of screening in the process of solution of the
Hubbard model (\ref{eqn:Hmanybody}) \cite{Imai,Ferdi04}.
The suppression of the transitions between $L$-bands in RPA
is similar to switching off all transfer integrals between
Wannier functions in the constrained DFT \cite{PRB06a}.\footnote{Indeed, suppose that we have
replaced $\{ \psi_i \}$ for the $L$-bands in (\ref{eqn:Polarization_Function}) by the
Wannier functions $\{ \tilde{W}_{\bf R}^\alpha \}$, and switched off all
transfer integrals. In the ground state of such an auxiliary system,
all occupation numbers are equal, $f_i$$=$$f_j$. Therefore, the polarization
function
(\ref{eqn:Polarization_Function}) will vanish.}
\item
Our goal is to evaluate the screening of the Coulomb interactions associated with the
relaxation of the Wannier functions $\{ \tilde{W}_{\bf R}^\alpha \}$
in the region of $L$-bands as well as of all other Kohn-Sham
orbitals $\{ \psi_i \}$ in the remaining part of the spectrum.
Suppose that we are working in a (flexible) basis of atomic orbitals $\{ \tilde{\chi}_{\bf R}^\alpha \}$,
and both $\{ \tilde{W}_{\bf R}^\alpha \}$ and $\{ \psi_i \}$ are expanded over this basis:
\begin{equation}
\left\{
\begin{array}{c}
\tilde{W}_{\bf R}^\alpha \\
\psi_i                   \\
\end{array}
\right\}
= \sum_{{\bf R}' \beta} c_{{\bf R}'}^\beta \tilde{\chi}_{{\bf R}'}^\beta.
\label{eqn:orbitals_relaxation}
\end{equation}
Then,
within linear response theory,
the relaxation of $\{ \tilde{W}_{\bf R}^\alpha \}$ and $\{ \psi_i \}$ will consist of
two parts: the relaxation of the basis functions $\{ \tilde{\chi}_{\bf R}^\alpha \}$ and
the relaxation of the coefficients $\{ c_{\bf R}^\alpha \}$ of the expansion over these
basis functions, which corresponds to the change of the hybridization in the process
of screening.
Therefore, it has been proposed (though somewhat heuristically) that
the screened Coulomb interactions in the $L$-band can be computed
in two steps \cite{PRB06a,PRB05}.
For the sake of clarity, let us assume the main contribution to the $L$-bands comes
from the $3d$ orbitals of the transition-metal sites.
\begin{enumerate}
\item
The first step takes into account the screening
caused by the relaxation of the basis functions $\{ \tilde{\chi}_{\bf R}^\alpha \}$.
It
can be easily evaluated
in the framework of constrained DFT. In this case, the Kohn-Sham equations are solved
numerically on a finite grid of points in the real space, and $\{ n_{{\bf R}\alpha} \}$ are
the constrained $3d$ occupation numbers inside certain atomic spheres \cite{Dederichs}.
On the other hand, since RPA is based on the perturbation theory expansion,
similar calculations in terms of the
polarization function (\ref{eqn:Polarization_Function}) would require an enormous number of states
$\{ \psi_i \}$, both in the occupied and unoccupied parts of the spectrum.
Probably, this is one of the reasons why the conventional RPA calculations converge very
slowly with the increase of the number of bands in the unoccupied part of the spectrum \cite{HybertsenLouie},
and are extremely sensitive to the treatment of the core states \cite{FerdiGunnarsson}.
The screening caused by
the relaxation of the Kohn-Sham orbitals $\{ \psi_i \}$ in the subspace orthogonal to $L$
(for example, the screening of localized $3d$ electrons by outer electrons in the
transition-metal compounds) can be also easily
taken into account in the
conventional constrained
DFT calculations \cite{Gunnarsson1989,AnisimovGunnarsson,PRB94a}.
\item
However, what the constrained DFT cannot do is to treat the screening
of Coulomb interactions amongst $3d$ electrons by \textit{the same} $3d$ electrons \cite{PRB05,PRL05}.\footnote{
The procedure implies that
the atomic $3d$ charges can be divided into two parts, where the first part is subjected to the
constraint conditions, while the second part is allowed to participate in the screening.
Although, it can be formally done within constrained DFT \cite{PRB96},
the actual computational scheme is rather laborious, and in many cases the procedure of dividing
the atomic charges
into the ``screened'' and ``screening'' parts
is not well defined.}
Since the atomic $3d$ occupation numbers are rigidly fixed by the constraint conditions,
the $3d$ electrons from other sites of the system cannot compensate the change of the
number of
the $3d$ electrons at the central site, and vise versa. However, such a channel of screening may exist.
Suppose that our $L$-band is mainly constructed from the transition-metal $3d$ orbitals
(the concrete example is the $t_{2g}$ band
in the transition-metal oxides), and there is
another, say  oxygen $2p$ band, which has an appreciable weight of
the atomic $3d$ orbitals (of both $t_{2g}$ and $e_g$ symmetry) coming from the
hybridization between oxygen and transition-metal sites (Figure \ref{fig.pdhybridization}).
Furthermore, the redistribution of the electron density in the $t_{2g}$ band, associated with
the reaction
($n_{{\bf R}\alpha}$,$n_{{\bf R}'\beta}$)$\rightleftharpoons$($n_{{\bf R}\alpha}$$+$$1$,$n_{{\bf R}'\beta}$$-$$1$),
will change the Coulomb potential around each transition-metal site.
If the number of $t_{2g}$ electrons is increased by the constraint conditions,
the Coulomb potential becomes more repulsive and vise versa. The more repulsive
Coulomb potential will additionally push the $3d$ states from the oxygen $2p$ band
to the higher-energy part of the spectrum. Therefore,
around certain transition-metal sites,
the change of
the number of $3d$ electrons in the $t_{2g}$ band will be partly compensated by the
change of the
$3d$-electron density in the region of oxygen $2p$ band.
This channel of screening can be easily evaluated in RPA, by rewriting
(\ref{eqn:Coulomb.RPA.Dyson}) in the matrix form
\begin{equation}
\hat{U} = \left[ 1- \hat{u} \hat{\cal P} \right]^{-1} \hat{u},
\label{eqn:Coulomb.RPA.Dyson.matrix}
\end{equation}
and assuming that all other channels of screening
are already
included in the definition the ``bare Coulomb interaction'' $\hat{u}$, derived
from the constrained DFT \cite{PRB06a}.
Since the polarization matrix $\hat{\cal P}$ in (\ref{eqn:Coulomb.RPA.Dyson.matrix})
is aimed to
describe the self-screening of the $3d$ electrons,
it should consist of the matrix elements of
(\ref{eqn:Polarization_Function}) in the basis of atomic $3d$ orbitals,
after the subtraction of
the unphysical metallic screening associated with RPA transitions
between $t_{2g}$ band.
\end{enumerate}
\end{enumerate}

\section{\label{sec:Solution}Solution of Model Hamiltonian}

  It is virtually impossible to provide a comprehensive analysis of all possible
methods of the solution of the low-energy model (\ref{eqn:Hmanybody}), and this is
definitely beyond the scopes of this review article.

  One option is the dynamical mean-field theory (DMFT) \cite{DMFT},
which becomes one of the popular low-energy solvers today.
The idea of DMFT is to map the many-body lattice problem to a single-site impurity
problem with effective parameters. The
vast majority of DMFT applications for realistic compounds
have been focusing on
the analysis of spectroscopic properties, especially in the context of the
metal-insulator transition, although some extensions for the ground-state properties,
such as
calculations of the
total energies and phonons, are also available today. Many examples of recent
applications of DMFT can be found in the review articles \cite{DMFTPhysToday,DMFTKotliar,DMFTHeld}.
The conventional DMFT becomes exact in the limit of infinite coordination
numbers or, equivalently, infinite dimensions, when all nonlocal correlations
vanish. In order to treat these nonlocal correlations, it is essential to go beyond
the single-site approximation. This is one of the challenging problems in DMFT, and
the recent progresses along this line can be found in \cite{DMFTMaier}.
Another limitation of DMFT is that, in order to be exact,
it is typically used in the combination with the Quantum Monte Carlo (QMC) method,
which provides an exact solution for the impurity model.
However, current applications of the QMC method are typically restricted
by rather high temperatures, which are substantially higher than, for example,
the magnetic transition temperatures in many strongly correlated compounds.
From this point of view, the method does not appear to be sufficiently useful
for studying the phenomena of
spin and orbital ordering. Probably, some of these difficulties may be
overcome by applying the projective QMC method \cite{PRLArita}.

  Unlike DMFT, the path-integral renormalization group (PIRG) method is
mainly oriented on the description of the ground-state properties of strongly
correlated systems. The entire procedure includes the following
steps \cite{ImadaKashima,KashimaImada,MizusakiImada}:
\begin{enumerate}
\item
The numerical construction of truncated basis of ${\cal L}$ Slater determinants, which
provides the best representation for the ground-state wavefunction;
\item
Calculation of the total energy and its variance in the obtained basis;
\item
Extrapolation
of the obtained results
to the full Hilbert space, which is achieved by a systematic increase of ${\cal L}$.
\end{enumerate}
The PIRG method has been recently applied
as the low-energy solver for studying the
correlation effects in the $t_{2g}$ bands of Sr$_2$VO$_4$ \cite{Imai} and YVO$_3$ \cite{OtsukaImada}.

  In the rest of this section we will discuss some details of the solution of the
model Hamiltonian (\ref{eqn:Hmanybody}), which will be directly
used for applications
considered in Section \ref{sec:applications}. We start with the simplest
Hartree-Fock method, which totally neglects the correlation effects.
Then, we consider simple corrections to the Hartree-Fock approximation, which
include some of these effects. One is the
perturbation theory for the total energy, and the other one is the
variational superexchange theory.

  All model calculations are performed in the basis of Wannier functions
$\{ \tilde{W}_{\bf R}^\alpha \}$, which have a finite weight
at the central transition-metal site
as well as the oxygen and other atomic sites located in its
neighborhood. In order to calculate the local quantities,
associated with the transition-metal atoms, such as spin and orbital magnetic
moments or the distribution of the electron density,
the Wannier
functions are expanded over the original basis $\{ \tilde{\chi}^\alpha_{\bf R} \}$.
Then, all
aforementioned quantities are calculated by integrating over appropriate regions
of the real space surrounding the
transition-metal sites, like the atomic spheres in the LMTO method \cite{LMTO1,LMTO2,LMTO3}.

\subsection{\label{sec:HFApproximation}Hartree-Fock Approximation}

  The Hartree-Fock method provides the simplest approximation to the many-body
problem (\ref{eqn:Hmanybody}). In this case, the trial
many-electron
wavefunction is searched in the form of a single Slater determinant
$|S\{ \varphi_k \} \rangle$, constructed from the one-electron orbitals
$\{ \varphi_k \}$.
In this notation, $k$ is a collective
index combining the momentum
${\bf k}$ of the first Brillouin zone, the band number, and
the spin ($s$$=$ $\uparrow$ or $\downarrow$) of the particle.
The one-electron orbitals are subjected to the variational principle and requested to minimize
the total energy
$$
E_{\rm HF}= \min_{\{ \varphi_k\}}
\langle S\{ \varphi_k \}|\hat{\cal H}| S\{ \varphi_k \} \rangle
$$
for a given number of particles $\cal{N}$.
This minimization is equivalent to the
solution of Hartree-Fock equations for $\{ \varphi_k \}$:
\begin{equation}
\left( \hat{h}_{\bf k} + \hat{\cal V} \right) | \varphi_{k} \rangle =
\varepsilon_k | \varphi_k \rangle,
\label{eqn:HFeq}
\end{equation}
where
$\hat{h}_{\bf k}$$\equiv$$\| h_{\bf k}^{\alpha \beta} \|$
is the one-electron part of the model Hamiltonian (\ref{eqn:Hmanybody}) in the reciprocal space,
$h_{\bf k}^{\alpha \beta}$$=
$$\sum_{{\bf R}'} h^{\alpha \beta}_{{\bf R}{\bf R}'} e^{-i {\bf k} \cdot ({\bf R}-{\bf R}')}$,
and $\hat{\cal V}$$\equiv$$\| {\cal V}_{\alpha \beta} \|$ is the Hartree-Fock potential,\footnote{
For the sake of simplicity, we drop the atomic index ${\bf R}$ in the notations of ${\cal V}_{\alpha \beta}$,
although such a dependence can take place (for example, in the case of inequivalent transition-metal
sites in the distorted perovskite structure), and was actually taken into account in
realistic calculations considered in Section \ref{sec:applications}.
}
\begin{equation}
{\cal V}_{\alpha \beta} = \sum_{\gamma \delta}
\left( U_{\alpha \beta \gamma \delta} - U_{\alpha \delta \gamma \beta} \right) n_{\gamma \delta}.
\label{eqn:HFpot}
\end{equation}
Equation (\ref{eqn:HFeq}) is solved self-consistently together with the equation
$$
\hat{n} = \sum_k^{occ} | \varphi_k \rangle \langle \varphi_k |
$$
for the
density matrix $\hat{n}$$\equiv$$\|n_{\alpha \beta}\|$ in the basis
of Wannier functions. After iterative solution of the Hartree-Fock equations,
the total energy
can be computed as
$$
E_{\rm HF} = \sum_k^{occ} \varepsilon_k -\frac{1}{2}
\sum_{\alpha \beta} {\cal V}_{\beta \alpha} n_{\alpha \beta}.
$$

  By knowing $\{ \varepsilon_k \}$ and $\{ \varphi_k  \}$,
one can construct the one-electron (retarded) Green function,
$$
\hat{\cal G}_{{\bf RR}'}(\omega) = \sum_k
\frac{ | \varphi_k \rangle \langle \varphi_k |}
{ \omega - \varepsilon_k + i\delta } e^{i {\bf k} \cdot ({\bf R}-{\bf R}')},
$$
which can be used for many applications. For example, the interatomic magnetic
interactions corresponding to infinitesimal rotations of
spin magnetic moments near the equilibrium can be computed as \cite{Liechtenstein,TRN}:
\begin{equation}
J_{{\bf RR}'} = \frac{1}{2 \pi} {\rm Im} \int_{-\infty}^{\varepsilon_{\rm F}}
d \omega {\rm Tr}_L \left\{ \hat{\cal G}_{{\bf RR}'}^\uparrow (\omega)
\Delta \hat{\cal V} \hat{\cal G}_{{\bf R}'{\bf R}}^\downarrow (\omega)
\Delta \hat{\cal V} \right\},
\label{eqn:JHeisenberg}
\end{equation}
where
$\hat{\cal G}_{{\bf RR}'}^{\uparrow, \downarrow}$$=$$\frac{1}{2} {\rm Tr}_S
\{ (\hat{1}$$\pm$$\hat{\sigma}_z ) \hat{\cal G}_{{\bf RR}'} \}$
is the projection of the Green function onto the majority ($\uparrow$)
and minority ($\downarrow$) spin states,
$\Delta \hat{\cal V}$$=$${\rm Tr}_S \{ \hat{\sigma}_z \hat{\cal V} \}$
is the magnetic (spin) part of the Hartree-Fock potential,
${\rm Tr}_S$ (${\rm Tr}_L$) denotes the trace over the spin (orbital) indices,
$\hat{1}$ and
$\hat{\sigma}_z$ is the unity and Pauli matrix, respectively,
and $\varepsilon_{\rm F}$ is the Fermi energy.\footnote{
According to  the definition (\ref{eqn:JHeisenberg}),
$J_{{\bf RR}'}$$>$$0$ ($<$$0$)
means that for a
given magnetic
state, the spin arrangement in the bond $\langle {\bf RR}' \rangle$
corresponds to the local minimum (maximum) of the total energy.
However, in the following we will use the universal notations,
according to which $J_{{\bf RR}'}$$>$$0$ and $<$$0$ will
stand
the ferromagnetic and antiferromagnetic
coupling, respectively.}

  The parameters $\{ J_{{\bf RR}'} \}$ are not universal, and
depend on the magnetic state in which they are calculated,
for example, through the change of the orbital ordering \cite{PRB06b}
or the change of the electronic structure by the
magnetic ordering \cite{Springer,PRL03}.

\subsection{\label{sec:2ndorder}Second Order Perturbation Theory for the Correlation Energy}

  The simplest way of going beyond the Hartree-Fock approximation is to include the correlation
interactions
in the second order of perturbation theory for the total energy \cite{Friedel,Kajzar,Treglia}.
It shares common
problems of the regular (nondegenerate) perturbation theory.
Nevertheless, by using this technique one can calculate
relatively easily the corrections to the total energy, starting from
the Hartree-Fock wavefunctions. This
method
is expected
to work well for the systems where the orbital degeneracy is lifted
(for example, by the crystal-field splitting)
and the ground state is described reasonably well
by a single Slater determinant, so that other corrections can be treated as a perturbation.

  The correlation interaction (or the interaction of fluctuations)
is defined as the difference between true many-body
Hamiltonian (\ref{eqn:Hmanybody}), and its one-electron
counterpart, obtained at the level of the Hartree-Fock approximation:
\begin{equation}
\hat{\cal{H}}_C = \sum_{\bf R} \left(
\frac{1}{2} \sum_{\alpha \beta \gamma \delta}
U_{\alpha \beta \gamma \delta}
\hat{c}^\dagger_{{\bf R}\alpha} \hat{c}^\dagger_{{\bf R}\gamma}
\hat{c}^{\phantom{\dagger}}_{{\bf R}\beta} \hat{c}^{\phantom{\dagger}}_{{\bf R}\delta} -
\sum_{\alpha \beta} {\cal V}_{\alpha \beta}
\hat{c}^\dagger_{{\bf R}\alpha} \hat{c}^{\phantom{\dagger}}_{{\bf R}\beta} \right).
\label{eqn:H2ndorder}
\end{equation}
It is important to note that although some of the matrix elements
$U_{\alpha \beta \gamma \delta}$ can be large, they also
contribute to the Hartree-Fock potentials ${\cal V}_{\alpha \beta}$.
Therefore, generally, one can expect some cancelation of contributions
in the first and second parts of (\ref{eqn:H2ndorder}), which formally extend the
applicability of the perturbation theory even for relatively large
$U_{\alpha \beta \gamma \delta}$. For example, in a number of cases
such a strategy can be applied even for the bare Coulomb interactions
in isolated atoms \cite{Veselov}.

  By treating $\hat{\cal{H}}_C$ as a perturbation, the correlation energy can be
easily
estimated as \cite{Friedel,Kajzar,Treglia}:
\begin{equation}
E_C^{(2)} = - \sum_{S} \frac{
\langle G | \hat{\cal{H}}_C | S \rangle \langle S | \hat{\cal{H}}_C | G \rangle }
{E_{\rm HF}(S) - E_{\rm HF}(G)},
\label{eqn:dE2ndorder}
\end{equation}
where $|G \rangle$ and $|S \rangle$ are the Slater determinants
corresponding to the low-energy ground state
(in the Hartree-Fock approximation), and the
excited state, respectively.
Due to the variational properties of the
Hartree-Fock method, the only processes that may contribute to
$E_C^{(2)}$ are the
two-particle excitations, for which each of
$|S \rangle$ is obtained from $|G \rangle$ by replacing
two one-electron orbitals, say $\varphi_{k_1}$
and $\varphi_{k_2}$, from the occupied part of the spectrum
by two unoccupied orbitals, say $\varphi_{k_3}$ and $\varphi_{k_4}$ \cite{Veselov}.
Hence, using the notations of Section \ref{sec:Model}, the matrix elements take the
following form:
\begin{equation}
\langle S | \hat{\cal{H}}_C | G \rangle =
\langle k_3 k_4 | v_{\rm scr} |
k_1 k_2 \rangle -
\langle k_3 k_4 | v_{\rm scr} |
k_2 k_1 \rangle.
\label{eqn:dHmelement}
\end{equation}
By employing further the approximation
of noninteracting quasiparticles,
the denominator in (\ref{eqn:dE2ndorder}) can be replaced
by the linear combination of Hartree-Fock  eigenvalues:
$E_{\rm HF}(S)$$-$$E_{\rm HF}(G) \approx \varepsilon_{k_3}$$+$$\varepsilon_{k_4}$$-
$$\varepsilon_{k_1}$$-$$\varepsilon_{k_2}$ \cite{Friedel,Kajzar,Treglia}.
The matrix elements (\ref{eqn:dHmelement})
satisfy the following condition:
$\langle S | \hat{\cal{H}}_C | G \rangle$$\sim$$\frac{1}{N}
\sum_{\bf R} e^{i({\bf k}_3+{\bf k}_4-{\bf k}_1-{\bf k}_2) \cdot {\bf R}}$
($N$ being the number of sites),
provided that
the effective Coulomb interactions are diagonal with respect to the site
indices. In the second-order perturbation theory
one can estimate relatively easily both on-site (${\bf R}$$=$$0$) and
intersite (${\bf R}$$\ne$$0$) contributions to $E_C^{(2)}$.
The ${\bf R}$$=$$0$ term corresponds to the commonly used
single-site approximation for the correlation interactions,
which becomes exact in the limit of infinite spacial dimensions \cite{DMFT}.

  In principle, one can go beyond the second order perturbation
theory
and consider, for example, the single-site approximation for the
$T$-matrix \cite{Kanamori}.
In this case, the expression for the energy
of electron-electron interactions
has the same form
as in the Hartree-Fock method, but with $\hat{U}$ being replaced by the effective $T$-matrix,
which takes into account the correlation effects.
The method has been employed
for the series of distorted transition-metal perovskite oxides \cite{JETP07}, where
the degeneracy of the Hartree-Fock ground state is lifted by the crystal field.
It that case,
the application of the $T$-matrix theory changed only some quantitative
conclusions, whereas the main trends for the correlation energy were
captured already by the second order perturbation theory.

\subsection{\label{sec:SEmethod}Atomic Multiplet Structure and
Superexchange Interactions}

  The variational superexchange theory
takes into account the multiplet structure of the excited atomic states.
By using this technique one can
study the effect of the electron correlations on the spin and orbital ordering.
However, it is limited by typical approximations made in the
theory of superexchange interactions, which treat all transfer integrals as
a perturbation.

  The superexchange interaction in the bond $\langle {\bf R}{\bf R}' \rangle$ is basically
the gain of the kinetic energy, which is acquired by an
electron at the center ${\bf R}$ in the
process of virtual hoppings into the subspace of unoccupied
orbitals at the center ${\bf R}'$, and vice versa \cite{PWA,KugelKhomskii}.
Therefore,
the energy gain caused by virtual hoppings
in the bond $\langle {\bf R}{\bf R}' \rangle$ can be found as \cite{PRB06b,KugelKhomskii,Oles05}:
\begin{equation}
\fl
{\cal T}_{{\bf R}{{\bf R}'}} = - \left\langle G
\left| \hat{h}_{{\bf R}{{\bf R}'}} \left( \sum_M \frac{{\hat{\mathscr{P}}}_{{\bf R}'}|{{\bf R}'} M
\rangle \langle {{\bf R}'} M|
{\hat{\mathscr{P}}}_{{\bf R}'}}{E_{{{\bf R}'}M}} \right) \hat{h}_{{{\bf R}'}{\bf R}} +
({\bf R} \leftrightarrow {{\bf R}'}) \right| G \right\rangle,
\label{eqn:egain}
\end{equation}
where
$G$ is the ground-state wavefunction of the lattice of isolated centers,\footnote{
In the present context, the ``lattice center'' means either isolated atomic site
or a molecule. An example of the molecular solid will be considered
in Section \ref{sec:KO2}.
}
each of which
accommodates $n$ electrons,
$E_{{{\bf R}'}M}$ and $|{{\bf R}'}M \rangle$ stand for the eigenvalues and eigenvectors of
the
excited ($n$$+$$1$ electron)
configurations of the center ${{\bf R}'}$, and $\hat{\mathscr{P}}_{{\bf R}'}$ is a projector
operator, which enforces the Pauli principle and suppresses any hoppings into
the subspace of occupied orbitals at the center ${{\bf R}'}$ \cite{PRB06b}.

  The formulation is extremely simple for the $n = 1$ compounds,
like YTiO$_3$ and LaTiO$_3$. In this case, there is only one
electron residing at each transition-metal site. This is essentially an
one-electron problem, where each atomic state is described by certain
one-electron orbital $\varphi_{\bf R}$ and $G$ is the single Slater
determinant constructed from $\{ \varphi_{\bf R} \}$ belonging to different
transition-metal sites \cite{PRB06b}.
A similar formulation can be performed for the hole
spin-orbitals $\{ \alpha_{\bf R} \}$ of compounds where at each
lattice center there is only one unbalanced hole. Such a situation
holds for the alkali hyperoxides, which will be considered
in Section \ref{sec:KO2}.

  The total energy of the system in the superexchange approximation
is obtained after summation over all bonds, which
should be combined with the site-diagonal elements, incorporating
the effects of the crystal-field splitting and the relativistic
spin-orbit interaction:
$$
E_{\rm SE}= \sum_{\bf R}
\langle \varphi_{\bf R} | \hat{h}_{\bf RR} | \varphi_{\bf R} \rangle +
\sum_{\langle {\bf RR}' \rangle}
{\cal T}_{{\bf R}{{\bf R}'}}.
$$
Finally, the set of occupied orbitals $\{ \varphi_{\bf R} \}$ is obtained
by minimizing
$E_{\rm SE}$. This can be done by using, for instance, the steepest descent method.

\section{\label{sec:applications}Examples and Applications for Realistic Compounds}

\subsection{Cubic Perovskites: SrVO$_3$}

\begin{figure}[h!]
\begin{center}
\resizebox{7cm}{!}{\includegraphics{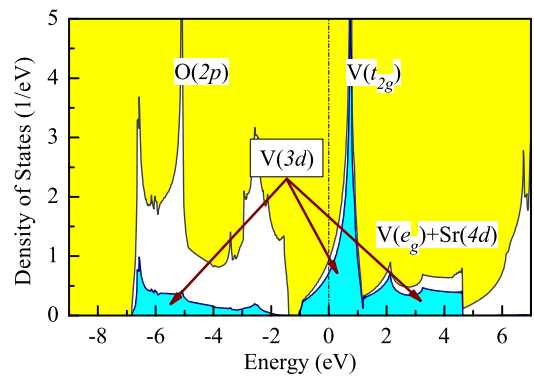}}
\end{center}
\caption{\label{fig.SrVO3DOS} Total and partial densities of states
of SrVO$_3$ in the local-density approximation.
The shaded area shows the contributions of the vanadium $3d$ states.
Other symbols show positions of the main bands.
The Fermi level is at zero energy.}
\end{figure}

  SrVO$_3$ is a rare example of perovskite compounds, which crystallizes in the
ideal cubic structure. It attracted a considerable attention in the connection
with the bandwidth control of the metal-insulator transition \cite{IFT,Liebsch2003}.
The region of interest is the $t_{2g}$ band, which is located near the Fermi level (Figure \ref{fig.SrVO3DOS}).

\subsubsection{Transfer integrals and Wannier functions.}

  For cubic compounds, the separation of the basis functions into
$\{  \tilde{\chi}_t \}$ and $\{ \tilde{\chi}_r \}$,
which is required in the
downfolding method,
is rather straightforward: three $t_{2g}$ orbitals centered at each vanadium site
of SrVO$_3$
are taken as the $\{ \tilde{\chi}_t \}$ orbitals, and the rest of the basis functions
is associated with the $\{ \tilde{\chi}_r \}$ orbitals.

\begin{figure}[h!]
\begin{center}
\resizebox{4.5cm}{!}{\includegraphics{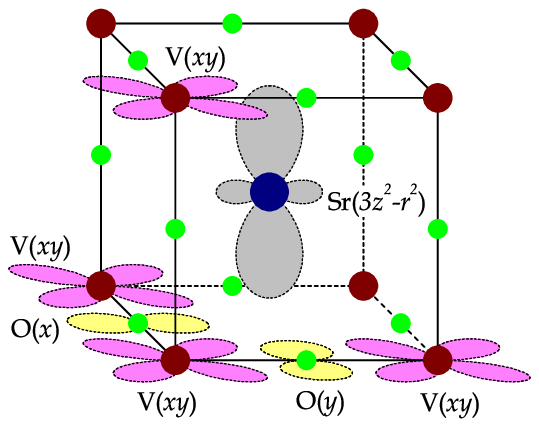}}
\resizebox{6.5cm}{!}{\includegraphics{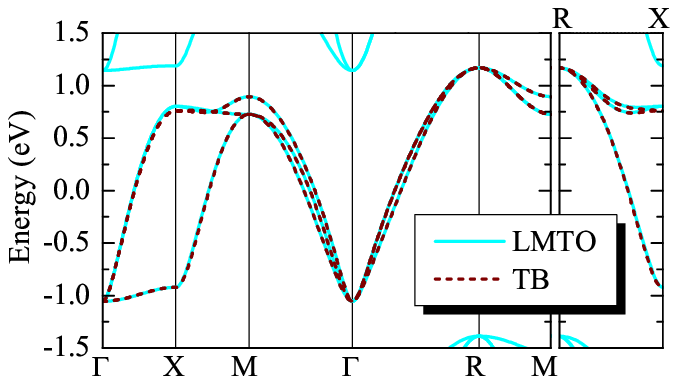}}
\end{center}
\caption{\label{fig.SrVO3bands}
Left panel: Crystal structure of cubic perovskites and atomic wavefunctions
mediating transfer interactions between vanadium $t_{2g}$ orbitals.
The standard V($xy$)-O($y$)-V($xy$) and V($xy$)-O($x$)-V($xy$) interactions
operate in the $x$- and $y$-directions, respectively. The
V($xy$)-Sr($3z^2$-$r^2$)-V($xy$) interaction operate in the ``forbidden''
$z$-direction.
Right panel: LDA energy bands of SrVO$_3$ obtained in the
original electronic-structure calculations using the
LMTO method and
after the ``tight-binding'' (TB) parametrization using the downfolding method.
Notations of the high-symmetry points of the Brillouin zone are
taken from \protect\cite{BradlayCracknell}.}
\end{figure}

  The downfolding procedure is nearly perfect and well reproduces the behavior
of three $t_{2g}$ bands (Figure \ref{fig.SrVO3bands}).
As expected for cubic compounds, the nearest-neighbor $dd\pi$-interactions mediated by the
oxygen $2p$ orbitals are the strongest (Table~\ref{tab.cubic_SrVO3}).
For the $xy$-orbitals, it operates in the $x$ and
$y$ directions.\footnote{
Similar dependencies for the $yz$ and $zx$ orbitals are obtained by the
cyclic permutation of the indices $x$, $y$, and $z$.}
However, there is also an appreciable $dd\delta$-interaction
operating in the ``forbidden'' direction (for example, the direction
$z$ in the case of the $xy$ orbitals). These interactions are mediated by the
strontium $4d$ orbitals and strongly depend on the proximity of the latter to the
Fermi level.
The transfer integrals connecting different $t_{2g}$ orbitals are small
and contribute only to the longer-range interactions separated by the vectors
$(a,a,0)$ and $(a,a,a)$, where $a$ is the cubic lattice parameter \cite{SlaterKoster}. Other interactions are
considerably smaller.
\begin{table}[h!]
\caption{\label{tab.cubic_SrVO3}
Transfer integrals (in meV)
between first, second, and third nearest neighbors
in SrVO$_3$, separated by the vectors
$(a,0,0)$, $(a,a,0)$, and $(a,a,a)$, respectively
($a$ being the cubic lattice parameter).}
\begin{indented}
\item[]\begin{tabular}{@{}cccc}
\br
  & $(a,0,0)$ & $(a,a,0)$ & $(a,a,a)$  \\
\mr
$
\begin{array}{c}
                     \\

  {\it xy}           \\
  {\it yz}           \\
  {\it zx}           \\
\end{array}
$
&
$
\begin{array}{rrr}
 {\it xy} & {\it yz} & {\it zx} \\
\hline
 -210     &  0       &  0       \\
  0       & -23      &  0       \\
  0       &  0       & -210     \\
\end{array}
$
&
$
\begin{array}{rrr}
 {\it xy} & {\it yz} & {\it zx} \\
\hline
 -84      &  0       &  0       \\
  0       &  11      &  11      \\
  0       &  11      &  11      \\
\end{array}
$
&
$
\begin{array}{rrr}
 {\it xy} & {\it yz} & {\it zx} \\
\hline
 -6       &  3       &  3       \\
  3       & -6       &  3       \\
  3       &  5       & -6      \\
\end{array}
$\\
\br
\end{tabular}
\end{indented}
\end{table}

\begin{figure}[h!]
\begin{center}
\resizebox{4cm}{!}{\includegraphics{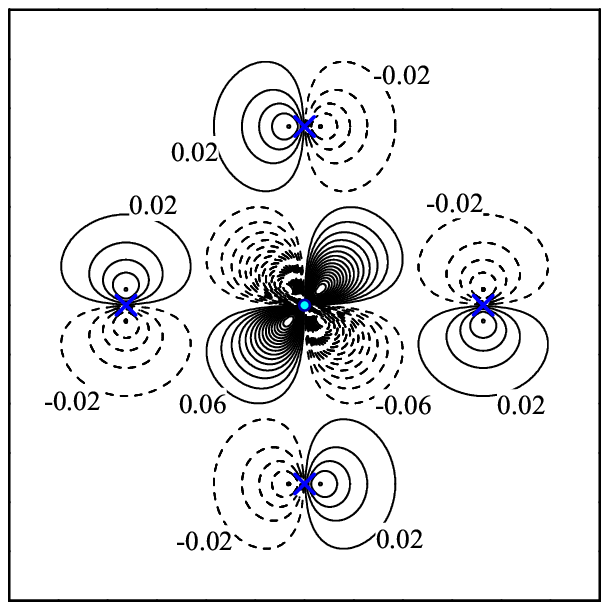}}
\resizebox{6cm}{!}{\includegraphics{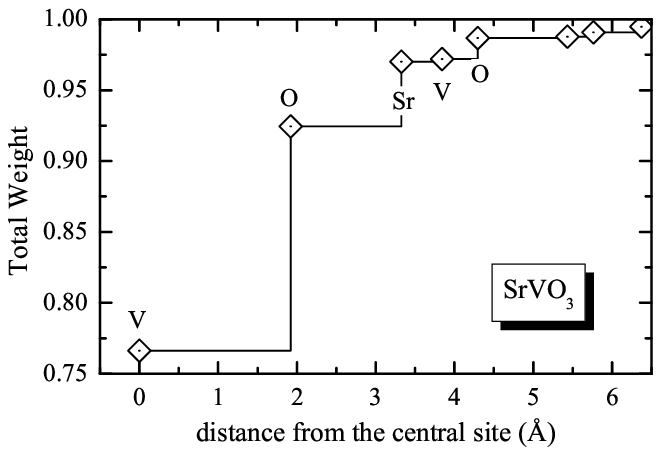}}
\end{center}
\caption{\label{fig.SrVO3WF}
Wannier function for the $t_{2g}$ band of SrVO$_3$ \protect\cite{PRB06a}.
Left panel shows the contour plot of the \textit{xy}-orbital
in the $(001)$ plane.
The solid and dashed lines correspond to the positive and negative values of the Wannier function.
The vanadium atom is located in the center of the plot, and other sites are four oxygen atoms located
in its nearest neighborhood.
Around each site, the Wannier function increases/decreses with the step $0.04$
from the values indicated on the graph.
Right panel shows
the total weight of the Wannier function
accumulated around the central vanadium site after adding every new sphere of
the neighboring sites.}
\end{figure}

  The shape of the Wannier functions is explained in Figure \ref{fig.SrVO3WF}.\footnote{
These Wannier functions have been reconstructed from the one-electron part
of the downfolded Hamiltonian using the ideas of the
LMTO method \cite{LMTO1,LMTO2,LMTO3}. The procedure has been explained in \cite{PRB06a}.}
Since $t_{2g}$ band is an \textit{antibonding} combination of the atomic
vanadium $3d$-$t_{2g}$ and oxygen $2p$ orbitals,
the Wannier function has nodes between vanadium and oxygen sites.
Right panel of Figure \ref{fig.SrVO3WF} illustrates the spacial extension of the Wannier functions.
It shows the weight of the Wannier function
accumulated around the central vanadium site after adding
every new sphere of the neighboring sites.
Since Wannier functions are normalized, their total weight is equal to one.
In the case of SrVO$_3$, 77\%
of the this weight belongs to the central vanadium site, 16\% is distributed
over four neighboring oxygen sites, about 5\% belongs to the next eight strontium sites,
and 1\% -- to the eight oxygen sites located in the fourth coordination sphere.
Other contributions are small.
Another quantity, characterizing the spread of the Wannier functions, is
the expectation value of square of the
position operator,
$\langle {\bf r}^2 \rangle$$=$$\langle \tilde{W}_{\bf R} | ({\bf r}$$-$${\bf R})^2 | \tilde{W}_{\bf R} \rangle$ \cite{MarzariVanderbilt},
which
in the case of SrVO$_3$ is about 2.37 \AA$^2$ \cite{PRB06a}.\footnote{
Somewhat smaller value (1.91 \AA$^2$) has been reported in \cite{Lechermann}.
Some overestimation of $\langle {\bf r}^2 \rangle$ is caused by
some additional approximations used in the process of
reconstruction of the Wannier functions from
the downfolded one-electron Hamiltonian, which has been employed in \cite{PRB06a}. Presumably,
the direct application of the projector-operator method can do a better job.
Nevertheless, as it was already pointed out in Section \ref{sec:downfoldingasprojector},
after the transformation (\ref{eqn:modifiedH}) of the Kohn-Sham Hamiltonian, the transfer integrals
derived from the downfolding method are totally equivalent to the ones obtained in the
projector-operator method. Thus, although the Wannier functions reported in \cite{PRB06a}
may suffer from some additional approximations, the transfer integrals are essentially correct.}

  For cubic perovskites, the transfer integrals
can be extracted from first-principles
electronic structure calculations
in several different ways.
For example, one can simply fit the LDA
band structure in terms of the Slater-Koster parameters \cite{SlaterKoster}.
However, the situation becomes increasingly complicated in materials with lower
crystal symmetry, like distorted perovskite oxides,
which will be considered below.
First, the number of the Slater-Koster parameters, permitted by the symmetry, increases dramatically.
Second, the form of these transfer integrals becomes more complicated and differs substantially
from cubic compounds.\footnote{One example is the
mixing of the $t_{2g}$ and $e_g$ orbitals by the orthorhombic distortion, which does not
occur in the cubic compounds.}
Therefore, it seems that for complex systems
the only way to proceed is to use
straightforward numerical algorithms, like the formal
downfolding method.

\subsubsection{\label{sec:SrVO3_U}Effective Interactions.}

  Applications of constrained DFT to the transition-metal
oxides have been widely discussed in the
literature \cite{Gunnarsson1989,AZA,NormanFreeman,McMahan,GunnarssonPostnikov}.
For example, the effective Coulomb interaction between $3d$ electrons in
SrVO$_3$ can be computed in the following way \cite{PRB06a}:
\begin{enumerate}
\item
In the supercell geometry, one can introduce the ``charge-density wave'',
describing the modulation of the atomic $3d$ occupations
around the ``ground-state'' configuration with $n = 1$,
$n^{\bf q}_{\bf R} = n$$+$$\delta n \cos ({\bf q} {\bf R})$,
where ${\bf q}$ is the propagation vector of the charge-density wave.
\item
Then, from the constrained DFT calculations, one can derive
the Kohn-Sham eigenvalues $\{ \varepsilon^{\bf q}_{\bf R} \}$,
corresponding to this charge-density wave,
and find the Fourier image of the effective Coulomb interaction as
$u_{\bf q} = d \varepsilon^{\bf q}_{\bf R} / d n^{\bf q}_{\bf R}$.
\item
Finally,
the parameters of Coulomb interaction in the real space are
obtained after the Fourier transformation of $u_{\bf q}$.\footnote{
For example, by considering only on-site ($u$) and nearest-neighbor intersite ($v$)
interactions, we would have
$u_{\bf q}=u-v \sum_{\bf R} \cos ({\bf q} {\bf R})$, etc.}
\end{enumerate}
For SrVO$_3$, this procedure yields the following parameters of the on-site Coulomb interaction
$u = 10.1$ eV and the nearest-neighbor Coulomb interaction $v = 1.2$ eV.
The intraatomic exchange interaction ($j$) can be derived by
constraining the $3d$ magnetization density \cite{Dederichs,AZA}.\footnote{
For example, if $m_{\bf R}$ is the $3d$-magnetization,
$m_{\bf R} = n^\uparrow_{\bf R}$$-$$n^\downarrow_{\bf R}$, and
$\varepsilon^\uparrow_{\bf R}$ and $\varepsilon^\downarrow_{\bf R}$ are
the Kohn-Sham eigenvalues for the majority- and minority-spin states,
respectively, the parameter of intraatomic exchange interaction
is given by $j = 2d(\varepsilon^\downarrow_{\bf R}$$-$$\varepsilon^\uparrow_{\bf R})/d m_{\bf R}$.
}
This yields $j = 1.0$ eV.
By knowing only $u$ and $j$ in the atomic limit,
one can reconstruct the full $5$$\times$$5$$\times$$5$$\times$$5$
matrix $\hat{u}$ of interactions between
the $3d$ electrons, as it is typically done in the LDA$+$$U$ method \cite{PRB94b}.
Some details of this procedure are explained in \ref{sec:appendixA}.

  In order to appreciate the magnitude of screening of different interaction
parameters
obtained in the constrained DFT, it is instructive to compare them with \textit{bare}
interactions.
For example, the values of bare Coulomb and exchange integrals,
calculated from $3d$ wavefunctions of the vanadium atoms, are 21.7 and 1.2 eV, respectively.
The bare Coulomb interaction between neighboring
vanadium sites, $e^2/a$, is about 3.7 eV.
Thus, in the constrained DFT,
the on-site Coulomb interaction is reduced by factor two,
the intersite Coulomb interaction is reduced by almost 70\%, and
the intra-atomic exchange interaction is
reduced by
20\%.
All these interactions are further reduced by
relaxation effects, related with
the change of the hybridization.

\begin{figure}[h!]
\begin{center}
\resizebox{12cm}{!}{\includegraphics{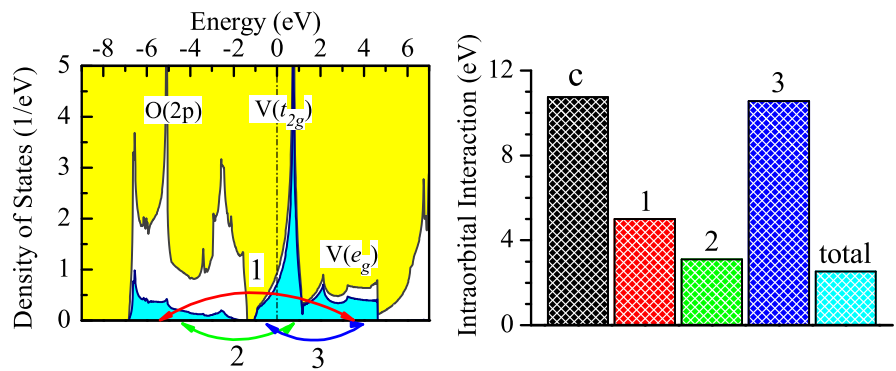}}
\end{center}
\caption{\label{fig.SrVO3DOSURPA}
Left panel shows the local density of states of SrVO$_3$ with the notation
of the main
interband
transitions, which contribute to the polarization function in RPA:
O($2p$)$\rightarrow$V($e_g$) (1), O$(2p$)$\rightarrow$V($t_{2g}$) (2), and
V($t_{2g}$)$\rightarrow$V($e_g$) (3). Right panel shows the intraorbital
Coulomb
interaction $\mathcal{U}$ as obtained in the
constrained DFT (denoted as ``c'')
and after including the screening caused by different interband transitions
in RPA. The RPA results show
the screening
corresponding to each type
of transitions in the polarization function as well as the final value of $\mathcal{U}$,
which incorporates the effect of all three transitions \protect\cite{PRB06a}.
}
\end{figure}

  As it was already pointed out in Section \ref{sec:cDFTRPA},
because of the
hybridization, the transition-metal $3d$ states may have a significant
weight in other bands. For example, in SrVO$_3$ besides the vanadium $t_{2g}$ band,
the $3d$ states contribute
to the vanadium $e_g$ as well as to the oxygen $2p$ bands (Figure \ref{fig.SrVO3DOSURPA}).
If the number of $t_{2g}$ electrons changes, it causes some change of the
Coulomb potential, which affects the distribution of the vanadium $3d$ states in other
parts of the spectrum. For example, if at certain vanadium site, the number of
$t_{2g}$ electrons increases, the Coulomb potential becomes more repulsive.
Therefore, the $3d$ states of this vanadium site will be pushed from the
oxygen $2p$ band to a higher energy region. This causes some change of the
coefficients $\{ c_{{\bf R}}^\beta \}$ of the expansion of the Kohn-Sham orbitals
over the basis functions (\ref{eqn:orbitals_relaxation})
or the change of the hybridization. This mechanism is responsible for an additional
channel of screening of Coulomb interactions, which can be evaluated within RPA.
In these calculations, the matrix $\hat{u}$, obtained in the constrained DFT method is used
as the starting point, while the RPA itself is employed in order to evaluate the
screening of $3d$ interactions in the vanadium $t_{2g}$ band by the same $3d$ states, which
contribute to other bands. Thus, the problem is reduced to evaluation of the
$3d$ matrix elements of the polarization function (\ref{eqn:Polarization_Function}).

  According to the electronic structure of SrVO$_3$,
one can identify three main contributions to the polarization function,
associated with the following interband transitions:
oxygen $2p$ $\rightarrow$ vanadium $e_g$, oxygen $2p$ $\rightarrow$ vanadium $t_{2g}$, and
vanadium $t_{2g}$ $\rightarrow$ vanadium $e_g$.

  The details of RPA screening are explained in Figure \ref{fig.SrVO3DOSURPA}.
For these purposes,
it is convenient to introduce three Kanamori parameters \cite{Kanamori}:
the intraorbital Coulomb interaction
$$
\mathcal{U} =
\int d{\bf r} \int d{\bf r}' \tilde{W}_{xy}^\dagger({\bf r}) \tilde{W}_{xy}({\bf r})
v_{\rm scr}({\bf r},{\bf r}') \tilde{W}_{xy}^\dagger({\bf r}') \tilde{W}_{xy}({\bf r}'),
$$
the interorbital Coulomb interaction
$$\mathcal{U}' =
\int d{\bf r} \int d{\bf r}' \tilde{W}_{xy}^\dagger({\bf r}) \tilde{W}_{xy}({\bf r})
v_{\rm scr}({\bf r},{\bf r}') \tilde{W}_{yz}^\dagger({\bf r}') \tilde{W}_{yz}({\bf r}'),
$$
and the exchange interaction
$$\mathcal{J} =
\int d{\bf r} \int d{\bf r}' \tilde{W}_{xy}^\dagger({\bf r}) \tilde{W}_{yz}({\bf r})
v_{\rm scr}({\bf r},{\bf r}') \tilde{W}_{xy}^\dagger({\bf r}') \tilde{W}_{yz}({\bf r}').
$$
In the atomic limit, all interactions between $t_{2g}$ electrons are reduced
to either ${\cal U}$, ${\cal U}'$, or ${\cal J}$, and there is no other
types of interactions connecting the $t_{2g}$ orbitals (see \ref{sec:appendixA}).
Below we will argue that similar property holds even after the RPA screening.

  In addition to the final value of $\mathcal{U}$,
Figure \ref{fig.SrVO3DOSURPA} shows the screened
interactions corresponding to each type of transitions
in the polarization function. The
screening
caused by the change of the hybridization is very efficient.
For example, in comparison with the constrained DFT,
the intraorbital interaction $\mathcal{U}$
is reduced from 11.2 to 2.5 eV (i.e., by more than factor four).
The main contribution to this screening comes from the
oxygen $2p$ $\rightarrow$ vanadium $e_g$ and oxygen $2p$ $\rightarrow$ vanadium $t_{2g}$
interband
transitions in the polarization functions.
Since the hybridization between vanadium $t_{2g}$ and $e_g$ orbitals is small
in perovskite compounds with the simple cubic structure,
the screening associated with the
transitions between vanadium $t_{2g}$ and $e_g$ bands is also small.

\begin{figure}[h!]
\begin{center}
\resizebox{8cm}{!}{\includegraphics{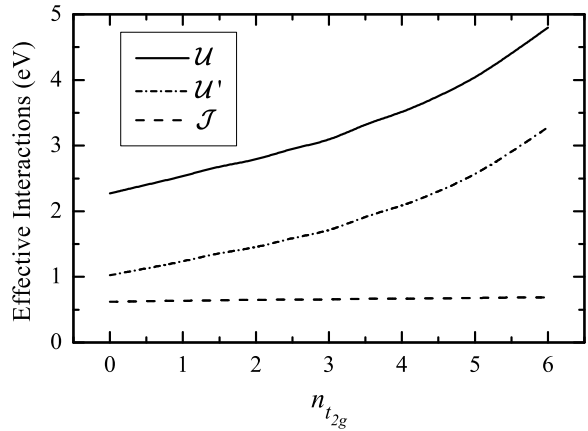}}
\end{center}
\caption{\label{fig.SrVO3RPADoping}
Doping-dependence of Kanamori parameters in SrVO$_3$ \protect\cite{PRB06a}:
the intraorbital Coulomb interaction $\mathcal{U}$,
the interorbital Coulomb interaction $\mathcal{U}'$,
and the exchange interaction $\mathcal{J}$
versus the number of electrons in the $t_{2g}$ band, $n_{t_{2g}}$.
}
\end{figure}

  The dependence of the screened Coulomb interactions
on the number of electrons, $n_{t_{2g}}$,
accommodated in the $t_{2g}$ band is shown in Figure \ref{fig.SrVO3RPADoping} \cite{PRB06a}.
The calculations have been performed in the rigid-band approximation
and using the electronic structure of SrVO$_3$.
Such an analysis may be useful for understanding the doping-dependence of the
effective Coulomb interactions.

  The Coulomb interactions reveal a monotonic behavior as the function
of doping. The screening is the most efficient when the whole $t_{2g}$ band is empty ($n_{t_{2g}}=0$).
The situation corresponds to SrTiO$_3$, where all
transitions
from the oxygen $2p$ to the transition-metal $t_{2g}$ band
can
contribute to the screening
(see Figure \ref{fig.SrVO3DOSURPA}).
This channel of screening is closed when the $t_{2g}$ band is filled ($n_{t_{2g}}=6$).
In the latter case, only the oxygen $2p$ $\rightarrow$ transition-metal $e_g$
interband transitions
may contribute to the screening. Hence, the effective Coulomb interaction becomes large.

  The screening of the exchange integral $\mathcal{J}$ practically does not depend on
the doping.
The Kanamori rule, $\mathcal{U} = \mathcal{U}' + 2 \mathcal{J}$,
which was originally established for the spherical environment in
isolated atoms, works well also for the $t_{2g}$
manifold in the cubic compounds, even after
the screening of $t_{2g}$ interactions by other electrons.

   This result support an old empirical rule suggesting that only
the Coulomb integral $\mathcal{U}$ is sensitive to the crystal environment in solids.
The nonspherical
interactions, which are also responsible for Hund's first and second rules,
appears to be much closer to their atomic values and practically insensitive
to the screening \cite{MarelSawatzky,Norman,Brooks}.

  It is important to note that the obtained values of effective Coulomb
interactions are substantially smaller than the experimental
parameters
derived from the
analysis of photoemission spectra \cite{ZaanenSawatzky,MizokawaFujimori}.
However, this is to be expected. Note that the photoemission spectra
are typically interpreted in the cluster model, which treats
\textit{explicitly} all transition-metal $3d$ as well as the oxygen $2p$ states.
However, in the model (\ref{eqn:Hmanybody}) we would like to keep only the transition-metal $t_{2g}$ bands
and include the effect of other bands \textit{implicitly}, i.e. through the
renormalization of interaction parameters in the $t_{2g}$ band.
Therefore, our parameters should be generally smaller in comparison with
the ones derived
from the cluster model.
As it was already discussed above, the transfer of an electron, associated with the reaction
($n_{{\bf R}\alpha}$,$n_{{\bf R}'\beta}$) $\rightleftharpoons$ ($n_{{\bf R}\alpha}$$+$$1$,$n_{{\bf R}'\beta}$$-$$1$)
will cause some change of the electronic structure in the region of oxygen $2p$ and
transition-metal $e_g$ bands, which tends to compensate
the change of the number of the $3d$ electrons in the $t_{2g}$ band.
Since the oxygen $2p$ and transition-metal $e_g$ bands are eliminated in our
$t_{2g}$ model, this change of the electronic structure is effectively
included into the screening of Coulomb interactions in the $t_{2g}$ band,
that naturally explains smaller values of the parameter $\mathcal{U}$.

  Finally, the obtained value of intraorbital Coulomb interaction
${\cal U} = 2.53$ eV is substantially smaller than ${\cal U} \approx 5.55$ eV,
which is typically used in DMFT calculations in order to reproduce the
experimental photoemission spectra \cite{Sekiyama}.
Recent full-potential RPA calculations based on the maximally
localized Wannier functions yielded ${\cal U} = 3.0$ eV \cite{MiyakeFerdi},
which is still too small in order to explain the photoemission spectra
in terms of conventional DMFT calculations for the $t_{2g}$ band.
This maybe a serious problem indicating that something is missing
in the current interpretation of the photoemission data.
Some of the missing ingredients may be the spacial correlations,
the explicit contribution of the oxygen states, or the frequency-dependence
of the effective Coulomb interaction in RPA \cite{Ferdi04}.
On the other hand,
the obtained value of the exchange interaction ${\cal J} = 0.64$ eV
is very close to ${\cal J} = 0.68$ eV, which is typically used in the
analysis of the photoemission spectra \cite{MizokawaFujimori}.

\subsection{Inversion-Symmetry Breaking and ``Forbidden'' Hoppings}

  In this small section we would like to consider two examples of
deformation of the ideal perovskite structure, which are related with
violation of the inversion symmetry around transition-metal sites.
One is the oxygen vacancy, and the other one is the $(001)$ surface of SrTiO$_3$.
Particularly,
we will argue that such an inversion-symmetry breaking may lead to
a number of new effects, and qualitatively change the character of
transfer integrals between Wannier orbitals.

\subsubsection{Oxygen Vacancy in SrTiO$_3$.}

   In cubic perovskites, such as SrTiO$_3$, the oxygen vacancy
creates a dimer of Ti atoms located in its first coordination sphere.
It also donates two electrons into the $t_{2g}$ band.\footnote{
Under certain conditions, such a situation may lead to the formation of the spin-singlet
bipolaronic state \cite{Kolodiazhnyi}.
}

\begin{figure}[h!]
\begin{center}
\resizebox{12.0cm}{!}{\includegraphics{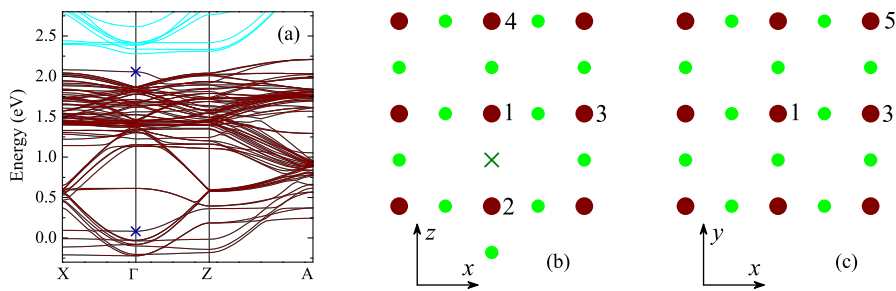}}
\end{center}
\caption{\label{fig.O_Vacancy_in_SrTiO3}
Left panel: LDA band structure of supercell Sr$_{27}$Ti$_{27}$O$_{80}$
corresponding to the oxygen-deficient SrTiO$_{2.963}$.
The (dark) brown curves show 83 bands located near the Fermi level
and
formed by $3$$\times$$27$$=$$81$ $t_{2g}$ Wannier orbitals
of all $27$ Ti atoms as well as
the $z^2$ orbitals of two Ti atoms located near the oxygen vacancy (denoted as `1' and `2'
in the right panel).
The position of these $z^2$ bands in the $\Gamma$-point are marked by the crosses.
Right panel explains the atomic positions around the oxygen vacancy, namely, in the $yz$ plane,
which contains the vacancy, and in the $xy$ plane, which is located just above the vacancy.
The Ti atoms are indicated by the big dark (brown) spheres, the oxygen atoms
are indicated by the small light grey (green) spheres, and the oxygen vacancy
is indicated the symbol $\times$.}
\end{figure}

  In order to study the effect of the
oxygen vacancy
on the electronic structure of SrTiO$_3$
we have used the
3$\times$3$\times$3 supercell, in which one of the
oxygen atoms has been replaced by the empty sphere.
Such a composition corresponds to the chemical formula SrTiO$_{2.963}$.
No lattice relaxation has been considered at this stage.
According to LDA calculations,
the electronic structure
of such a supercell
near the Fermi level
is formed by 83 bands, which are well isolated from the rest of the
spectrum (Figure \ref{fig.O_Vacancy_in_SrTiO3}).
Among them,
$3$$\times$$27$$=$$81$ bands
are the regular $t_{2g}$ bands, whereas
two additional bands are formed predominantly by
$z^2$ orbitals of two Ti atoms located near the oxygen vacancy.
The $t_{2g}$ and $z^2$ bands are strongly mixed.

\begin{figure}[h!]
\begin{center}
\resizebox{10.0cm}{!}{\includegraphics{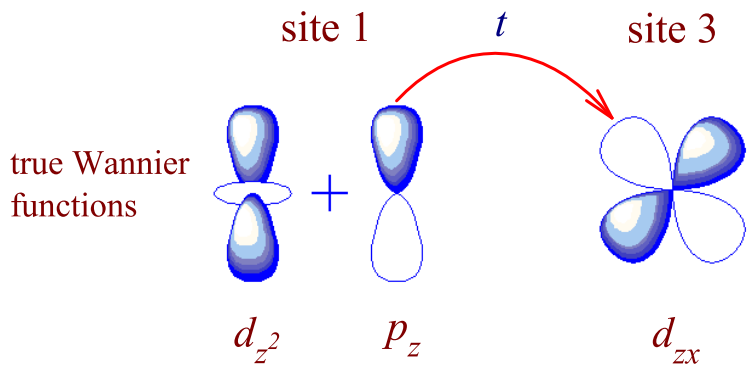}}
\end{center}
\caption{\label{fig.hopping_cartoon}
Cartoon picture explaining the appearance of the ``forbidden hoppings'' near the
points of the inversion symmetry breaking. The notations of the atomic sites are
explained in Figure \protect\ref{fig.O_Vacancy_in_SrTiO3}. Due to the local inversion-symmetry breaking
(or the parity violation), the atomic orbital $d_{z^2}$ at the site $1$ is allowed to
mix with the atomic orbital $p_z$, belonging to the same site. The combination of these
orbitals will form a true Wannier function, centeerd at the site $1$. Then, the electron
can transfer from the $p_z$ orbital of the site 1 to the $d_{zx}$ orbital of the site $3$
(and vice versa).}
\end{figure}

  Therefore, it is clear that the minimal model near the Fermi level should
be constructed in the basis of
four Wannier orbitals
(nominally, $z^2$, $xy$, $yz$, and $zx$) of two Ti atoms located near the vacancy,
and three Wannier orbitals (nominally, $xy$, $yz$, and $zx$)
of all remaining Ti atoms located in the next coordination spheres.
The atomic wavefunctions of these types can be used as the trial functions in the
downfolding method.
The behavior of transfer integrals
and the crystal-field splitting
obtained after the downfolding is
explained in Table \ref{tab.O_Vacancy_in_SrTiO3}.
\begin{table}[h!]
\caption{\label{tab.O_Vacancy_in_SrTiO3}
Crystal-field splitting (1-1) and transfer integrals in the bonds 1-2, 1-3,
1-4, and 1-5
around the oxygen vacancy in SrTiO$_3$.
All energies are measured in meV.
The atomic positions are explained in Figure \protect\ref{fig.O_Vacancy_in_SrTiO3}.}
\begin{indented}
\item[]\begin{tabular}{@{}ccc}
\br
 & 1-1 & 1-2 \\
\mr
$
\begin{array}{c}
           \\

 {\it z}^2 \\
  {\it xy} \\
  {\it yz} \\
  {\it zx} \\
\end{array}
$
&
$
\begin{array}{rrrr}
 {\it z}^2 & {\it xy} & {\it yz} & {\it zx} \\
\hline
  164      &  0       &  0       &  0       \\
  0        &  124     &  0       &  0       \\
  0        &  0       & -144     &  0       \\
  0        &  0       &  0       & -144     \\
\end{array}
$
&
$
\begin{array}{rrrr}
 {\it z}^2 & {\it xy} & {\it yz} & {\it zx} \\
\hline
 -709      &  0       &  0       &  0       \\
  0        & -25      &  0       &  0       \\
  0        &  0       &  37      &  0       \\
  0        &  0       &  0       &  37      \\
\end{array}
$\\
\br
\end{tabular}
\end{indented}

\begin{indented}
\item[]\begin{tabular}{@{}cccc}
\br
 & 1-3 & 1-4 & 1-5 \\
\mr
$
\begin{array}{c}
           \\

 {\it z}^2 \\
  {\it xy} \\
  {\it yz} \\
  {\it zx} \\
\end{array}
$
&
$
\begin{array}{rrr}
 {\it xy} & {\it yz} & {\it zx} \\
\hline
  0       &  0       &  -145    \\
 -224     &  0       &  0       \\
 -9       & -22      &  0       \\
  0       &  0       & -269     \\
\end{array}
$
&
$
\begin{array}{rrr}
 {\it xy} & {\it yz} & {\it zx} \\
\hline
  0       &  0       &  0       \\
  -28     &  0       &  0       \\
  0       & -219     &  0       \\
  0       &  0       & -219     \\
\end{array}
$
&
$
\begin{array}{rrr}
 {\it xy} & {\it yz} & {\it zx} \\
\hline
 -24      &  7       &  7       \\
 -82      &  0       &  0       \\
  5       &  -1      &  7       \\
  5       &  7       &  -1      \\
\end{array}
$\\
\br
\end{tabular}
\end{indented}
\end{table}
There is a number of interesting effects related with the
presence of the oxygen vacancy.
\begin{enumerate}
\item
The oxygen vacancy breaks the cubic symmetry and splits the
$t_{2g}$ levels of two Ti atoms located next to it.
The splitting is about 270 meV.
However, already in the next coordination sphere,
the $t_{2g}$-level splitting is greatly reduced,\footnote{
For example, for the titanium atoms $3$ and $4$ depicted in Figure \ref{fig.O_Vacancy_in_SrTiO3},
the $t_{2g}$-level splitting is only 37 meV and 40 meV, respectively.}
and the situation becomes close to the perfect cubic environment.
On the other hand,
the position of the impurity $z^2$ level is lowered
due to the missing Ti-O bond. As a result, the
$z^2$ levels become close to the $t_{2g}$ ones.\footnote{
Note that the impurity $z^2$ level is an atibonding combination of the atomic
oxigen $2p$ and titanium $3z^2$ orbitals. Therefore,
the lack of one of the Ti-O bond formed by the Ti atom near
the vacancy
will shift the $z^2$ level to the low-energy region.
}
For example, the atomic splitting between the $z^2$ and $xy$ levels is only $40$ meV.\footnote{
For comparison, the $t_{2g}$-$e_g$ splitting in the perfect perovskites is about 3 eV
(Figures \ref{fig.PerovskiteDOS} and \ref{fig.SrVO3DOS}).
}
\item
The behavior of transfer integrals
across the vacancy
(the bond 1-2)
is fundamentally different from the conventional case, when they are mediated by the
oxygen $2p$ states (for example, in the bond 1-4):
the transfer integrals between all three $t_{2g}$ orbitals are negligibly small,
while the main interaction occurs between $z^2$ orbitals.
\item
The lack of the inversion symmetry leads to the mixing of the atomic
$3d$ and $4p$ orbitals at the same Ti site. For example,
the Wannier function, which is nominally denoted as $z^2$,
besides the conventional $3d_{z^2}$ atomic orbitals will have some weight of the $4p_z$ orbitals.
Since the $4p$ orbitals are rather extended in the real space,
such a mixing may change the form of the transfer integrals
and even lead to the appearence of new interactions.
The most striking example is the large transfer integral
occuring between
neighboring
$z^2$ and $zx$ Wanner orbitals in the bond $1$-$3$
near the vacancy (Figure \ref{fig.hopping_cartoon}).
Such an interaction would vanishe in the perfect
cubic environment.
\end{enumerate}

  The distortion of the perfect cubic environment by the oxygen vacancy
will affect not only the one-electron part of the
model Hamiltonian (\ref{eqn:Hmanybody}), but also the Coulomb
interactions (Table \ref{tab.O_Vacancy_in_SrTiO3_U}).
\begin{table}[h!]
\caption{\label{tab.O_Vacancy_in_SrTiO3_U}
Intraorbital Coulomb interactions associated with different Ti sites
around the oxygen vacancy in SrTiO$_3$.
All energies are measured in eV.
The atomic positions are explained in Figure \protect\ref{fig.O_Vacancy_in_SrTiO3}.}
\begin{indented}
\item[]\begin{tabular}{@{}ccccc}
\br
 orbital    & site 1 & site 3 & site 4 & site 5 \\
\mr
{\it xy}    &  2.61  & 2.72   & 2.71   & 2.70   \\
{\it yz}    &  2.57  & 2.71   & 2.77   & 2.70   \\
{\it zx}    &  2.57  & 2.75   & 2.77   & 2.70   \\
{\it z$^2$} &  2.67  &  -     & -      & -      \\
\br
\end{tabular}
\end{indented}
\end{table}
For example, around the vacancy, the Coulomb interactions associated
with different $t_{2g}$ orbitals are clearly different. This effect
is captured by the RPA screening.
The cubic symmetry of Coulomb interactions is
practically restored is the fourth coordination sphere
(site 5 in Figure \ref{fig.O_Vacancy_in_SrTiO3}).

\subsubsection{Surface states in SrTiO$_3$.}

\begin{figure}[h!]
\begin{center}
\resizebox{10.0cm}{!}{\includegraphics{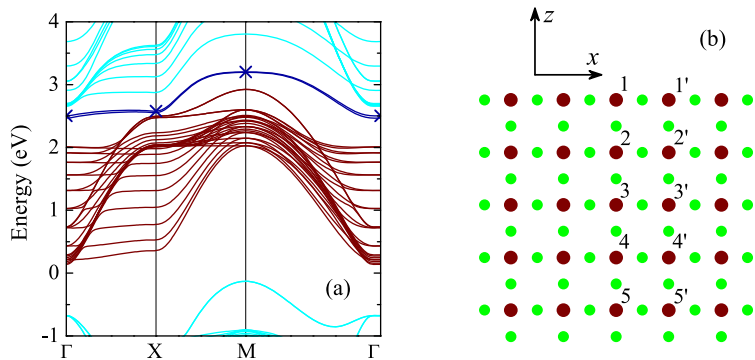}}
\end{center}
\caption{\label{fig.surface_SrTiO3}
Left panel: LDA band structure of the slab (TiO$_2$)$_9$(SrO)$_8$.
The (dark) brown curves are the $t_{2g}$ bands.
The ``surface'' $z^2$ bands are depicted by
the symbols $\times$.
Right panel: The atomic positions near the (TiO$_2$)-terminated
surface of SrTiO$_3$.
The Ti atoms are indicated by the big dark (brown) spheres, and the oxygen atoms
are indicated by the small light grey (green) spheres.}
\end{figure}

  Another example of the inversion symmetry breaking is the
TiO$_2$ terminated
surface of SrTiO$_3$. The surface states has been studied in the
slab geometry. Each slab contained nine TiO$_2$ layers, which were separated
by the
SrO layers. Hence, the chemical formula of the slab was
(TiO$_2$)$_9$(SrO)$_8$.\footnote{
The considered geometry can been obtained from the bulk
SrTiO$_3$ by
cutting the slab (TiO$_2$)$_9$(SrO)$_8$ and
replacing the next (TiO$_2$)$_2$(SrO)$_3$
layers by empty spheres.
The considered region of empty spheres
was sufficient to make the interaction between
different slabs negligibly small.
}
According to the adopted notations, the first TiO$_2$ layer corresponds to the surface,
while the fifth TiO$_2$ layer corresponds to the bulk of SrTiO$_3$ (Figure \ref{fig.surface_SrTiO3}).
No lattice relaxation has been considered at this stage.

  The electronic structure of (TiO$_2$)$_9$(SrO)$_8$ near the Fermi level
consists of the $3$$\times$$9$$=$$27$ $t_{2g}$ bands and two $z^2$ bands,
which are mainly formed by the surface TiO$_2$ layers.
Therefore, the minimal model can be constructed in the Wannier basis of
(nominally)
$z^2$, $xy$, $yz$, and $zx$ orbitals centered at the surface Ti sites
and
the
$xy$, $yz$, and $zx$ orbitals representing the remaining (``bulk'') sites.
Thus, there is a direct analogy with the case of the
oxygen vacancy in SrTiO$_3$. In both cases, the Ti atoms located next to the
``defect'' (the surface, in the present case) acquire an additional
$z^2$ orbital, whereas other Ti sites are described in the standard
$t_{2g}$ basis.

  The one-electron part of the model Hamiltonian is explained in Table \ref{tab.surface_SrTiO3}.
At the surface, there is a huge crystal-field splitting, which even
exceeds the crystal-field splitting near the single oxygen vacancy.
\begin{table}[h!]
\caption{\label{tab.surface_SrTiO3}
Crystal-field splitting (1-1) and transfer integrals at the surface and in the bulk of SrTiO$_3$.
All energies are measured in meV.
The atomic positions are explained in Figure \protect\ref{fig.surface_SrTiO3}.}
\begin{indented}
\item[]\begin{tabular}{@{}ccc}
\br
 & $1$-$1$ & $1$-$1'$ \\
\mr
$
\begin{array}{c}
           \\

 {\it z}^2 \\
  {\it xy} \\
  {\it yz} \\
  {\it zx} \\
\end{array}
$
&
$
\begin{array}{rrrr}
 {\it z}^2 & {\it xy} & {\it yz} & {\it zx} \\
\hline
  617      &  0       &  0       &  0       \\
  0        &  18      &  0       &  0       \\
  0        &  0       & -317     &  0       \\
  0        &  0       &  0       & -317     \\
\end{array}
$
&
$
\begin{array}{rrrr}
 {\it z}^2 & {\it xy} & {\it yz} & {\it zx} \\
\hline
 -97       &  0       &  0       & -80      \\
  0        & -234     & -10      &  0       \\
  0        &  10      & -19      &  0       \\
  80       &  0       &  0       & -325     \\
\end{array}
$\\
\br
\end{tabular}
\end{indented}

\begin{indented}
\item[]\begin{tabular}{@{}ccc}
\br
 & $1$-$2$ & $5$-$5'$ \\
\mr
$
\begin{array}{c}
           \\

  {\it xy} \\
  {\it yz} \\
  {\it zx} \\
\end{array}
$
&
$
\begin{array}{rrr}
 {\it xy} & {\it yz} & {\it zx} \\
\hline
 -29      &  0       &  0       \\
  0       & -207     &  0       \\
  0       &  0       & -207     \\
\end{array}
$
&
$
\begin{array}{rrr}
 {\it xy} & {\it yz} & {\it zx} \\
\hline
 -234     &  0       &  0       \\
  0       & -30      &  0       \\
  0       &  0       & -234     \\
\end{array}
$\\
\br
\end{tabular}
\end{indented}
\end{table}
Due to the inversion-symmetry breaking, there is
an appreciable ``forbidden'' hopping between the
$z^2$ and $zx$ Wannier orbitals operating in the surface bond $1$-$1'$.
The transfer integrals operating between $t_{2g}$ orbitals
near the surface (the bonds $1$-$1'$ and $1$-$2$)
are also different from the ones in the bulk (the bond $5$-$5'$).
Thus, the effect of the surface on the electronic structure
of the transition-metal perovskite oxides is not only
in the narrowing of the $yz$ and $zx$ bands,
caused by the reduced number of bonds available for the hoppings \cite{Liebsch2003}.\footnote{
Note that the orbital $zx$ is perpendicular to the surface. Therefore, it can be involved
in the hoppings in the directions $\pm$$x$ and $-$$z$ (in the geometry shown in Figure \ref{fig.surface_SrTiO3}).
Similar situation holds for the $yz$ orbitals. On the contrary, the $xy$ orbital is
involved in the hoppings in all four directions $\pm$$x$ and $\pm$$y$.
}
Even more serious consequences can be caused by the
crystal-field splitting and the ``forbidden hoppings''.\footnote{
Note that in addition to the hopping, the $z^2$ and $zx$ orbitals are
coupled at the same transition-metal site by the spin-orbit interactions.
If the surface were magnetic, this type of coupling would lead to
the Dzyaloshinsky-Moriya interactions between the spins \cite{Dzyaloshinsky,Moriya}.
Thus, the ``forbidden hoppings'' provide a microscopic basis for the appearence of these
inetractions.}

  The Coulomb interactions at the surface of SrTiO$_3$ are also considerably distorted
in comparison with the bulk (Table \ref{tab.surface_SrTiO3_U}).
\begin{table}[h!]
\caption{\label{tab.surface_SrTiO3_U}
Intraorbital Coulomb interactions associated with different Ti-sites
in the slab (TiO$_2$)$_9$(SrO)$_8$.
All energies are measured in eV.
The atomic positions are explained in Figure \protect\ref{fig.surface_SrTiO3}.}
\begin{indented}
\item[]\begin{tabular}{@{}cccccc}
\br
 orbital    & site 1 & site 2 & site 3 & site 4 & site 5 \\
\mr
{\it xy}    &  2.43  & 2.65   & 2.66   & 2.65   & 2.65   \\
{\it yz}    &  2.52  & 2.64   & 2.65   & 2.65   & 2.65   \\
{\it z$^2$} &  2.69  &  -     & -      & -      & -      \\
\br
\end{tabular}
\end{indented}
\end{table}
However,
the bulk-like behavior is practically restored already in the second
TiO$_2$ layer.
Similar to the single oxygen vacancy, the surface breaks the cubic symmetry
of the Coulomb interactions. Moreover, the Coulomb interactions between
$t_{2g}$ orbitals are somewhat smaller at the surface of SrTiO$_3$
then in the bulk. Like in the case of the single oxygen vacancy, this dependence
of the effective Coulomb interactions
on the local environment of the transition-metal sites is captured by the RPA screening.

\subsection{\label{sec:DistortedPerovskites} Distorted Perovskite Oxides}

  The transition-metal perovskite oxides $AB$O$_3$ (where $A$$=$ Y or La, and $B$$=$ Ti or V)
are regarded as some of the
key materials for understanding the strong coupling among spin, orbital, and lattice
degrees of freedom in correlated electron systems \cite{IFT,TokuraNagaosa}.

  All compounds have distorted perovskite structure.
The distortion can be either orthorhombic (LaTiO$_3$, YTiO$_3$, and YVO$_3$ below 77 K)
or monoclinic (LaVO$_3$ and YVO$_3$ above 77 K).
The space group of the orthorhombic phase is $D^{16}_{2h}$
(in Sch\"{o}nflies notations or $Pbnm$ in the Hermann-Maguin notations,
No.~62 in the International Tables).
The space group of the monoclinic phase is $C^5_{2h}$
($P2_1/a$, No.~14 in the International Tables).\footnote{
There are certain indications that the actual symmetry
can be even lower
than $P2_1/a$ \cite{Tsvetkov}.
}
The magnitude of the distortion is controlled by the size of the $A^{3+}$ ions.
Typically, the smaller is the size of the $A^{3+}$ ions, the larger is the distortion.
For example, the Y$B$O$_3$ oxides are always more distorted than the La$B$O$_3$ oxides.\footnote{
For example, the Ti-O-Ti angle is 157-158$^\circ$ in LaTiO$_3$ \cite{Cwik},
and only 140-144$^\circ$ in YTiO$_3$ \cite{Maclean}.
}

  According to electronic structure calculations in the local-density approximation,
all compounds have a common
transition-metal $t_{2g}$ band, located near the Fermi level
and well isolated from the rest of the spectrum (Figure \ref{fig.PerovskiteDOS}).
The number of electrons donated by each transition-metal site into
the $t_{2g}$ band is one and two for $A$TiO$_3$ and $A$VO$_3$, respectively.
These electrons
are subjected to strong Coulomb correlations, beyond the
local-density approximation, and the
systems are classified as Mott insulators \cite{IFT}.

\begin{figure}[h!]
\begin{center}
\resizebox{10.0cm}{!}{\includegraphics{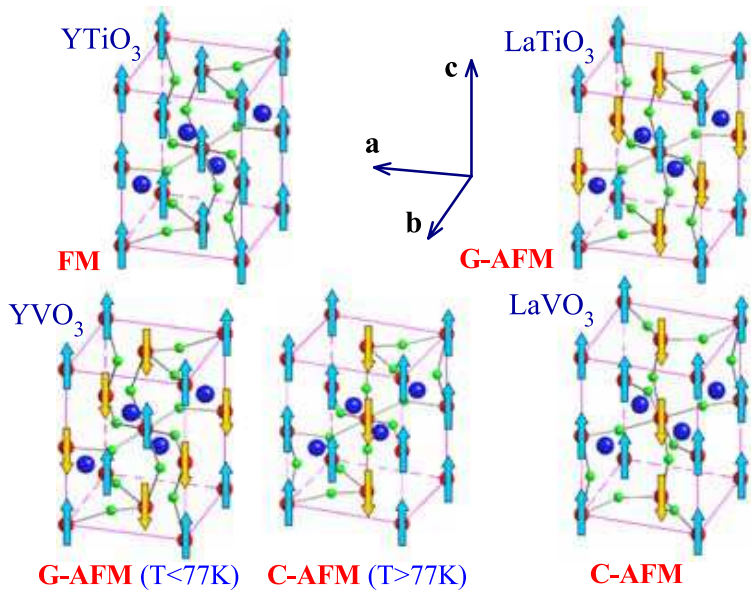}}
\end{center}
\caption{\label{fig.PerovskiteStructures}
Crystal and magnetic structure of distorted
perovskite oxides. The La and Y atoms are indicated by the big (blue) dark spheres,
the Ti and V atoms are indicated by the medium (red) dark grey spheres, and the oxygen
atoms are indicated by the small
(green) light grey spheres. The vectors ${\bf a}$, ${\bf b}$, and ${\bf c}$ show the directions
of orthorhombic translations. The directions of the magnetic moments are shown by arrows.}
\end{figure}

  Furthermore, the Coulomb correlations interplay with the lattice distortions,
leading to a peculiar phase diagram,
where each compound has a distinct magnetic structure (Figure \ref{fig.PerovskiteStructures}).
For example,
YTiO$_3$ is a
ferromagnet \cite{Maclean,Itoh,Akimitsu,Ulrich2002}. LaTiO$_3$ is
a three-dimensional (G-type) antiferromagnet \cite{Cwik,Keimer}.
At the low temperature,
YVO$_3$ has G-type antiferromagnetic structure,
which transforms to a chain-like (C-type)
antiferromagnetic structure at around $77$ K \cite{Tsvetkov,Ren,Blake,Ulrich2003}.
The magnetic transition coincides with the
structural orthorhombic-to-monoclinic transition.
On the other hand,
LaVO$_3$ is the C-type antiferromagnet
in the whole temperature range below the
magnetic transition point \cite{Zubkov,Bordet}.
The understanding of these magnetic properties will be the main goal
of this section.

  It is worth noticing that on the theoretical side the magnetic behavior of these ``$t_{2g}$''
compounds has been and continues to be the subject of numerous controversies.
So far, the theoretical interpretation of the unusual magnetic properties has been based
on two different standpoints, which practically exclude each other. One is the picture
of the orbital liquid, which implies that the effect of the crystal distortion
on the electronic structure
is small
and the (quasi-) degeneracy of the atomic $t_{2g}$ levels is preserved even in the real
crystal environment \cite{KhaliullinMaekawa,Khaliullin}.
Another one is the theory of crystal field, which lifts
the orbital degeneracy and stabilizes certain orbital structure
being compatible with
the observed magnetic ground state \cite{MochizukiImada,MochizukiImadaNJP,Schmitz}.

\subsubsection{Results of Downfolding for the One-Electron Part and
Wannier Functions.}
\begin{figure}[h!]
\begin{center}
\resizebox{9cm}{!}{\includegraphics{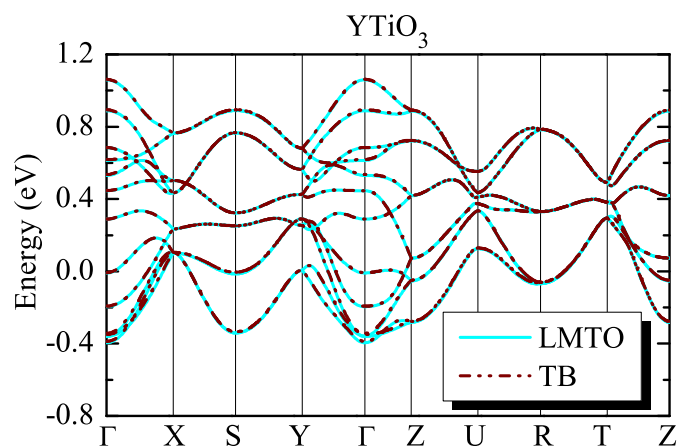}}
\end{center}
\caption{\label{fig.YTiO3bands}
LDA bands for YTiO$_3$ as obtained in
the original
electronic structure calculations using the
LMTO method and
after the tight-binding (TB) parametrization using the downfolding method \protect\cite{PRB06a}.
Notations of the high-symmetry points of the Brillouin zone are taken
from \protect\cite{BradlayCracknell}.
}
\end{figure}

  An example of the electronic structure in the region of the $t_{2g}$ bands for YTiO$_3$
is shown in Figure \ref{fig.YTiO3bands}.
It reveals an excellent agreement between
results of the original LMTO calculations and their tight-binding parametrization
obtained in the downfolding method.
Since cubic $t_{2g}$ and $e_g$ orbitals are mixed by the
crystal distortion,
the trial functions $\{ \tilde{\chi}^t_{\bf R} \}$
of the downfolding method cannot be longer chosen
from geometrical considerations.
Generally, such a choice is not unique.
Nevertheless, for these purposes one can use the eigenvectors
obtained from the diagonalization
of the density matrix. As it was already
pointed out in Section \ref{sec:downfoldingasprojector},
such a procedure should guarantee a good
degree of localization of the Wannier functions
as well as of the parameters of the model Hamiltonian in the real space.

\begin{figure}[h!]
\begin{center}
\resizebox{6.5cm}{!}{\includegraphics{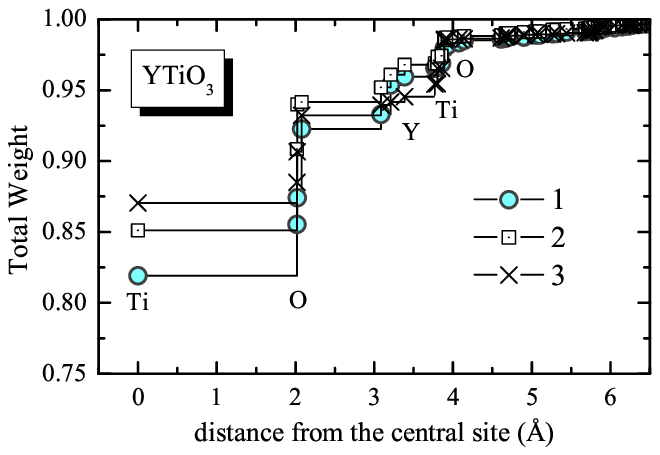}}
\resizebox{6.5cm}{!}{\includegraphics{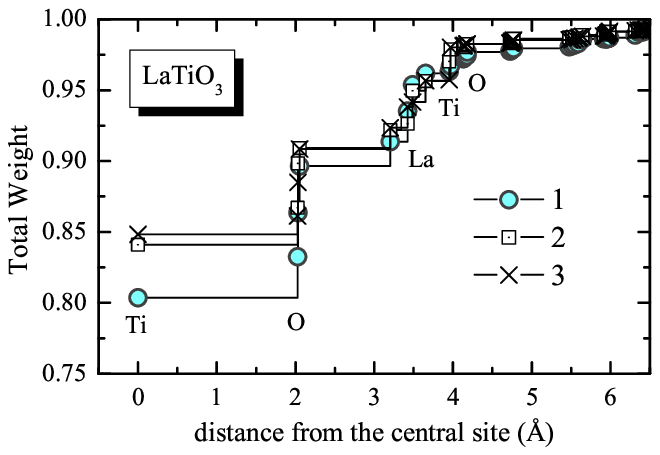}}
\end{center}
\caption{\label{fig.YTO_LTO_WF}
Spacial extension of Wannier functions for YTiO$_3$ (left) and LaTiO$_3$ (right).
The figure shows the total weight of the
Wannier function
accumulated around the central Ti site after adding every new sphere of
neighboring atomic sites. The symbols `1', `2', and `3' denote the Wannier
functions obtained after projection onto the eigenvectors of the density matrix,
where `1', `2', and `3' correspond to the largest, second largest, and
third largest eigenvalues, respectively.}
\end{figure}

  The spacial extension
of the Wannier functions constructed for YTiO$_3$ and LaTiO$_3$
is illustrated in Figure \ref{fig.YTO_LTO_WF}.
In LaTiO$_3$, about 80-85\% of the total weight
of the Wannier function
is accumulated at the central
Ti site, 5-9 \% belong to
six oxygen sites surrounding the central Ti atom, and about 10 \% is distributed over
La, Ti, and O sites located in next coordination spheres.
In YTiO$_3$, the same distribution parameters are
82-87\%, 6-10\%, and 5\%, correspondingly for the central Ti site, its neighboring oxygen sites,
and
Y, Ti, and O sites located in the
next coordination spheres.
Another measure of localization is
the expectation value of square of the position operator:
$\langle {\bf r}^2 \rangle_\alpha$$=$$\langle \tilde{W}_{\bf R}^\alpha |
({\bf r}$$-$${\bf R})^2 | \tilde{W}_{\bf R}^\alpha \rangle$ \cite{MarzariVanderbilt},
which
yields
$\langle {\bf r}^2 \rangle_\alpha$$=$ $2.68$, $2.36$, and
$2.37$ \AA$^2$ for LaTiO$_3$, and
$\langle {\bf r}^2 \rangle_\alpha$$=$ $2.28$, $1.90$, and $2.05$ \AA$^2$ for YTiO$_3$.
Thus,
the Wannier functions for LaTiO$_3$ and SrVO$_3$ are less localized in comparison
with the more distorted YTiO$_3$.
However, this is to be expected.
One reason is the lattice distortion, which increases in the direction
SrVO$_3$ $\rightarrow$ LaTiO$_3$  $\rightarrow$ YTiO$_3$. Another reason is the
proximity of transition-metal $t_{2g}$ and oxygen $2p$ bands.\footnote{
For example, the distance between the transition-metal $t_{2g}$ and oxygen $2p$ bands is $3.2$ eV in YTiO$_3$,
$2.7$ eV in LaTiO$_3$,
and only $0.3$ eV in SrVO$_3$. Therefore in SrVO$_3$, the mixing between the $t_{2g}$ and $2p$
states is stronger, and the Wannier functions have a
larger weight at the oxygen sites. To smaller extent the same is true for LaTiO$_3$.}
The degree of the localization in SrVO$_3$ and LaTiO$_3$ is very similar, despite the distortion
which takes place in LaTiO$_3$. One reason may be the proximity of the lanthanum $5d$ states
to the Fermi level, which leads to stronger hybridization with the transition-metal
$t_{2g}$ states \cite{PRB06b,PavariniNJP}.

\subsubsection{Transfer Integrals and Crystal-Field Splitting.}

\begin{figure}[h!]
\begin{center}
\resizebox{11cm}{!}{\includegraphics{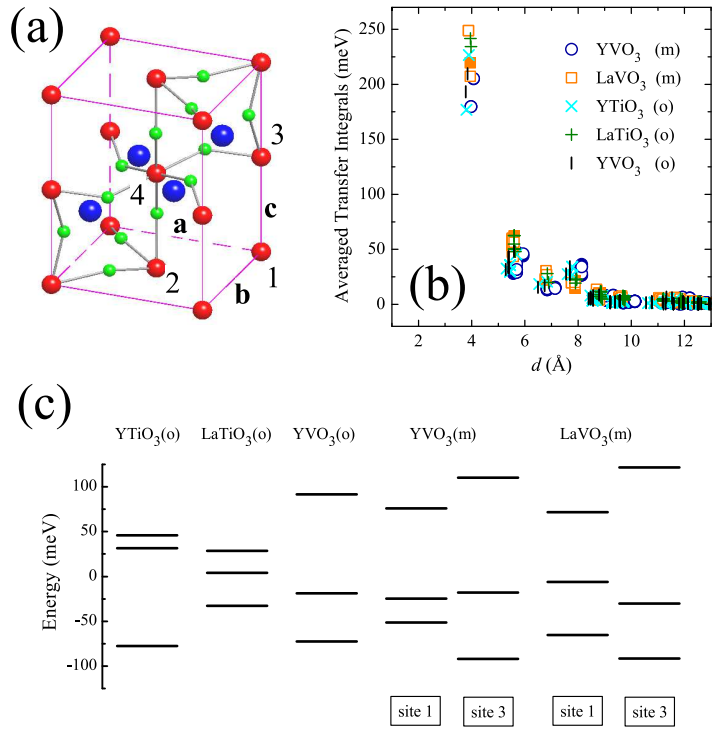}}
\end{center}
\caption{\label{fig.Perovskites_CFandHoppings}
(a): Crystal structure of the distorted perovskite oxides with the notation of
the positions
of four transition-metal sites in the unit cell. In orthorhombic compounds all sites
are equivalent and can be transformed to each other by the symmetry operations of the
$D_{2h}^{16}$ group. In monoclinic compounds, there are two inequivalent pairs of
the transition-metal sites: (1,2) and (3,4).
(b): Distance dependence of averaged transfer integrals.
In the orthorhombic (o) structure, all sublattices are equivalent and shown by a single symbol.
In the monoclinic (m) structure, the transfer integrals around two
inequivalent transition-metal sites are shown by closed and open symbols.
(c): $t_{2g}$-level splitting.
The notations `site 1'
and `site 3' stand for two nonequivalent transition-metal sites in
the monoclinic structure \protect\cite{PRB06b}.
}
\end{figure}

  The behavior of parameters of the one-electron part of the model Hamiltonian
is explained in Figure \ref{fig.Perovskites_CFandHoppings}.
Because of the lattice distortion, there is an appreciable crystal-field splitting,
which is larger for the more distorted YTiO$_3$ and YVO$_3$, and
substantially smaller for the least distorted LaTiO$_3$.
Furthermore, there is a clear correlation between the
number of $t_{2g}$ electrons and the form of the crystal-field splitting. As the rule,
the crystal-field splitting tends to quench the orbital degrees of freedom.
For example, in YTiO$_3$, it splits off one $t_{2g}$ level to the low-energy
part of the spectrum, that is just enough to accommodate one $t_{2g}$ electron.
The gap separating the
lowest $t_{2g}$ level from the middle one is about
109 meV.
On the contrary,
the crystal-field splitting
in YVO$_3$
lowers the energies of simultaneously two $t_{2g}$ levels, that is again consistent with the
number of $t_{2g}$ electrons per one vanadium site.
The distance between middle and highest $t_{2g}$ levels in the orthorhombic phase of YVO$_3$
is about 111 meV.

  The monoclinic distortion
in YVO$_3$ and LaVO$_3$
creates two inequivalent types of vanadium atoms,
which lie in different ${\bf ab}$-planes and are denoted as
(1,2) and (3,4) in Figure \ref{fig.Perovskites_CFandHoppings}.
Typically, the less distorted planes (1,2) are alternated with the more
distorted planes (3,4). The magnitude of the crystal-field splitting depends
on the compound. For example, in YVO$_3$, the energy splitting between middle and
highest $t_{2g}$ levels
in the planes (1,2) and (3,4)
is rather similar:
101 meV and 128 meV, respectively.
However,
in LaVO$_3$, the plane (1,2) appears to be much less distorted in comparison
with the plane (3,4), resulting in different crystal-field splittings:
$78$ meV and $152$ meV, respectively.
Not only the energies, but also the directions of the crystal-field splitting
are different in the orthorhombic and monoclinic phases, that immediately
follows from the symmetry considerations.\footnote{
For example, in the orthorhombic phase, the sites 1 and 3 (Figure \ref{fig.Perovskites_CFandHoppings})
can be transformed to each other by the 180$^\circ$ rotations around the
${\bf c}$-axis associated with the translation by ${\bf c}/2$.
In the monoclinic phase, such a symmetry operation is no longer available.
}
In Section \ref{sec:ApplicationsPRK}, we will argue that such a difference is directly
related
with the type of the orbital
ordering, which is reflected in the magnetic properties of the distorted perovskite oxides.
It is important to note that nonsphericity of the Madelung potential (\ref{eqn:CFEI})
is crucial for reproducing the correct
magnetic ground state of YVO$_3$ and LaVO$_3$, contrary to the
conventional atomic-spheres approximation \cite{PRB06b}.

   Because of the complexity of transfer integrals in the distorted
perovskite structure, it is
practically impossible to discuss the behavior of individual
matrix elements of $\| h^{\alpha \beta}_{{\bf RR}'} \|$.
Nevertheless,
some useful information can be obtained from the analysis of
\textit{averaged} parameters
$$
\bar{h}_{{\bf RR}'}(d) = \left( \sum_{\alpha \beta} h^{\alpha \beta}_{{\bf RR}'}
h^{\beta \alpha}_{{\bf R}'{\bf R}} \right)^{1/2},
$$
where
$d$ is the distance between transition-metal sites ${\bf R}$ and ${\bf R}'$.
All transfer integrals are well localized and practically restricted by the nearest
neighbors, located at around 4\AA~(Figure \ref{fig.Perovskites_CFandHoppings}b).
As expected,
the transfer integrals between the nearest neighbors are generally larger
for the less distorted LaTiO$_3$ and LaVO$_3$, and smaller
for the more distorted YTiO$_3$ and YVO$_3$.

\subsubsection{\label{sec:RMO3_U}Effective Interactions.}

  Matrix elements of the effective Coulomb interaction in the
$t_{2g}$ band can be computed by applying the combined constrained DFT
plus RPA approach, which has been explained in details in Section \ref{sec:Coulomb}.
At each transition-metal site, it yields the
$3$$\times$$3$$\times$$3$$\times 3$
$\hat{U}$-matrix of screened Coulomb interactions in the basis of
Wannier orbitals. The intersite interactions are considerably smaller
and can be neglected \cite{PRB06a}.
For the explanatory purposes, the $\hat{U}$-matrix
can be fit
in terms of
two Kanamori parameters: the intraorbital Coulomb interaction ${\cal U}$
and the intraatomic exchange interaction ${\cal J}$ \cite{Kanamori}.
The results of such fitting are shown in Table \ref{tab:Kanamori}.
\begin{table}[h!]
\caption{Results of fitting of the
effective Coulomb interactions
for the distorted perovskite oxides
in terms of two Kanamori parameters: the intraorbital Coulomb interaction ${\cal U}$
and the exchange interaction ${\cal J}$ \protect\cite{PRB06b}.
All energies are measured in eV.
The symbols `o' and `m' stand for the orthorhombic and monoclinic phase, respectively.
The positions of the transition-metal sites are
explained in Figure \protect\ref{fig.Perovskites_CFandHoppings}.
Generally, the site `1' is located in less distorted environment while the
site `3' is located in more distorted environment.}
\label{tab:Kanamori}
\begin{indented}
\item[]\begin{tabular}{@{}rccccc}
\br
 compound  & phase & site & ${\cal U}$ & ${\cal J}$ \\
\mr
 YTiO$_3$  &  o    & 1    & $3.45$     & $0.62$     \\
 LaTiO$_3$ &  o    & 1    & $3.20$     & $0.61$     \\
 YVO$_3$   &  o    & 1    & $3.27$     & $0.63$     \\
 YVO$_3$   &  m    & 1    & $3.19$     & $0.63$     \\
           &       & 3    & $3.26$     & $0.63$     \\
 LaVO$_3$  &  m    & 1    & $3.11$     & $0.62$     \\
           &       & 3    & $3.12$     & $0.62$     \\
\br
\end{tabular}
\end{indented}
\end{table}
There is certain dependence of the parameter ${\cal U}$ on the
local environment in solid, which is captured by RPA calculations \cite{PRB06a}.
For example, the value of ${\cal U}$
appears to be larger for the more distorted Y$B$O$_3$ compounds.
There is also a clear correlation between the value of ${\cal U}$ and the magnitude of
the
local distortion around two inequivalent transition-metal sites in the monoclinic structure:
the sites experiencing larger distortion (according to the magnitude
of the crystal-field splitting in Figure \ref{fig.Perovskites_CFandHoppings}c)
have larger ${\cal U}$, and vice-versa.
On the other hand, the exchange interaction ${\cal J}$ is less sensitive to the local environment in solids.

  Some of these trends can be rationalized by considering some additional
approximations for the screening of Coulomb interactions in the $t_{2g}$
band by the oxygen $2p$ band (\ref{sec:appendixB}).
Nevertheless, we would like to emphasize that it is quite possible that
these arguments are applicable only for this particular class of compounds.
In other systems, other channels of screening may prevail
and the situation can be generally different.

\begin{figure}[h!]
\begin{center}
\resizebox{12.0cm}{!}{\includegraphics{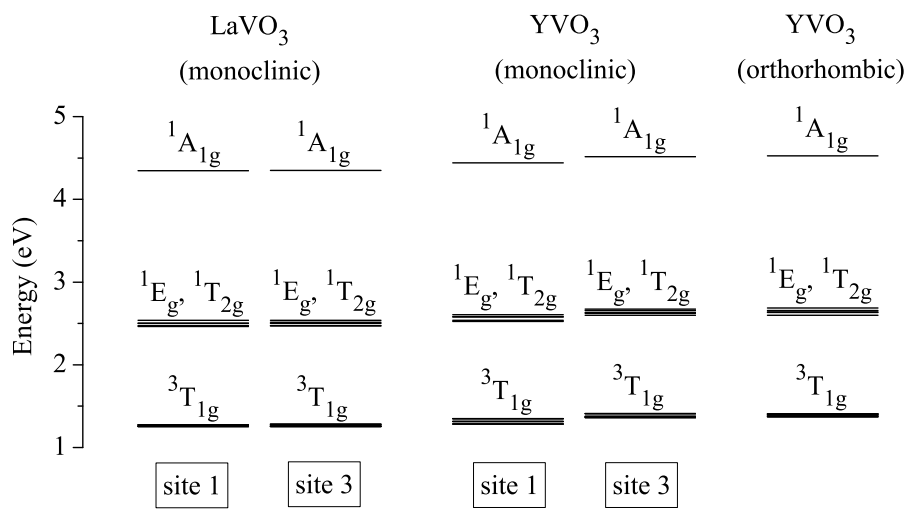}}
\end{center}
\caption{\label{fig.multiplet}
Atomic
multiplet structure of two interacting $t_{2g}$ electrons in the ``ground state''
of vanadates.
The positions of the transition-metal sites in the monoclinic structure are
explained in Figure \protect\ref{fig.Perovskites_CFandHoppings}.
Generally, the sites `1' are located in less distorted environment, while the
sites `3' are located in more distorted environment.}
\end{figure}

  Thus,
even the effective Coulomb interaction in the $t_{2g}$ band
can be affected by the crystal distortion, and this effect is clearly
captured by realistic calculations of the model parameters.
Besides two Kanamori parameters, ${\cal U}$ and ${\cal J}$,
the crystal distortion affects the fine structure of the entire
$\hat{U}$-matrix and
may even change its symmetry.
Generally, such a symmetry is no longer cubic.
This effect is clearly seen in the atomic multiplet structure,
corresponding to
two interacting $t_{2g}$ electrons in the case of vanadates:
because of the crystal distortions the levels $^3{\rm T}_{1g}$,
$^1{\rm E}_g$, and $^1{\rm T}_{2g}$ are slightly split
(Figure \ref{fig.multiplet}).
The effect is not particularly strong.
However, it can probably contribute to some delicate magnetic properties
of the distorted perovskite oxides, such as the orbital
magnetization, magnetocrystalline anisotropy, Dzyaloshinsky-Moriya interactions, etc.

  Finally, it is important to note that the obtained parameters of on-site interactions
are substantially smaller than
the value of Coulomb integral
${\cal U} \sim 5$ eV derived from the photoemission data \cite{MizokawaFujimori},
which is typically used in calculations based on the
dynamical mean-field theory \cite{PavariniNJP,Pavarini,Raychaudhury}.
As it was already pointed out in Section \ref{sec:SrVO3_U}, the reason for such a
difference is rather transparent and related with the additional screening
coming from the oxygen $2p$ as well as the transition-metal $e_g$ bands, which are not explicitly
included into the low-energy model (\ref{eqn:Hmanybody}). The
application of the dynamical mean-field theory with the parameters reported
in Table \ref{tab:Kanamori} would apparently lead to a metallic behavior for
all considered compounds \cite{PavariniNJP}, being in straight contrast
with the experimental data. Apparently, this contradiction can be
resolved by going beyond the single-site approximation and
considering the spacial correlations \cite{OtsukaImada}.

\subsubsection{\label{sec:ApplicationsPRK}Orbital Ordering and Magnetic Properties.}

\begin{figure}[h!]
\begin{center}
\resizebox{10cm}{!}{\includegraphics{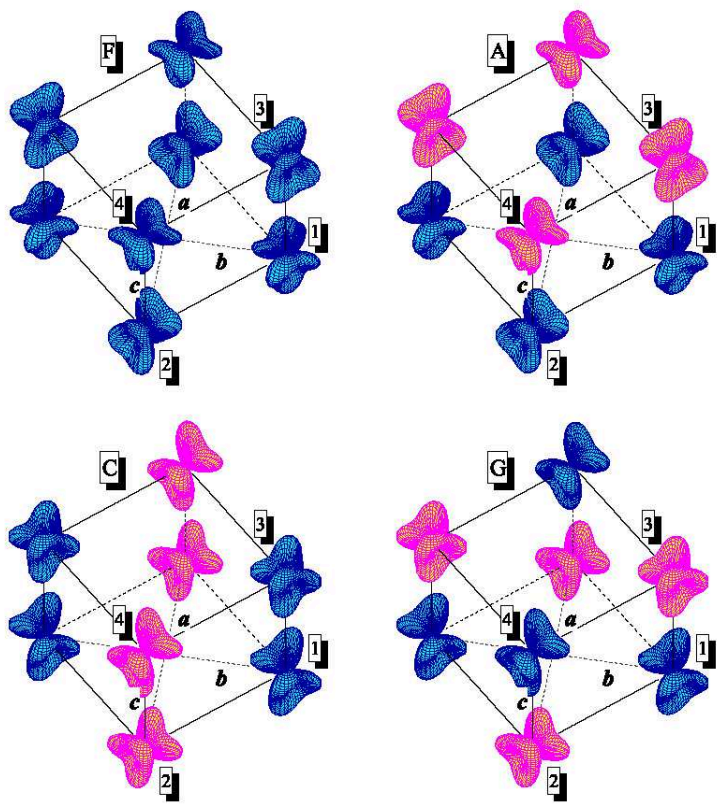}}
\end{center}
\caption{\label{fig.YVO3o_OO}
Distribution of the electron density around vanadium sites
as obtained in the Hartree-Fock calculations for the
ferromagnetic (F) and A-, C-, and G-type antiferromagnetic alignment
in the orthorhombic phase of YVO$_3$ ($T<77$ K) \protect\cite{PRB06b}.
Different magnetic sublattices are shown by different colors.
}
\end{figure}

  It is convenient to start the discussion with the low-temperature orthorhombic
phase of YVO$_3$, which is one of the most distorted systems among the
considered ones. The distribution of the $t_{2g}$ electron density around the
vanadium sites (the so-called orbital ordering) obtained in the Hartree-Fock calculations
is shown in Figure \ref{fig.YVO3o_OO}.
The type of the orbital ordering can be identified as ``C'', meaning that the
orientation of the orbital clouds relative to each other
is nearly orthogonal in the ${\bf ab}$ plane, and nearly parallel
along the ${\bf c}$ axis, i.e. in the close analogy with the C-type
antiferromagnetic spin ordering (Figure \ref{fig.PerovskiteStructures}).

  The orbital ordering is very robust, and practically does not depend on the type
of the magnetic arrangement. This means that the orbital degrees of freedom are
quenched by the crystal-field splitting and the superexchange processes, which may adjust
the orbital ordering in order to minimize the total energy of the system \cite{KugelKhomskii},
are considerably weaker.

  Generally, the orbital ordering of the C-type is compatible with the G-type
antiferromagnetic spin ordering \cite{MizokawaFujimori,Mizokawa99,SawadaTerakura}.
Therefore, it is not surprising that the experimentally observed G-type antiferromagnetic
ground state of orthorhombic YVO$_3$ is reproduced
already at the level of mean-field
Hartree-Fock calculations (Table \ref{tab:TotalEnergies}).
\begin{table}[h!]
\caption{Total energies of $AB$O$_3$ compounds in the ferromagnetic (F)
and
A-, C-, and G-type antiferromagnetic states measured
in meV per one formula unit relative to the most stable
state (marked by the symbol $\times$) as obtained in the Hartree-Fock calculations
and after
taking into account the correlation energy in the second order perturbation theory
(shown in the parenthesis).
The correlation energy includes both on-site and intersite contributions.
The symbols `o' and `m' stand for the orthorhombic and monoclinic phases, respectively.}
\label{tab:TotalEnergies}
\begin{indented}
\item[]\begin{tabular}{@{}rccccc}
\br
 compound  & phase & F-state     & A-state     & C-state     &  G-state       \\
\mr
 YVO$_3$   &  o    & 21.7 (26.8) & 14.6 (17.2) & 10.1 (11.7) &  $\times$         \\
 YVO$_3$   &  m    & 11.7 (17.3) & 14.0 (17.0) & $\times$    & ~6.6 (~7.2) \\
 LaVO$_3$  &  m    & 21.0 (30.7) & 20.6 (24.3) & $\times$    & ~7.6 (11.5) \\
 YTiO$_3$  &  o    & $\times$    & ~2.1 (~0.9) & 14.4 (11.3) & 16.2 (12.9) \\
 LaTiO$_3$ &  o    & ~5.0 (16.9) & $\times$    & 19.6 (26.5) & 11.5 (11.3) \\
\br
\end{tabular}
\end{indented}
\end{table}
The conclusion is totally consistent with results of all-electron LDA+$U$ calculations
\cite{SawadaTerakura,FangNagaosa}, and provides a transparent physical explanation for them.

\begin{figure}[h!]
\begin{center}
\resizebox{10cm}{!}{\includegraphics{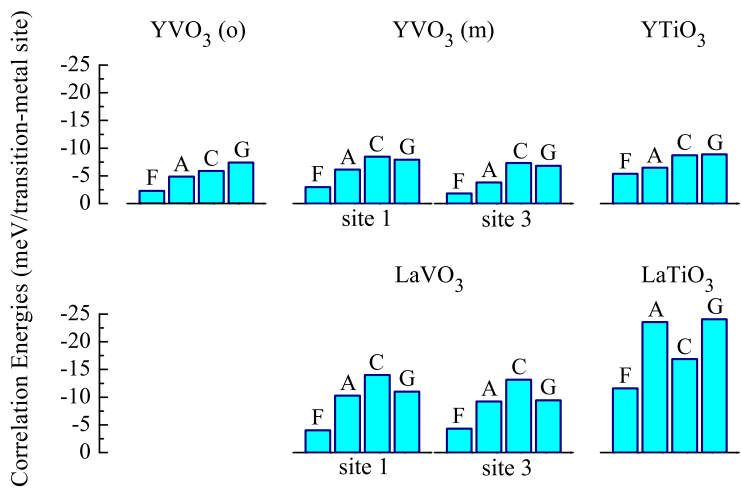}}
\end{center}
\caption{\label{fig.CorrelationEnergies}
Correlation energies in the distorted perovskite oxides
as obtained in the second order perturbation theory starting from the
Hartree-Fock ground state for the
ferromagnetic (F) and A-, C-, and G-type antiferromagnetic alignment.
Only on-site contributions are shown.
The symbols `o' and `m' stand for the orthorhombic and monoclinic phases, respectively.
The positions of the transition-metal sites in the monoclinic structure are
explained in Figure \protect\ref{fig.Perovskites_CFandHoppings}.
Generally, the sites `1' are located in less distorted environment, while the
sites `3' are located in more distorted environment.
}
\end{figure}

  The correlation effects beyond the mean-field Hartree-Fock approximation play a
very important role and additionally stabilize the G-type antiferromagnetic
ground state (Table \ref{tab:TotalEnergies}). The values of
correlation energy obtained in the second order perturbation theory are
summarized in Figure \ref{fig.CorrelationEnergies}.
One can clearly see that
for the orthorhombic phase of YVO$_3$, the correlation energy is clearly the largest in the
G-type antiferromagnetic state among all considered magnetic structures.

  The behavior of interatomic magnetic interactions has been discussed in \cite{PRB06b}.
In the case of orthorhombic YVO$_3$, there is a good agreement with experimental data concerning both the
form (nearly isotropic three-dimensional antiferromagnetic network) and the absolute values of these
interactions \cite{Ulrich2002}.

\begin{figure}[h!]
\begin{center}
\resizebox{10cm}{!}{\includegraphics{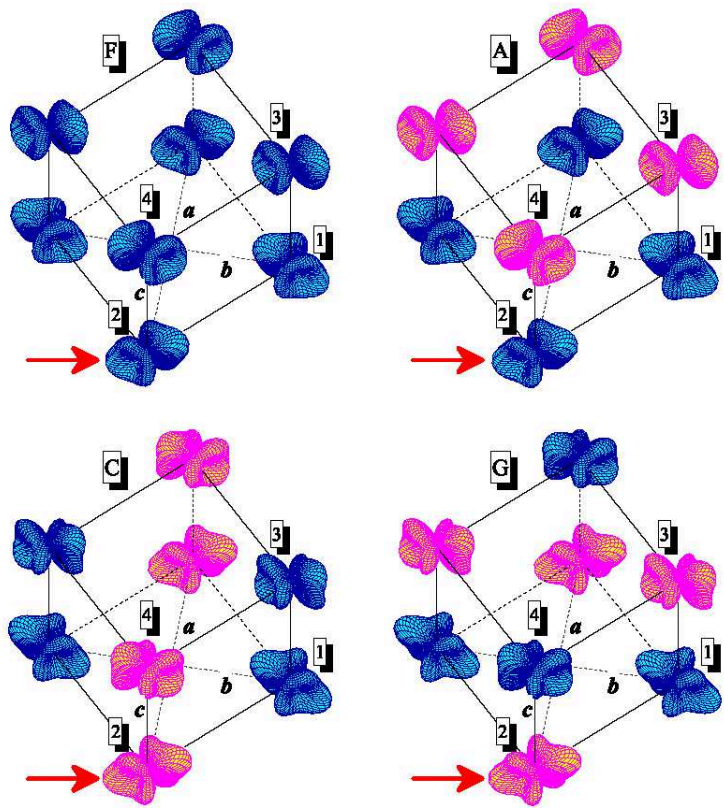}}
\end{center}
\caption{\label{fig.YVO3m_OO}
Distribution of the electron density around vanadium sites
as obtained in the Hartree-Fock calculations for the
ferromagnetic (F) and A-, C-, and G-type antiferromagnetic alignment
in the monoclinic phase of YVO$_3$ ($T>77$ K) \protect\cite{PRB06b}.
Different magnetic sublattices are shown by different colors.
The arrows point at the orbital clouds in the $(1,2)$ planes, which slightly change
their shape depending on the magnetic ordering. On the contrary, the
orbitals in the $(3,4)$ planes are well quenched by the crystal
distortion.
}
\end{figure}

  The transition to the higher-temperature monoclinic phase of YVO$_3$ is associated with
an abrupt change of the orbital ordering patters, which can be identified as the ``G-type''
and
corresponds to nearly orthogonal orientation of the orbital clouds both in the
${\bf ab}$-plane and along the ${\bf c}$-axis (Figure \ref{fig.YVO3m_OO}).
Therefore, it is reasonable to expect the emergence of the C-type
antiferromagnetic structure, which is generally compatible with the
orbital ordering of the G-type \cite{Mizokawa99}. In this sense, there is a close analogy
with the low-temperature orthorhombic phase, where the type of the magnetic ordering
is strictly determined by the type of the orbital ordering. However, there is also
a difference. In the monoclinic phase, there are two types of vanadium atoms,
which are located in the planes (1,2) and (3,4), respectively. The orbital degrees of freedom in the
(3,4) plane are almost rigidly quenched by the large crystal-field splitting, similar to the
orthorhombic phase. However, in the (1,2) plane this quenching is not complete, leaving some
room for the orbital fluctuations. This effect can be seen even visually in Figure \ref{fig.YVO3m_OO},
where the shape of the orbital clouds in the (1,2) plane slightly change depending on the
magnetic state in order to
additionally
minimize the energy of superexchange interactions \cite{KugelKhomskii}.

  Nevertheless, the mean-field Hartree-Fock approach appears to be a good starting point
also for the monoclinic phase of YVO$_3$. It correctly reproduces the C-type antiferromagnetic
ground state (Table \ref{tab:TotalEnergies}), in agreement with all-electron
LDA+$U$ calculations \cite{FangNagaosa}.
The correlation energy additionally stabilizes the C-type antiferromagnetic phase relative to
other magnetic structures. It is also important that the orthorhombic-to-monoclinic transition
in YVO$_3$ is well reflected in the behavior of the correlation energy, which was the largest
for the G-type antiferromagnetic state in the orthorhombic phase and
becomes the largest for the
C-type antiferromagnetic state in the monoclinic phase
(Figure \ref{fig.CorrelationEnergies}). In the monoclinic phase, there is certain dependence
of the correlation energy on the local environment of the vanadium sites:
generally, the site `1' with less distorted environment has larger correlation energy
(and vise versa), though the effect is not particularly strong.

\begin{figure}[h!]
\begin{center}
\resizebox{10cm}{!}{\includegraphics{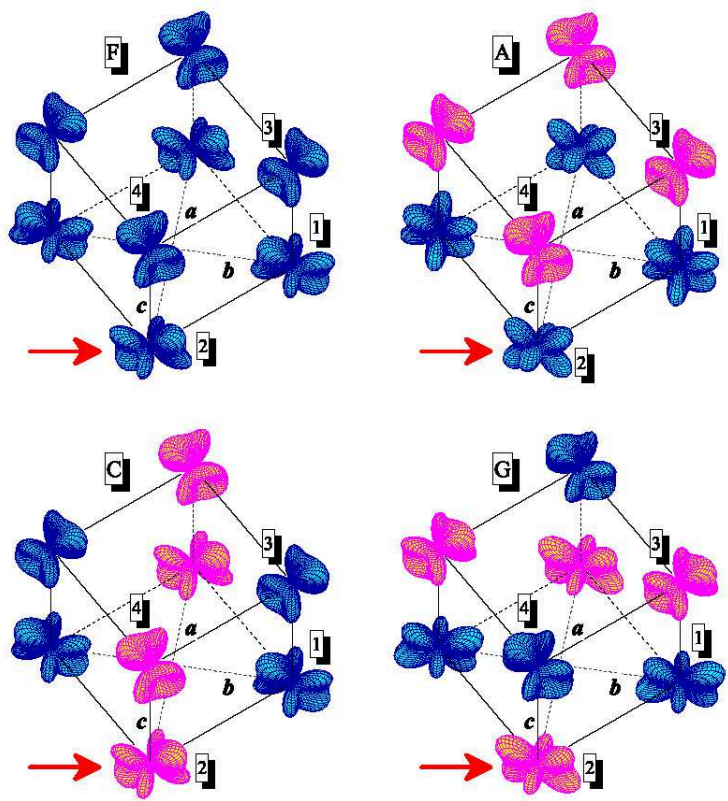}}
\end{center}
\caption{\label{fig.LaVO3_OO}
Distribution of the electron density around vanadium sites
as obtained in the Hartree-Fock calculations for the
ferromagnetic (F) and A-, C-, and G-type antiferromagnetic alignment
in LaVO$_3$ \protect\cite{PRB06b}.
Different magnetic sublattices are shown by different colors.
The arrows point at the orbital clouds in the $(1,2)$ planes, which change
their shape depending on the magnetic ordering. On the contrary, the
orbitals in the $(3,4)$ planes are well quenched by the crystal
distortion.
}
\end{figure}

  Similar to the monoclinic phase of YVO$_3$, the orbital ordering in LaVO$_3$ tends to
stabilize the C-type antiferromagnetic ground state. The orbital degrees of freedom are
quenched in the (3,4) plane and retain enough flexibility in the (1,2) plane.
For example, each change of the magnetic state is accompanied by the substantial
reconstruction of the orbital ordering patter in the (1,2) plane, which is shown by arrows
in Figure \ref{fig.LaVO3_OO}.
Thus, the importance of the orbital fluctuations should rise in the direction
orthorhombic YVO$_3$ $\rightarrow$ monoclinic YVO$_3$ $\rightarrow$ (monoclinic) LaVO$_3$.
Apparently, the application of the mean-field Hartree-Fock theory to the latter
compound is already rather critical. Nevertheless, it still
provides a consistent explanation for the number of properties of LaVO$_3$.
For example, the C-type antiferromagnetic ground state is
successfully reproduced by the Hartree-Fock calculations and additionally
stabilized by correlation effects treated in the second order
perturbation theory (Table \ref{tab:TotalEnergies}).
The absolute values of correlation energy are larger than in YVO$_3$ (Figure \ref{fig.CorrelationEnergies}),
but still substantially smaller than the crystal-field splitting.
Apparently, this is one of the reasons why the Hartree-Fock theory is still applicable.

\begin{figure}[h!]
\begin{center}
\resizebox{10cm}{!}{\includegraphics{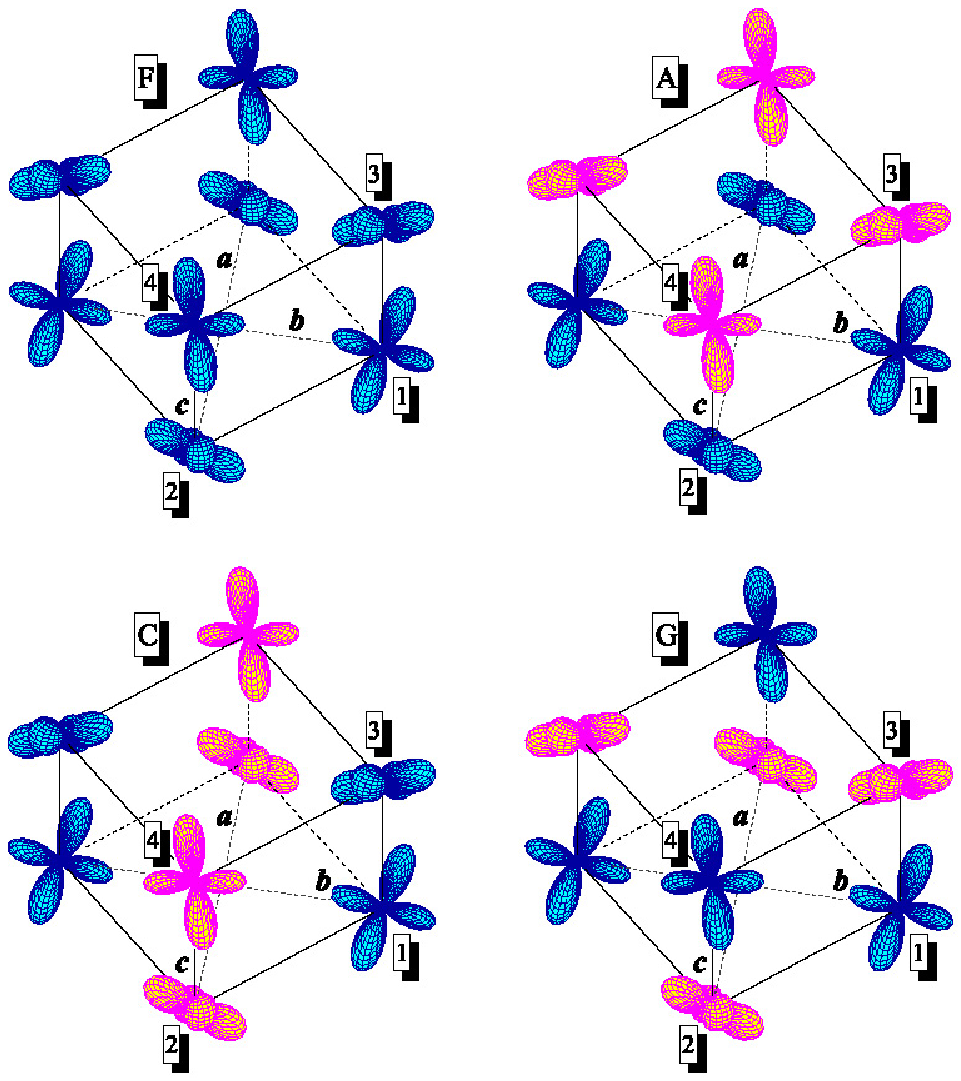}}
\end{center}
\caption{\label{fig.YTiO3_OO}
Distribution of the electron density around titanium sites
as obtained in the Hartree-Fock calculations for the
ferromagnetic (F) and A-, C-, and G-type antiferromagnetic alignment
in YTiO$_3$ \protect\cite{PRB06b}.
Different magnetic sublattices are shown by different colors.
}
\end{figure}

  The situation with YTiO$_3$ is similar to the orthorhombic YVO$_3$. The orbital degrees of freedom
are frozen in some particular configuration by the crystal-field splitting. This orbital configuration
practically does not depend on the magnetic state (Figure \ref{fig.YTiO3_OO}).
The type of the orbital ordering is compatible with the ferromagnetic ground state,
which can be reproduced at the Hartree-Fock level. The same conclusion has been
drawn in all-electron LDA+$U$ calculations \cite{SawadaTerakura,Okatov}.
The correlation effects tend to destabilize the ferromagnetic ground state
(Table \ref{tab:TotalEnergies} and Figure \ref{fig.CorrelationEnergies}).
Although the ferromagnetic state has the lowest energy even after taking into account the
correlation effects, the energy separation
from the next A-type antiferromagnetic state is very fragile.\footnote{
More precisely, the on-site correlations tends to destabilize the ferromagnetic ground state,
as it is clearly seen from Figure \ref{fig.CorrelationEnergies}. Small intersite correlations
partially compensate this trend, again in the favor of the ferromagnetic alignment \cite{JETP07}.
}
However, this seems to be consistent with the relatively low Curie temperature
($T_C \approx 30$ K \cite{Akimitsu}) observed in YTiO$_3$ \cite{PRB06b}.

\begin{figure}[h!]
\begin{center}
\resizebox{10cm}{!}{\includegraphics{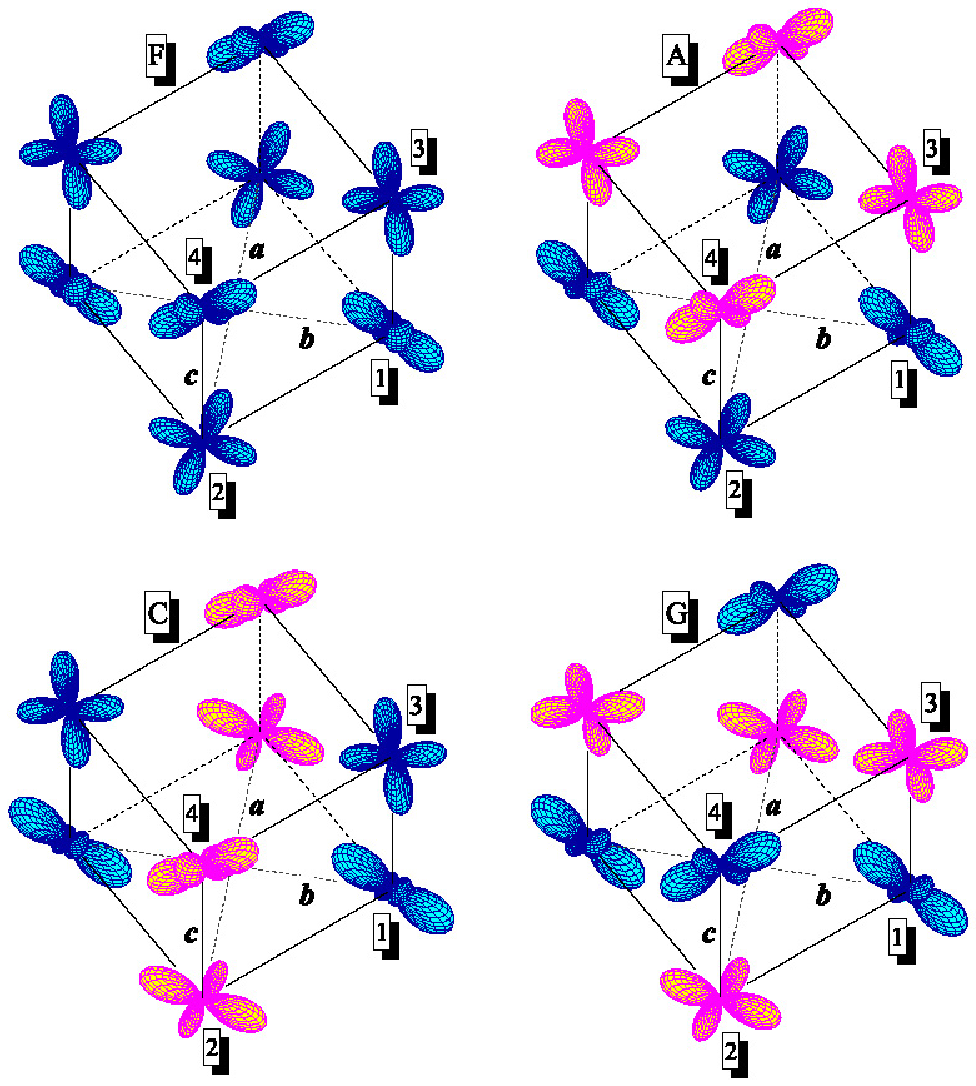}}
\end{center}
\caption{\label{fig.LaTiO3_OO}
Distribution of the electron density around titanium sites
as obtained in the Hartree-Fock calculations for the
ferromagnetic (F) and A-, C-, and G-type antiferromagnetic alignment
in orthorhombic phase of LaTiO$_3$ \protect\cite{PRB06b}.
Different magnetic sublattices are shown by different colors.
}
\end{figure}

  LaTiO$_3$ is clearly an exception. The orbital degrees of freedom are rather flexible
and not completely quenched by the crystal distortion (Figure \ref{fig.LaTiO3_OO}).
The mean-field theory yields an
incorrect magnetic ground state
(antiferromagnetic A-type, instead of the G-type, Table \ref{tab:TotalEnergies}).
The correlation interactions treated as a perturbation to the Hartree-Fock solution
do not change the overall picture, and the G-type antiferromagnetic state
remain unstable relative to the A-state.\footnote{
Effects of the higher-order perturbation theory expansion have been considered in \cite{JETP07},
in the framework of the $T$-matrix theory. Despite some quantitative differences from the second-order
perturbation theory, the main conclusion was the same and the
G-type antiferromagnetic state was always higher in energy than the A-type antiferromagnetic state.
}

  Thus, the origin of the G-type antiferromagnetic ground state in LaTiO$_3$ remains a challenging
problem for the future theories. Apparently, one of the basic assumptions about nondegeneracy
of the Hartree-Fock ground state breaks down in the case of LaTiO$_3$, and the true magnetic ground state
cannot be continuously approached through the series of corrections applied to the
Hartree-Fock ground state.
One intuitive reason could be seen already in Figure \ref{fig.CorrelationEnergies}:
the correlation energies, derived as perturbations to the Hartree-Fock solutions,
are exceptionally large in the case of LaTiO$_3$, so that some of them become comparable with the
crystal-field splitting (Figure \ref{fig.Perovskites_CFandHoppings}). This means that
there is certain inconsistency in our theoretical treatment: the perturbation becomes comparable
with the basic energy splitting, that was used in order to justify this
perturbation-theory expansion.

  In summary, the crystal-field theory can indeed explain many properties of the
distorted perovskite oxides, but apparently not all.
Particularly, the decrease of the crystal distortion in the case of lanthanides leaves
some room for the orbital fluctuations. LaTiO$_3$ is definitely the most difficult case,
where the crystal-field theory,
if
supplemented with realistic values of the model parameters,
fails to reproduce the correct magnetic ground state.

  The main conclusions of this section are based on the second-order
perturbation theory for the correlation interactions, which may be questionable
because the values of the Coulomb repulsion ${\cal U}$ (Table \ref{tab:Kanamori})
and not particularly small and it is reasonable to expect that
higher-order effects may modify some of these conclusions.
The role of higher-orders of the perturbation theory expansion
for the correlation energy was investigated in \cite{JETP07}, in the frameworks
of the $T$-matrix theory. Generally, the higher-order effects tend to reduce
the correlation energy. However, the amount of this reduction strongly
depends on the magnetic state. For example, the correlation energy
in the ferromagnetic state is overestimated in the second-order
perturbation theory by less than 15\% in comparison with the
$T$-matrix method. For the antiferromagnetic configurations, the effect
is more pronounced and the correlation energy can be overestimated
by about 50\%. Nevertheless, the main trends in the behavior of the
correlation energy are well captured already by the second-order
perturbation theory. For example, if the absolute value of the
correlation energy was the largest for certain magnetic configuration
in the second-order perturbation theory, the same tendency is clearly
seen in the $T$-matrix theory, etc. In this sense we believe that
the main conclusions of the present Section are valid.
Nevertheless, the correlation interactions in the distorted
perovskite oxides is certainly one of the most important and
interesting problems, which deserves a thorough investigation.
The analysis based on the second-order perturbation theory,
considered in this Section, can be regarded as only the first step in this
direction.

\subsubsection{Effects of Spin-Orbit Interaction.}

  The spin-orbit interaction in distorted perovskite structures generally leads
to a noncollinear magnetic alignment, which obeys certain
symmetry rules \cite{Treves,Moskvin,Yamaguchi,PRL96}.
The spin magnetic moments aligned along one of the orthorhombic axes are subjected to certain
rotational forces originating from anisotropic and Dzyaloshinsky-Moriya interactions \cite{Dzyaloshinsky,Moriya},
which lead to the reorientation (or canting) of these magnetic moments.
Hence, in the equilibrium
magnetic structure we will generally have all three projections of the magnetic moments
onto the orthorhombic axes.
Furthermore, the type of the magnetic ordering for these three projections
will be generally different.
Thus, each magnetic structure can be generally abbreviated as
X-Y-Z, where X, Y, and Z is the type of the magnetic ordering
(F, A, C, or G) formed by the projections of the spin magnetic moments
onto the orthorhombic axes ${\bf a}$, ${\bf b}$, and ${\bf c}$,
respectively.
The orbital magnetic structure has the same symmetry, although it may have a
different origin of the canting, which arises mainly from the minimization
of the single-ion anisotropy energy at each transition-metal site.
Generally, the spin and orbital magnetic moments are not collinear to each other \cite{PRL96,PRB97}.

\begin{figure}[h!]
\begin{center}
\resizebox{12cm}{!}{\includegraphics{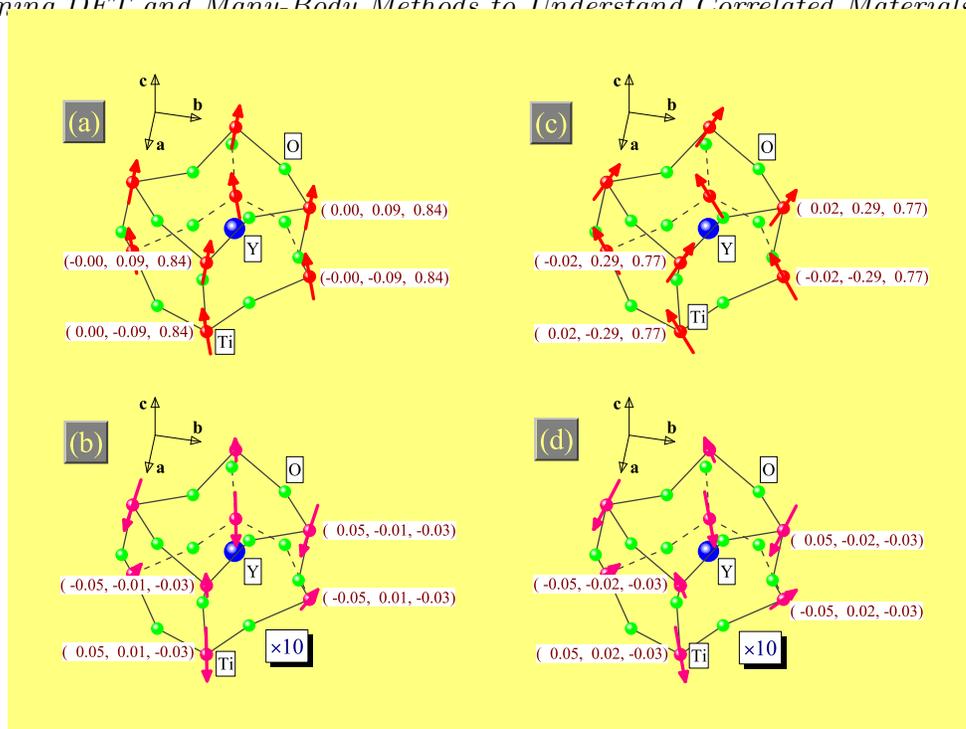}}
\end{center}
\caption{\label{fig.YTiO3_magnetic_structure}
Spin (a and c) and orbital (b and d) magnetic structures of
YTiO$_3$ in the Hartree-Fock approximation (a and b) and
after taking into account the correlation effects
in the frameworks of variational superexchange theory (c and d).
The directions of the magnetic moments at different titanium sites
are shown by arrows. The size of each arrow is proportional
to the size of the magnetic moment.
For the clarity,
the arrows for the orbital magnetic moments
have been additionally scaled by the factor ten.
Corresponding values of spin and orbital magnetic moments,
measured in Bohr magnetons,
are given in the parenthesis.
}
\end{figure}

  For example, the true magnetic ground state of YTiO$_3$ is G-A-F (Figure \ref{fig.YTiO3_magnetic_structure}),
where in addition to the main ferromagnetic components along the ${\bf c}$ axis, there
will be two antiferromagnetic
components parallel to the ${\bf a}$ and ${\bf b}$ axes.
In this case, one can expect rather interesting consequence of the correlation effects
on the noncollinear magnetic alignment \cite{PRB06b}.
As it was already pointed out in previous Section,
without spin-orbit interaction,
the correlation effects in YTiO$_3$
favor the antiferromagnetic interatomic coupling and systematically
lower the energies of all antiferromagnetic structures
relative to the ferromagnetic one. Then, it is reasonable to
expect that after taking into consideration the spin-orbit interaction, the correlation effects
will systematically increase the weight of the antiferromagnetic components in the ground state
configuration
and result in the additional spin canting away from the collinear ferromagnetic state.
Such an effect is clearly seen in Figure \ref{fig.YTiO3_magnetic_structure}:
after taking into account the correlation effects
in the frameworks of variational superexchange theory,
about 8\% of the spin magnetization-density
is transferred from the ferromagnetic part parallel to the ${\bf c}$ axis
to the antiferromagnetic part lying in the ${\bf ab}$ plane.
The distribution of the orbital magnetization-density is less sensitive to the correlation effects.
We also note that the correlation effects readily explain the experimental values
reported for the F and G components of the magnetic moments \cite{Ulrich2002}.

\begin{figure}[h!]
\begin{center}
\resizebox{8cm}{!}{\includegraphics{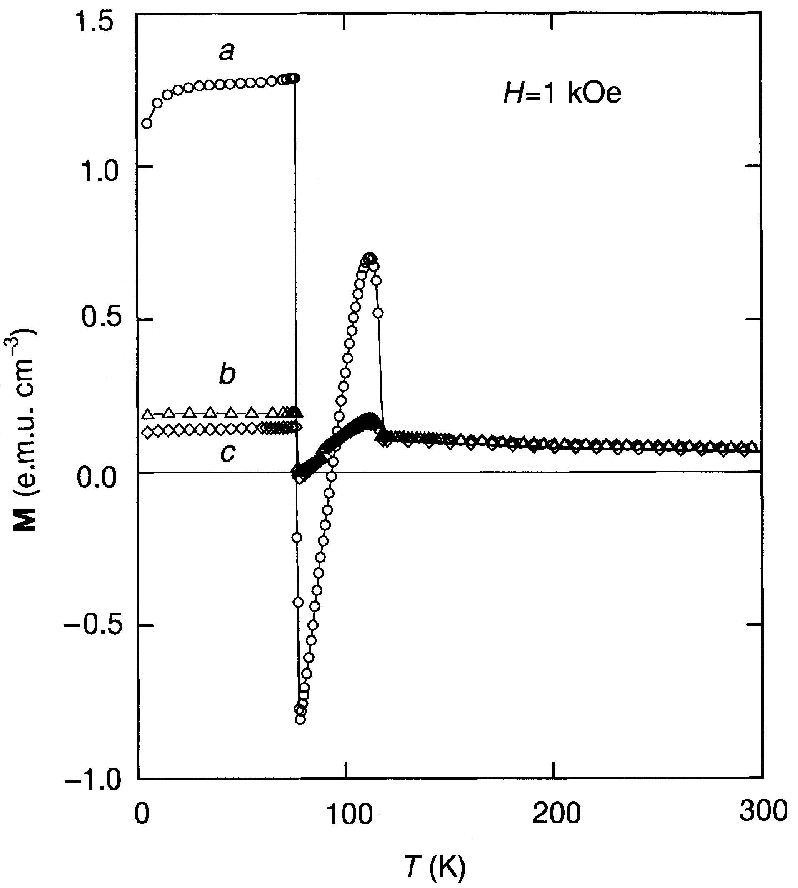}}
\end{center}
\caption{\label{fig.YVO3_reversal}
Temperature dependence of the magnetization in YVO$_3$ (from \protect\cite{Ren}).
The magnetic field of 1 kOe is applied along the
${\bf a}$, ${\bf b}$, and ${\bf c}$ axes, respectively. In the monoclinic
phase, realized between 116 K and 77 K, the magnetization after first increasing
start decreasing and crosses zero at around 95 K to a large negative value.
With further cooling, it jumps to a large positive value at the point of
monoclinic-to-orthorhombic transition. The orientation of the
${\bf a}$, ${\bf b}$, and ${\bf c}$ axes corresponds to the $Pbnm$ group, both in orthorhombic and
monoclinic phases.
}
\end{figure}

  Another interesting phenomenon, which is directly related with the spin-orbit interaction,
is the temperature-induced magnetization reversal behavior observed in
YVO$_3$ (Figure \ref{fig.YVO3_reversal}) \cite{Ren}.
Briefly,
upon cooling,
the magnetization parallel to the ${\bf a}$ axis (in the notations of the $Pbnm$ group)
changes the sign two times: first continuously, in the monoclinic regions, and then discontinuously,
at the point of monoclinic-to-orthorhombic transition.
Although the discontinuity of the magnetization is certainly expected for the
first order monoclinic-to-orthorhombic transition, its behavior in the monoclinic phase
is somewhat surprising.
Nevertheless,
it is important to note that
a similar effect is know to occur in some ferrimagnetic materials, like
Co$_2$VO$_4$,
consisting of two or more types of antiferromagnetically ordered magnetic ions \cite{Menyuk}.
Although there are several additional complications in comparison with Co$_2$VO$_4$,
this may be a clue for understanding the unusual behavior of YVO$_3$ \cite{PRB06b}.
\begin{enumerate}
\item
As it was already pointed out in Section \ref{sec:ApplicationsPRK}, although all vanadium atoms
are \textit{chemically} equivalent,
in the monoclinic phase
they are located in \textit{different crystallographic environments} and one can clearly distinguish
two types of vanadium atoms lying in
the planes
(1,2) and (3,4) (Figure \ref{fig.YVO3m_OO}).
\begin{table}[h!]
\caption{Interatomic magnetic interactions (in meV) in the orthorhombic and monoclinic phases
of YVO$_3$, as obtained in the Hartree-Fock calculations without spin-orbit interaction \protect\cite{PRB06b}.
The orthorhombic phase is stabilized below 77 K, while the monoclinic phase is stabilized above 77 K.
The atomic positions are explained in Figure \protect\ref{fig.Perovskites_CFandHoppings}.
The magnetic interactions were computed using formula (\protect\ref{eqn:JHeisenberg}) for
infinitesimal magnetic rotations near the antiferromagnetic ground state
of the G-type in the case of the orthorhombic phase and the C-type in the case of
the monoclinic phase.}
\label{tab:YVO3_Jij}
\begin{indented}
\item[]\begin{tabular}{@{}lcccc}
\br
  phase           & $J_{12}$ & $J_{13}$         & $J_{24}$         &  $J_{34}$ \\
\mr
  orthorhombic    & $-4.4$   & $-4.8$           & $-4.8$           &  $-4.4$   \\
  monoclinic      & $-0.9$   & $\phantom{-}2.2$ & $\phantom{-}2.2$ &  $-4.5$   \\
\br
\end{tabular}
\end{indented}
\end{table}
The magnetic interactions in these two planes
are also different: while $J_{34}$ is strongly antiferromagnetic, similar to the orthorhombic phase,
$J_{12}$ is considerably weaker.\footnote{
Note that the existence of two magnetic sublattices also leads to a splitting of
the
magnon spectrum into acoustic and optical branches \cite{PRB06b,FangNagaosa},
which was clearly seen in the
experiment \cite{Ulrich2003}.
}
Thus, there is a clear analogy with two magnetic sublattices
of Co and V
existing in Co$_2$VO$_4$. However, in the monoclinic
phase of YVO$_3$, this difference is entirely related
with the local crystal distortions, which create two inequivalent types of vanadium sites.
\item
An additional complication comes from the fact that YVO$_3$ is an antiferromagnet, and no net magnetic
moment is expected for the neither C- nor G-type antiferromagnetic ordering,
realized in the monoclinic and orthorhombic phases, respectively. Nevertheless,
it is reasonable to expect a
weak ferromagnetism arising from the spin-orbit interaction in the
distorted perovskite structure. In the orthorhombic phase, the weak ferromagnetic moment is
indeed aligned along the
${\bf a}$ axis (in the notations of the $Pbmn$ group), that directly follows from the
symmetry considerations (Figure \ref{fig.YVO3_magnetic_structure}).
\begin{figure}[h!]
\begin{center}
\resizebox{12cm}{!}{\includegraphics{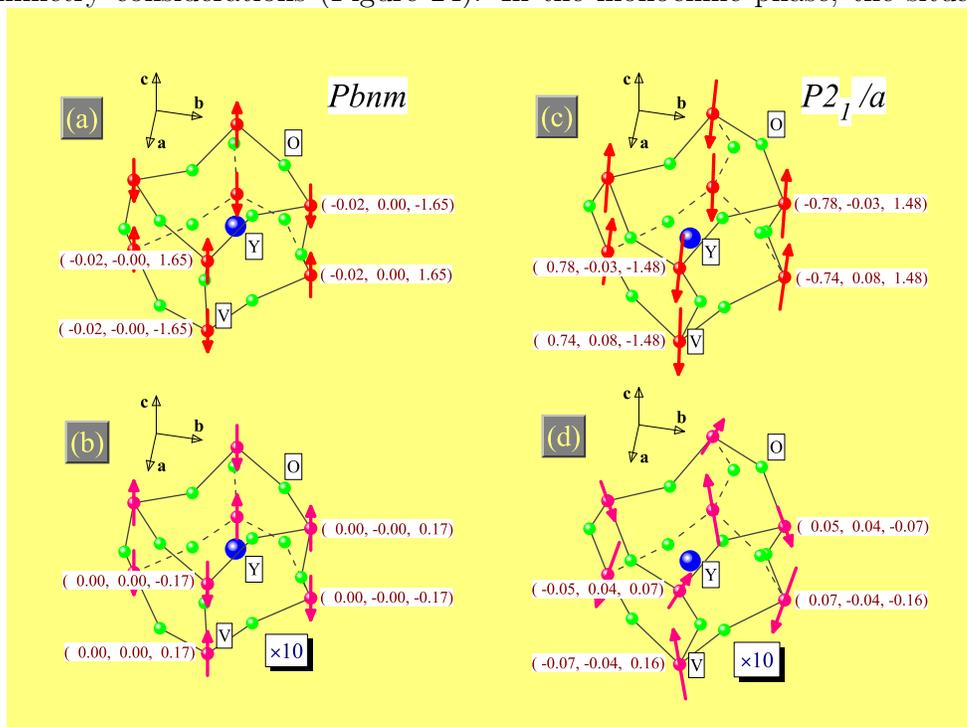}}
\end{center}
\caption{\label{fig.YVO3_magnetic_structure}
Spin (a and c) and orbital (b and d) magnetic structures
realized in the orthorhombic (a and b) and
monoclinic (c and d) phases of
YVO$_3$ in the Hartree-Fock approximation.
The directions of the magnetic moments at different vanadium sites
are shown by arrows. The size of each arrow is proportional
to the size of the magnetic moment.
For the clarity,
the arrows for the orbital magnetic moments
have been additionally scaled by the factor ten.
Corresponding values of spin and orbital magnetic moments,
which are measured in Bohr magnetons,
are given in the parenthesis.
The orientation of the
${\bf a}$, ${\bf b}$, and ${\bf c}$ axes corresponds to the $Pbnm$ group
in the orthorhombic phase and the $P2_1/a$ group in the
monoclinic phases.
The $P2_1/a$ notations can be transformed to the $Pbnm$ notations by
interchanging the axes ${\bf a}$ and ${\bf b}$ \protect\cite{Blake}.
}
\end{figure}
In the monoclinic phase, the situation is a little bit complicated because there is a
large cancelation for all three projections of the magnetic moments. However, since
the planes (1,2) and (3,4) are inequivalent, this cancelation is not complete
and the net magnetic moment can be expected along the ${\bf b}$ direction in the notations of the
$P2_1/a$ group, which corresponds to the ${\bf a}$ direction in the notations of the
$Pbmn$ group.
\end{enumerate}

  Thus, there is at least a qualitative consistency with the experimental data \cite{Ren}.
A quantitative theory of the magnetization reversal behavior in YVO$_3$ is still missing,
and would be an interesting step to do.

\subsection{\label{sec:KO2} Hyperoxide KO$_2$}

  As a final example, we would like to consider a molecular analog of ``strongly-correlated'' system
without any transition-metal or rare-earth elements.

\begin{figure}[h!]
\begin{center}
\resizebox{12cm}{!}{\includegraphics{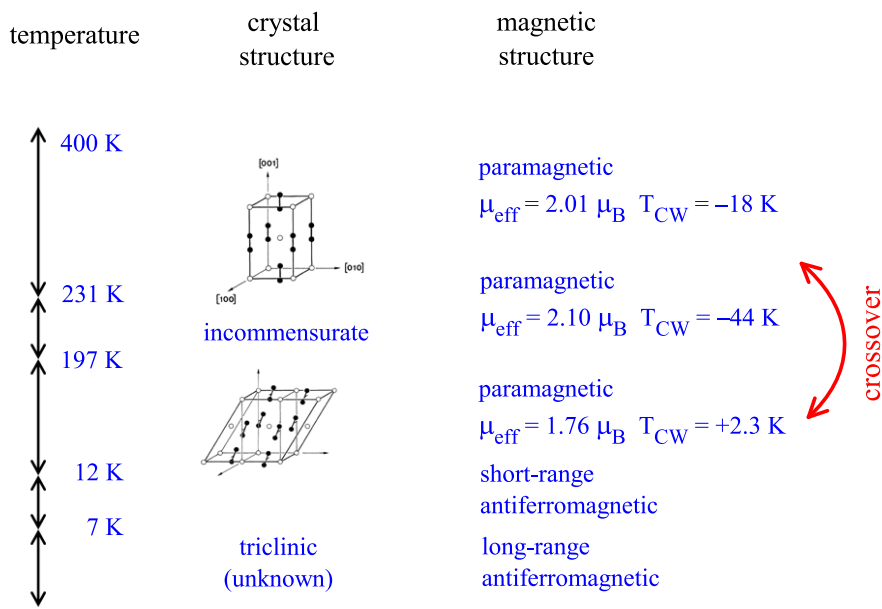}}
\end{center}
\caption{\label{fig.KO2_PhaseDiagram}
Brief summary of structural and magnetic properties of KO$_2$ \protect\cite{Kanzig,Labhart}.
The oxygen molecules are shown as black dimers and potassium atoms are indicated by white symbols.
$\mu_{\rm eff}$ is the effective magnetic moment and $T_{CW}$ is the Curie-Weiss temperature.
Both are
derived from the analysis of magnetic susceptibility data in different temperature regimes.
The arrow shows the region of ``ferromagnetic-antiferromagnetic crossover'', where the Curie-Weiss temperature
changes the sign.}
\end{figure}

  The simplest molecule, whose ground state
is different from the conventional spin singlet is O$_2$.\footnote{
The O$_2$ molecule has two unpaired electrons in the doubly degenerate $\pi_g$ level,
resulting in the spin-triplet ground state, as it is required by the first Hund rule.
}
If these molecules form a crystal (by either cooling or pressurizing), it
may become magnetic. Such a situation is indeed realized in the solid oxygen \cite{Serra,Goncharenko}.
Since
O$_2$ is a good oxidizer and can easily take an additional electron when
it brought in contact with alkali elements,
there is another way of making crystalline arrays of the oxygen molecules,
in the form of ionic crystals.
Such materials do exist.
One typical example is KO$_2$, which was used as the chemical oxygen generator
in rebreathers, spacecrafts, and life support systems.
The magnetic properties of KO$_2$ and other alkali hyperoxides were intensively
studied in 1970s \cite{Ziegler,Kanzig,Labhart,Boesch,Lines,Zumsteg}.
However, many details of their behavior remain largely unknown, even today.\footnote{
Evidently, the early progress was severely hampered by many objective
difficulties related with the preparation of rigidly held samples
and separation of intrinsic properties of KO$_2$ from
inhomogeneity effects.
}
KO$_2$ has six crystallographic modifications, apparently related with
different orientations of the oxygen molecules (Figure \ref{fig.KO2_PhaseDiagram}).
However, the details are known only for the body-centered tetragonal ($bct$) phase
stabilized between 231 K and 400 K. Below 7K, KO$_2$ develops a long-range antiferromagnetic order \cite{Kanzig,Smith}.
However, its type is unknown. In the paramagnetic region, the Curie-Weiss temperature ($T_{CW}$) changes
the sign at around 230 K, indicating at some change of interatomic magnetic interactions. Formally, this
change may be related with one of the structural phase transitions. However, no profound change
of the magnetic susceptibility has been observed around this transition, indicating that there might be a
more fundamental reason for the change of $T_{CW}$.
Finally, the effective magnetic moment in the paramagnetic region ($\mu_{\rm eff}$) is about 2 $\mu_B$,
meaning that the large orbital contribution (about 1 $\mu_B$) is not quenched by the crystal distortion and
persists down to 12 K. This suggests that the actual crystal distortion
related with the reorientation of the oxygen molecules near the $bct$ phase is quite week,
and the $bct$ phase itself can be regarded as a good starting point for the analysis of the magnetic properties of KO$_2$
in the wide paramagnetic region \cite{KO2_condmat}.

\begin{figure}[h!]
\begin{center}
\resizebox{12cm}{!}{\includegraphics{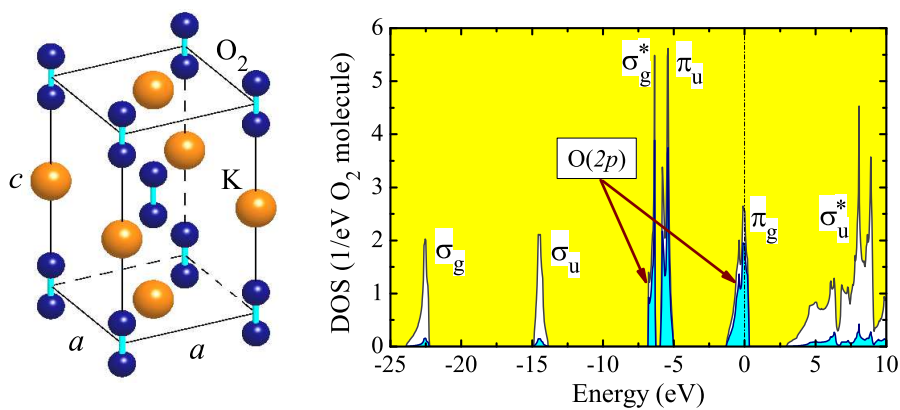}}
\end{center}
\caption{\label{fig.KO2_DOS}
Crystal structure and LDA density of states of the \textit{bct} phase of KO$_2$.
The shaded area shows the contributions of the oxygen $2p$ states.
Other symbols show the positions of the main bands.
Fermi level is at zero energy.}
\end{figure}

  The oxygen molecule appears to be the building block of not only the crystal,
but also of the electronic structure of KO$_2$
in the local-density approximation.
The strong hybridization within the
molecule leads to the formation of the
molecular levels. The interaction between the
molecules is considerably weaker, so that the molecular orbitals form a group
of narrow nonoverlapping bands
(Figure \ref{fig.KO2_DOS}).
Thus, there is a
clear analogy with the atomic limit in the physics of strongly-correlated systems \cite{KugelKhomskii,TokuraNagaosa},
except that now,
the localized electrons (or holes) reside on
the \textit{molecular orbitals}, which are distributed between two atomic sites.
The doubly-degenerate
$\pi_g$ band located near the Fermi level is formed by antibonding
molecular $p_x$ and $p_y$ orbitals.
In comparison with the solid oxygen, the potassium atom donates an extra electron
into the $\pi_g$ band. Therefore, the band is $\frac{3}{4}$ filled.
Due to the peculiar $\frac{3}{4}$ filling, not only spin but also orbital degrees of freedom appear to be active
and contribute to spin and lattice dynamics of KO$_2$.
In this sense,
KO$_2$ can be regarded as a molecular analog of correlated electron systems, comprising of
orbitally degenerate magnetic O$_2^-$ ions \cite{KO2_condmat}.

\subsubsection{Construction and Parameters of Model Hamiltonian.}

  Obviously that the minimal model for KO$_2$ should
be constructed in the basis of $\pi_g$ bands
and take into account the Coulomb correlations beyond
conventional LDA.
The basic idea of the construction of such a model is to relate each lattice point
to the single oxygen molecule and formulate the problem in the Wannier basis
corresponding to antibonding molecular $p_x$ and $p_y$ orbitals.

\begin{figure}[h!]
\begin{center}
\resizebox{10cm}{!}{\includegraphics{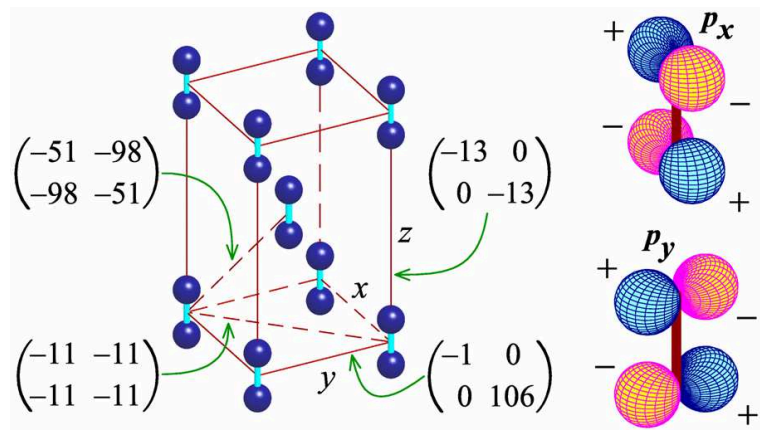}}
\end{center}
\caption{\label{fig.KO2_transfer_integrals}
Transfer integrals (in meV)
associated with different bonds
in the $bct$ phase of KO$_2$ \protect\cite{KO2_condmat}.
The antibonding molecular
orbitals
are shown in the right part of the figure.
The order of orbitals is $p_x$ and $p_y$.
}
\end{figure}

  The transfer integrals
operating between such \textit{molecular Wannier orbitals}
can be derived by using formal downfolding procedure described in Section \ref{sec:downfoldingasprojector}.
The behavior of these transfer integrals
in the \textit{bct} lattice
is explained in Figure \ref{fig.KO2_transfer_integrals}.
Basically, all interactions are restricted by four nearest neighbors.
Other interactions are considerably smaller.
The on-site part of the one-electron Hamiltonian $\| h^{\alpha \beta}_{{\bf R}{\bf R}'} \|$
incorporates the relativistic spin-orbit interaction,
$$
\hat{h}_{{\bf R}{\bf R}} =
\frac{\xi}{2}
\left(
\begin{array}{cccc}
  0  & -i  &  0  &  0  \\
  i  &  0  &  0  &  0  \\
  0  &  0  &  0  &  i  \\
  0  &  0  & -i  &  0  \\
\end{array}
\right),
$$
in the basis of
$| p_x$$\uparrow \rangle$,
$| p_y$$\uparrow \rangle$,
$| p_x$$\downarrow \rangle$,
and $| p_y$$\downarrow \rangle$ orbitals,
where
the arrows indicate the spin state.
The
parameter of spin-orbit interaction $\xi$ is about 17 meV \cite{KO2_condmat}.

  The screened Coulomb interactions in the $\pi_g$ band are
computed in two steps~\cite{PRB06a}.
First,
the interaction parameters between \textit{atomic} $2p$ orbitals
are derived
by using the constrained DFT method.
It yields
the intraatomic Coulomb interaction $u \approx 11.4$ eV,
the interatomic intramolecular Coulomb interaction $v \approx 1.3$ eV,
and the intraatomic exchange interaction $j \approx 2.3$ eV.\footnote{
For comparison, the parameters of bare interactions are
$u \approx 27.8$ eV, $v \approx 11.0$ eV, and $j \approx 2.9$ eV.
Thus, similar to the cubic perovskites considered in Section \ref{sec:SrVO3_U}
the interatomic Coulomb interactions $v$ appears to be the most screened,
while the intraatomic exchange interaction $j$ is the least screened.
}
After that, it is necessary to consider the
additional screening of these interactions in the
$\pi_g$ band by other bands,\footnote{
Namely, the $\pi_u$ band, that is a bonding
combination of the $p_x$ and $p_y$ orbitals, as well as the
$\sigma_g^*$ and $\sigma_u^*$ bands, constructed
from the $p_z$ orbitals (Figure \ref{fig.KO2_DOS}).
}
and derive
the full matrix
$\hat{U}$$=$$\| U_{\alpha \beta \gamma \delta} \|$
of screened Coulomb interactions
between \textit{antibonding molecular orbitals}
$p_x$ and $p_y$, which is fully specified by
the parameters of
intraorbital Coulomb interaction ${\cal U} \approx 3.6$ eV
and the exchange interaction ${\cal J} \approx 0.6$ eV.
This part can be done in the random-phase approximation,
by
starting from interaction parameters obtained in the constrained DFT.
It is also important to remember that
the interorbital Coulomb interaction ${\cal U}'$
is related with ${\cal U}$ and ${\cal J}$ by the identity
${\cal U}'={\cal U}-2{\cal J}$.

\subsubsection{Implications to the Properties of KO$_2$.}

Since the Coulomb interactions ${\cal U}$ and ${\cal U}'$ are clearly the largest
parameters in the problem, the Hubbard model (\ref{eqn:Hmanybody})
can be further converted into a spin-orbital superexchange model
by starting from the limit of isolated oxygen molecules and treating
all transfer integrals as a perturbation \cite{KugelKhomskii}.
Due to the $\frac{3}{4}$ filling, it is convenient to use
the hole representation and ascribe to each molecular site
a single hole spin-orbital
\begin{equation}
| \alpha \rangle= ab_\uparrow | p_x \uparrow \rangle +
a\tilde{b}_\uparrow | p_y \uparrow \rangle +
\tilde{a}b_\downarrow | p_x \downarrow \rangle +
\tilde{a}\tilde{b}_\downarrow | p_y \downarrow \rangle,
\label{eqn:HoleOrbital}
\end{equation}
where
$|a|^2$$+$$|\tilde{a}|^2 =
|b_\uparrow|^2$$+$$|\tilde{b}_\uparrow|^2 =
|b_\downarrow|^2$$+$$|\tilde{b}_\downarrow|^2 = 1$.
Then, the energy gain
${\cal T}_{{\bf R}{\bf R}'} \equiv {\cal T}(\alpha_{\bf R},\alpha_{{\bf R}'})$
caused by the virtual hoppings
in the bond $\langle {\bf RR}' \rangle$
can be computed
using the formula (\ref{eqn:egain}). In this case,
$G \equiv G(\alpha_{\bf R},\alpha_{{\bf R}'})$
is the Slater determinant constructed from the hole orbitals
at the molecular centers ${\bf R}$ and ${\bf R}'$,
$E_{{\bf R}'M}$ and $|{\bf R}'M \rangle$ are the eigenvalues and eigenstates of
the
excited two-hole configurations at the molecular center ${\bf R}'$, and $\hat{\mathscr{P}}_{{\bf R}'}$ is a projector
operator enforcing the Pauli principle and preventing any hoppings of holes into $\alpha_{{\bf R}'}$.

  The problem can be further simplified by eliminating the orbital
degrees of freedom, described by the $b$-vasriable in (\ref{eqn:HoleOrbital}),
and constructing an effective
spin model separately for each temperature regime.
This can be generally done by averaging (\ref{eqn:egain})
with some distribution function
${\cal D}(b_\uparrow,\tilde{b}_\uparrow,b_\downarrow,\tilde{b}_\downarrow,T)$ \cite{KugelKhomskii}.
The formulation is especially simple for two limiting cases:
$T \rightarrow 0$, corresponding to an orbitally ordered state, and
$T \rightarrow \infty$, corresponding to the complete orbital disorder.

  In the \textit{low-temperature} limit
$k_B T$$\ll$$\xi$ (note that $\xi/k_B \approx 200$ K),
${\cal D}(b_\uparrow,\tilde{b}_\uparrow,b_\downarrow,\tilde{b}_\downarrow,T)$
is fully controlled by the relativistic spin-orbit interaction, which picks up
a linear combination of two spin-orbitals,
$|p_+$$\uparrow$$\rangle = -$$(|p_x$$\uparrow$$\rangle$$-$$i|p_y$$\uparrow$$\rangle)/\sqrt{2}$
and
$|p_-$$\uparrow$$\rangle = (|p_x$$\uparrow$$\rangle$$+$$i|p_y$$\uparrow$$\rangle)/\sqrt{2}$,
minimizing the
spin-orbit interaction energy.
Since each hole-orbital $\alpha$ is confined in the two-dimensional subspace spanned
by $|p_+$$\uparrow$$\rangle$ and $|p_-$$\uparrow$$\rangle$,
the energies (\ref{eqn:egain})
can be further mapped
onto the anisotropic Heisenberg model with pseudospin $1/2$ \cite{KO2_condmat}:
\begin{equation}
\hat{H}_{\rm Heis} = - \frac{1}{2} \sum_{{\bf R}{\bf R}'}
\left\{
\left( \hat{\tau}_{\bf R}^x \hat{\tau}_{{\bf R}'}^x + \hat{\tau}_{\bf R}^y \hat{\tau}_{{\bf R}'}^y \right)
J^\perp_{{\bf R}{\bf R}'} +
\hat{\tau}_{\bf R}^z \hat{\tau}_{{\bf R}'}^z J^\parallel_{{\bf R}{\bf R}'}
\right\},
\label{eqn:AHeisenberg}
\end{equation}
where
$$
\hat{\tau}_x =
\left(
\begin{array}{rr}
0 & 1 \\
1 & 0 \\
\end{array}
\right),
\quad
\hat{\tau}_y =
\left(
\begin{array}{rr}
0 & -i \\
i &  0 \\
\end{array}
\right),
\quad
\textrm{and}
\quad
\hat{\tau}_z =
\left(
\begin{array}{rr}
1 &  0 \\
0 & -1 \\
\end{array}
\right)
$$
are the Pauli matrices in the basis
of $|p_+$$\uparrow \rangle$ and $|p_-$$\downarrow \rangle$ orbitals.
The parameters
of superexchange interactions,
$J^\perp_{{\bf R}{\bf R}'}$ and $J^\parallel_{{\bf R}{\bf R}'}$
can be expressed through the pair-interaction energies
${\cal T}_{{\bf R}{\bf R}'} \equiv {\cal T}(\alpha_{\bf R},\alpha_{{\bf R}'})$
by considering different combinations of $\alpha_{\bf R}$ and $\alpha_{{\bf R}'}$.\footnote{
Namely,
by denoting as
$| \tau_x^\pm \rangle = ( |p_+$$\uparrow$$\rangle$$\pm$$|p_-$$\downarrow$$\rangle )/\sqrt{2}$,
$| \tau_y^\pm \rangle = ( |p_+$$\uparrow$$\rangle $$\pm$$ i|p_-$$\downarrow$$\rangle )/\sqrt{2}$,
$| \tau_z^+ \rangle = |p_+$$\uparrow$$\rangle$, and $| \tau_z^- \rangle = |p_-$$\downarrow$$\rangle$
the pseudospin states corresponding to the positive ($+$) and negative ($-$) directions
parallel to the $x$-, $y$-, and $z$-exes, the parameters of superexchange interactions can be found as
$2J^\perp_{{\bf R}{\bf R}'} =
{\cal T}(\tau_{{\bf R}x}^+,\tau_{{{\bf R}'}x}^-) - {\cal T}(\tau_{{\bf R}x}^+,\tau_{{{\bf R}'}x}^+) =
{\cal T}(\tau_{{\bf R}y}^+,\tau_{{{\bf R}'}y}^-) - {\cal T}(\tau_{{\bf R}y}^+,\tau_{{{\bf R}'}y}^+)$, and
$2J^\parallel_{{\bf R}{\bf R}'} =
{\cal T}(\tau_{{\bf R}z}^+,\tau_{{{\bf R}'}z}^-) - {\cal T}(\tau_{{\bf R}z}^+,\tau_{{{\bf R}'}z}^+)$.
}
They are summarized in Table \ref{tab:KO2}.
\begin{table}[h!]
\caption{\label{tab:KO2} Parameters of
Heisenberg model (in meV) for the \textit{bct} phase of KO$_2$
in different temperature regimes \protect\cite{KO2_condmat}.
$J^\parallel_{{\bf R}{\bf R}'}$ and $J^\perp_{{\bf R}{\bf R}'}$ correspond to the
low-temperature limit, while $\bar{J}_{{\bf R}{\bf R}'}$
corresponds to the high-temperature limit.
The vectors separating two oxygen molecules,
${\bf b} = {\bf R}' - {\bf R}$,
are denoted as
${\bf b}_1$$=$$(0,a,0)$,
${\bf b}_2$$=$$(\frac{a}{2},\frac{a}{2},\frac{c}{2})$,
${\bf b}_3$$=$$(a,a,0)$, and
${\bf b}_4$$=$$(0,0,c)$ (see Figure \protect\ref{fig.KO2_DOS} for the notations).}
\begin{indented}
\item[]\begin{tabular}{@{}crrr}
\br
vector              & $J^\parallel_{{\bf R}{\bf R}'}$ & $J^\perp_{{\bf R}{\bf R}'}$  &  $\bar{J}_{{\bf R}{\bf R}'}$ \\
\mr
${\bf b}_1$         & $-$$0.43$     &  $-$$1.15$ &  $-$$0.47$ \\
${\bf b}_2$         &    $1.31$     &  $-$$1.09$ &  $-$$1.01$ \\
${\bf b}_3$         & $-$$0.02$     &  $-$$0.05$ &  $-$$0.02$ \\
${\bf b}_4$         & $-$$0.07$     &  $-$$0.07$ &  $-$$0.01$ \\
\br
\end{tabular}
\end{indented}
\end{table}
The second-neighbor coupling
$J^\parallel_{{\bf R}{\bf R}+{\bf b}_2}$ stabilizes the easy-axis ferromagnetic state.
Other interactions are antiferromagnetic and frustrated on the \textit{bct} lattice.
Then, the Curie temperature can be estimated using the renormalized spin-wave theory \cite{Tyablikov}.
It yields $T_C \sim 70$ K, which does not seem to be fully consistent with the
experimental data, because no long-range magnetic order has been observed
down to the very low temperature \cite{Ziegler,Kanzig}.
However, there is also a number of factors, which may further affect this theoretical picture \cite{KO2_condmat}.
One is the deformation of the relativistic spin-orbitals
$|p_+$$\uparrow$$\rangle$ and $|p_-$$\downarrow$$\rangle$ caused
by superexchange interactions. Another one is the reorientation of the oxygen molecules
(which probably takes place but not precisely known from early experimental studies \cite{Kanzig,Labhart}).
For example, one can suggest that above 12 K, the reorientation of the oxygen molecules
is not particularly large and leads only to some quantitative redefinition of parameters
of the anisotropic Heisenberg model \ref{eqn:AHeisenberg},
which systematically decreases $T_C$ \cite{KO2_condmat}.
However, it seems that below 12 K the situation changes dramatically. Apparently,
the transition to the new crystallographic phase is accompanied by
the large reorientation
of the oxygen molecules, which
not only quenches the orbital magnetic moment, but also
plays a decisive role in the
formation of the long-range antiferromagnetic order \cite{KO2_condmat}.
The complete quantitative theory describing these reorientation effects
is missing at the present stage but would certainly be an interesting
step to develop in the future.

  Nevertheless,
the intrigue of KO$_2$ is that there is another region of antiferromagnetic interactions,
which is
realized in the high-temperature \textit{bct} phase, as it is clearly
manifested in the behavior of
inverse magnetic susceptibility \cite{Ziegler,Kanzig}. The origin of this
``high-temperature antiferromagnetism'' may be directly
related with the orbital disorder.
Indeed,
in the \textit{high-temperature limit} $k_BT$$\gg$$\xi$, the thermal fluctuations will
eventually destroy the relativistic coupling between spin and orbital degrees of freedom.
Therefore, it is reasonable to assume that all orbital configurations are realized with
equal probabilities.
In this case,
the spin system becomes fully isotropic, and the
parameters of the spin Hamiltonian,
$2 \bar{J}_{{\bf R}{\bf R}'} = \bar{\cal T}_{{\bf R}{\bf R}'}^{\uparrow \downarrow} -
\bar{\cal T}_{{\bf R}{\bf R}'}^{\uparrow \uparrow}$,
can be derived by averaging the energies (\ref{eqn:egain}) for antiferromagnetic ($\uparrow$$\downarrow$)
and ferromagnetic ($\uparrow$$\uparrow$) configurations of spins in each bond over all
combinations of orbital variables \cite{KugelKhomskii,KO2_condmat}. The orbital disorder
gives rise to antiferromagnetic interactions
$\bar{J}_{{\bf R}{\bf R}'}$, which are also summarized in Table \ref{tab:KO2}.
The corresponding Curie-Weiss temperature can be
estimated using renormalizes spin-wave theory. It yields $T_{CW} \sim -$$80$ K,
which is
comparable with the experimental data \cite{Ziegler,Kanzig}.

  In summary,
the magnetic properties of KO$_2$ provide an example of
spin-orbital superexchange physics realized in the molecular solid.
The properties largely depend on the orbital state of the O$_2^-$ ions.
In the paramagnetic region,
the character of intermolecular interactions gradually
changes from mainly ferromagnetic,
and driven by the relativistic spin-orbit interaction,
to antiferromagnetic, and
corresponding to the picture of an independent spin
and orbital disorder.
All these features seem to be consistent with the experimental behavior of
KO$_2$.
Finally, it is important to note that
the geometry of molecular orbitals
can be used as an additional degree of freedom, which controls
the properties of superexchange interactions.
Definitely, it
adds many new functionalities
into the classical problem of superexchange, which deserve further
exploration.

\section{\label{sec:Summary}Summary and Concluding Remarks}

  The goal of this review article was to outline the main
ideas and scopes of new developing direction
for the strongly correlated systems
which can be called
as
the ``realistic modeling''.
The primary purpose of this project
is to make a
bridge
between methods of first-principles
electronic structure calculations, based on the density-functional theory,
and many-body models, describing properties of strongly correlated systems
in terms of a limited number of the most relevant model parameters
and including information about all
remaining electronic structure implicitly, through the
renormalization of these model parameters.

  The realistic modeling has all potentials to become a useful tool of
electronic structure calculations for the strongly correlated systems.
It is true that we still have to rely on a number of approximations, particularly
in the process of calculation of screened Coulomb interactions, which are inevitable
in any approach.
However, it is very important that, apart from these approximations,
the entire procedure is parameter-free.
Namely, we do not have to deal anymore with
adjustable parameters and we do need to resolve numerous ambiguities
with the choice of these parameters.
Instead, the realistic modeling brings the state of the discussion to a qualitatively
new level: how to improve the approximations used for the definition and calculation
of these model parameters.

  One undeniable advantage of realistic modeling is that
it allows us to combine the accuracy and predictable power of first-principles
electronic structure
calculations with flexibility and insights of the model analysis.
This idea was illustrated on the series of examples,
for which we were able to consider
the whole chain of actions
starting from
conventional electronic structure calculations in the local-density approximation,
which was followed by the
construction of an appropriate low-energy model, motivated by these calculations,
and finally --
by the
solution of this model
and by
the analysis of properties of strongly correlated systems in terms of
these model categories and trends.
The first applications are indeed very encouraging and we would like to hope that in
future the ideas of
realistic modeling will
continue to develop in order to become a powerful
tool for theoretical analysis, design, and control
of properties of strongly correlated materials.

\ack
This work is partly supported by Grant-in-Aids
for Scientific Research in Priority Area ``Anomalous Quantum Materials''
from the Ministry of Education, Culture, Sport, Science and Technology of Japan.

\appendix
\section{\label{sec:appendixA}Properties of Coulomb Interactions
in the Atomic Limit}

  In the atomic limit, the $5$$\times$$5$$\times$$5$$\times$$5$
matrix $\hat{u}$ of interactions between $d$ electrons
is totally specified by three radial Slater's integral $F^0$,
$F^2$, and $F^4$, which are related with
the parameters
$u$ and $j$, obtained in the constrained DFT, by the identities
\begin{equation}
u = F^0
\label{eqn:appendixdu}
\end{equation}
and
\begin{equation}
j = \frac{1}{14} \left( F^2 + F^4 \right).
\label{eqn:appendixdj1}
\end{equation}

  In order to derive all three Slater's integrals by knowing
only two parameters $u$ and $j$, extracted from
the constrained DFT, one can use
the additional relation
\begin{equation}
F^4/F^2 \simeq 0.63,
\label{eqn:appendixF4F2}
\end{equation}
which holds approximately
in the atomic limit.

  We would like to emphasize that the parameter $j$,
defined as (\ref{eqn:appendixdj1}),
is the measure of the exchange splitting
corresponding to the
spherically averaged electron densities
for
the majority- and
minority-spin states in LDA.
In the literature one can find other definitions of the
exchange integrals, which sometimes cause certain confusion.
For example, Mizokawa and Fujimori \cite{MizokawaFujimori} defined
the exchange integral as
\begin{equation}
j_{\rm MF} = \frac{5}{2}B + C,
\label{eqn:appendixdj2}
\end{equation}
in terms of the two Racah parameters:
$B=(9F^2-5F^4)/441$ and $C=5F^4/63$. Then, it is easy to verify
the validity of the following relation between $j$ and $j_{\rm MF}$:
\begin{equation}
j_{\rm MF} = \frac{5}{7}j \approx 0.71 j.
\label{eqn:appendixdj3}
\end{equation}
Thus, $j_{\rm MF}$ \textit{is always smaller than} $j$.

  Alternatively, one can define parameters of Coulomb
interactions only for the $t_{2g}$ manifold. There are three types
of interactions, which are sometimes called as the
Kanamori parameters \cite{Kanamori}:
\begin{equation}
u_t =
\int d{\bf r} \int d{\bf r}' \tilde{W}_{xy}^\dagger({\bf r}) \tilde{W}_{xy}({\bf r})
|{\bf r}-{\bf r}'|^{-1} \tilde{W}_{xy}^\dagger({\bf r}') \tilde{W}_{xy}({\bf r}'),
\label{eqn:appendixtuintra1}
\end{equation}
\begin{equation}
u_t' =
\int d{\bf r} \int d{\bf r}' \tilde{W}_{xy}^\dagger({\bf r}) \tilde{W}_{xy}({\bf r})
|{\bf r}-{\bf r}'|^{-1} \tilde{W}_{yz}^\dagger({\bf r}') \tilde{W}_{yz}({\bf r}'),
\label{eqn:appendixtuinter1}
\end{equation}
and
\begin{equation}
j_t =
\int d{\bf r} \int d{\bf r}' \tilde{W}_{xy}^\dagger({\bf r}) \tilde{W}_{yz}({\bf r})
|{\bf r}-{\bf r}'|^{-1} \tilde{W}_{xy}^\dagger({\bf r}') \tilde{W}_{yz}({\bf r}').
\label{eqn:appendixtj1}
\end{equation}
In the atomic limit,
they can be expressed in terms of radial Slater's integrals as
\begin{equation}
u_t = F^0 + \frac{4}{49}F^2 + \frac{4}{49}F^4,
\label{eqn:appendixtuintra2}
\end{equation}
\begin{equation}
u_t' = F^0 - \frac{2}{49}F^2 - \frac{4}{441}F^4,
\label{eqn:appendixtuinter2}
\end{equation}
and
\begin{equation}
j_t = \frac{3}{49}F^2 + \frac{20}{441}F^4.
\label{eqn:appendixtj2}
\end{equation}
Other types of interactions between $t_{2g}$ electrons vanish.
In fact, there are only two independent interactions because
$u_t$, $u_t'$, and $j_t$ are related by the identity:
\begin{equation}
u_t = u_t' + 2j_t.
\label{eqn:appendixidentity}
\end{equation}
It is also straightforward to show that
\begin{equation}
u_t = u + \frac{8}{7}j.
\label{eqn:appendixtuintra3}
\end{equation}
The parameter $j_t$ can be expressed through $j$ using the
approximate relation between Slater's integrals (\ref{eqn:appendixF4F2}),
which yields
\begin{equation}
j_t \approx 0.77 j.
\label{eqn:appendixtj3}
\end{equation}
Thus, generally we have the following inequality for the exchange integrals
defined in three different ways:
\begin{equation}
j_{\rm MF} < j_t < j.
\label{eqn:appendixtj4}
\end{equation}

  After taking into account the RPA screening, the parameters
$u_t$, $u_t'$, and $j_t$ correspond to the parameters
$\mathcal{U}$, $\mathcal{U}'$, and $\mathcal{J}$
considered in Sections \ref{sec:SrVO3_U} and \ref{sec:RMO3_U}.

  For the $f$-shell, the $7$$\times$$7$$\times$$7$$\times$$7$ matrix $\hat{u}$
can be reconstructed from $F^0$,
$F^2$, $F^4$, and $F^6$, by using the following identities
in the atomic limit \cite{PRB94b}:
\begin{equation}
u = F^0,
\label{eqn:appendixfu}
\end{equation}
\begin{equation}
j = \frac{1}{3} \left( \frac{2}{15}F^2 + \frac{1}{11}F^4 + \frac{50}{429}F^4 \right),
\label{eqn:appendixfj1}
\end{equation}
$F^4/F^2 \simeq 451/675$, and $F^6/F^2 \simeq 1001/2025$.

\section{\label{sec:appendixB}Correlations Between $2p$-$3d$ Hybridization
and Screening of Coulomb Interactions in the $t_{2g}$ band of
transition-metal perovskite oxides}

  In this appendix we derive some approximate expression for the static
RPA screening of Coulomb interactions in the $t_{2g}$ band of
transition-metal oxides by the oxygen $2p$ band.

  First, we assume that the band dispersion is considerably smaller than
the charge-transfer energy $\Delta_{\rm CT}$,
which is the energy difference between the centers of gravity of the
oxygen $2p$ band and the transition-metal $t_{2g}$ and $e_g$ bands
(Figure \ref{fig.PerovskiteDOS}). Then, for the static ($\omega = 0$)
screening caused by the oxygen $2p$ band, the denominator of the
polarization function (\ref{eqn:Polarization_Function}) can be
replaced by $\Delta_{\rm CT}$ and one can perform separate summation
over the occupied and empty states.

  Then, we focus on the self-screening caused by the atomic $3d$ states,
which contribute to the oxygen $2p$ band due to the hybridization effects
and consider the matrix elements of (\ref{eqn:Polarization_Function})
in the basis of atomic $3d$ orbitals. These matrix elements can be
expressed through the density matrices of the $3d$ states calculated
separately in the occupied oxygen $2p$ band and in the empty part of the
spectrum (correspondingly $\hat{n}^o$ and $\hat{n}^e$). Finally, we
assume that $n^{o(e)}_{\alpha \beta} \sim N^{o(e)} \delta_{\alpha \beta}$
($N^o$ and $N^e$ being the total number of $3d$ electrons in the oxygen
$2p$ band and in the empty part of the spectrum, respectively)
and consider the diagonal matrix elements of (\ref{eqn:Polarization_Function}),
which mainly contribute to the screening of the intraorbital Coulomb
interaction ${\cal U}$. In this case, one can to derive the
following (approximate) expression for the diagonal matrix elements
of the polarization:
$$
{\cal P} \sim \frac{N^o N^e}{\Delta_{\rm CT}}.
$$
Moreover, since the total number of $3d$ electrons is conserved,
$N^o$$+$$N^e$ is a constant, which depends only on the number of
$3d$ electrons in the occupied part of the $t_{2g}$ band.
Therefore, it is reasonable to expect the following rules
for compounds having the same number of $t_{2g}$ electrons:
\begin{enumerate}
\item
the larger is the weight of the transition-metal $3d$ states
in the oxygen $2p$ band, $N^o$, the stronger is the screening
of Coulomb interactions in the $t_{2g}$ band;
\item
$N^o$ is controlled by the hybridization between oxygen $2p$
and transition-metal $3d$ states, which in turn depends on the
crystal distortion (particularly, the buckling of the
Ti-O-Ti and V-O-V bonds). Therefore, stronger distortion
will tend to reduce the screening;
\item
smaller $\Delta_{\rm CT}$ in the case of SrVO$_3$ will
additionally increase the screening of Coulomb interactions
in the $t_{2g}$ band.
\end{enumerate}

  All these trends are clearly seen in Table \ref{tab:AppendixB},
which shows the correlation between $N^o$ and ${\cal U}$
in different transition-metal perovskite oxides.
\begin{table}[h!]
\caption{Correlation between
number of $3d$ electrons in the oxygen $2p$ band ($N^o$)
and the value of screened Coulomb interaction (${\cal U}$, in eV)
in the transition-metal $t_{2g}$ band.
The symbols `c', `o', and `m' stand for the cubic, orthorhombic,
and monoclinic structure, respectively.
The positions of the transition-metal sites are
explained in Figure \protect\ref{fig.Perovskites_CFandHoppings}.
Generally, the site `1' is located in less distorted environment while the
site `3' is located in more distorted environment.}
\label{tab:AppendixB}
\begin{indented}
\item[]\begin{tabular}{@{}rccccc}
\br
 compound  & phase & site & $N^o$  & ${\cal U}$ \\
\mr
 YTiO$_3$  &  o    & 1    & $0.66$ & $3.45$     \\
 LaTiO$_3$ &  o    & 1    & $0.73$ & $3.20$     \\
 SrVO$_3$  &  c    & 1    & $1.19$ & $2.53$     \\
 YVO$_3$   &  o    & 1    & $0.74$ & $3.27$     \\
 YVO$_3$   &  m    & 1    & $0.76$ & $3.19$     \\
           &       & 3    & $0.72$ & $3.26$     \\
 LaVO$_3$  &  m    & 1    & $0.81$ & $3.11$     \\
           &       & 3    & $0.80$ & $3.12$     \\
\br
\end{tabular}
\end{indented}
\end{table}
For example, more distorted Y$B$O$_3$ compounds are characterized by
smaller $N^o$ and, therefore,
by somewhat larger
${\cal U}$ in comparison with the less distorted La$B$O$_3$ compounds.
The same tendency holds for different transition-metal sites in the
monoclinic structure: the sites with more distorted environment
have large ${\cal U}$ and vice versa.

\end{document}